\documentclass[11pt,a4paper]{article}
\pdfoutput=1

\usepackage{jcappub}
\usepackage{slashed}

\usepackage{multirow}
\usepackage{amssymb}
\usepackage{ulem}
\usepackage{cancel}
\usepackage{verbatim}
\usepackage{caption} 
\usepackage{subcaption}
\usepackage[utf8]{inputenc}
\usepackage{xcolor}
\usepackage{amsmath}
\usepackage{mathtools}
\usepackage{mathrsfs}
\usepackage{array}
\newcolumntype{L}[1]{>{\raggedright\let\newline\\\arraybackslash\hspace{0pt}}m{#1}}
\newcolumntype{C}[1]{>{\centering\let\newline\\\arraybackslash\hspace{0pt}}m{#1}}
\newcolumntype{R}[1]{>{\raggedleft\let\newline\\\arraybackslash\hspace{0pt}}m{#1}}

\allowdisplaybreaks

\newcommand{\be}{\begin{equation}}
\newcommand{\ee}{\end{equation}}

\definecolor{darkred}{rgb}{0.5,0,0}
\definecolor{darkgreen}{rgb}{0,0.5,0}
\definecolor{darkblue}{rgb}{0,0,0.5}

\usepackage{hyperref}
\hypersetup{ colorlinks,
linkcolor=darkblue,
filecolor=darkgreen,
urlcolor=darkred,
citecolor=darkblue }

\newcommand{\inspire}[1]{\href{http://inspirehep.net/search?p=find+J+#1}
 {{\color{black}[{\color{blue} {\small in}SPIRE}]}}}
\newcommand{\book}[1]{\href{http://inspirehep.net/search?p=#1}
 {{\color{black}[{\color{blue} {\small in}SPIRE}]}}}

\newcommand{\inspired}[1]{\href{http://inspirehep.net/search?p=#1}
 {{\color{black}[{\color{blue} {\small in}SPIRE}]}}}

\newcommand{\CLASS}{\textsc{class}}

\begin{document}

\title{Invisible neutrino decay in precision cosmology}

\date{\today}

\author[a]{Gabriela Barenboim,}
\author[b]{Joe Zhiyu Chen,}
\author[c]{Steen Hannestad,}
\author[a]{Isabel~M.~Oldengott,}
\author[c]{Thomas Tram,}
\author[b]{and Yvonne~Y.~Y.~Wong}
\emailAdd{gabriela.barenboim@uv.es, zhiyu.chen@unsw.edu.au, sth@phys.au.dk, isabel.oldengott@uv.es, thomas.tram@phys.au.dk, yvonne.y.wong@unsw.edu.au}

\affiliation[a]{Departament de Fisica Te\`{o}rica and IFIC, CSIC-Universitat de Val\`{e}ncia, 46100 Burjassot, Spain}
\affiliation[b]{Sydney Consortium for Particle Physics  and Cosmology, School of Physics, The University of New South Wales, Sydney NSW 2052, Australia}
\affiliation[c]{Department of Physics and Astronomy, University of Aarhus, DK-8000 Aarhus C, Denmark}

\abstract{We revisit the topic of invisible neutrino decay in the precision cosmological context, via a first-principles approach to understanding the cosmic microwave background and large-scale structure phenomenology of such a non-standard physics scenario.  Assuming an effective Lagrangian in which a heavier standard-model neutrino $\nu_H$ couples to a lighter one $\nu_l$ and a massless scalar particle $\phi$ via a Yukawa interaction, we derive from first principles the complete set of Boltzmann equations, at both the spatially homogeneous and the first-order inhomogeneous levels,  for the phase space densities of $\nu_H$, $\nu_l$, and $\phi$ in the presence of the relevant decay and inverse decay processes.  With this set of equations in hand, we perform a critical survey of recent works on cosmological invisible neutrino decay in both limits of decay while $\nu_H$ is ultra-relativistic and non-relativistic. Our two main findings are: (i)~in the non-relativistic limit, the effective equations of motion used to describe perturbations in the neutrino--scalar system in the existing literature formally violate momentum conservation and gauge invariance, and (ii)~in the ultra-relativistic limit, exponential damping of the anisotropic stress does not occur at the commonly-used rate $\Gamma_{\rm T} =(1/\tau_0) (m_{\nu H}/E_{\nu H})^3$, but at a rate $\sim (1/\tau_0) (m_{\nu H}/E_{\nu H})^5$.  Both results are model-independent. The impact of the former finding  on the cosmology of invisible neutrino decay is likely small.
The latter, however, implies a significant revision of the cosmological limit on the neutrino lifetime~$\tau_0$ from  $\tau_0^{\rm old} \gtrsim 1.2 \times 10^9\, {\rm s}\, (m_{\nu H}/50\, {\rm meV})^3$ to
  $\tau_0 \gtrsim (4 \times 10^5 \to 4 \times 10^6)\, {\rm  s}\,  (m_{\nu H}/50 \, {\rm meV})^5$.}

\begin{flushright}
	{\large \tt CPPC-2020-06}
\end{flushright}	


\arxivnumber{2011.01502}

\maketitle   

\flushbottom
%%%%%%%%%%%%%%%%%%%%%%%%%%%%%%%%%%%%%%%%%%%%%%%%%%%%%%%%%%

\section{Introduction}
Among the many unexplained phenomena in fundamental physics, the observation of neutrino oscillations is perhaps the most direct hint towards the existence of physics beyond the standard model of particle physics~\cite{Zyla:2020zbs}. While neutrino oscillations imply that neutrinos are massive particles, the standard model assumes them to be massless, a direct consequence of the assumption that only left-handed neutrinos exist.  Thus, any attempt to incorporate neutrino masses into the theory must require the existence of new particles and/or new interactions within the neutrino sector. 

Meanwhile, cosmology, particularly precision cosmological observations in the past two decades, has proven to be a very useful tool in probing the fundamental properties of neutrinos and hence in shedding light on the neutrino sector. 
The most prominent example thereof are cosmological constraints on the absolute neutrino mass scale; measurements of the cosmic microwave background (CMB) anisotropies and the large-scale structure (LSS) of the universe
currently constrain the neutrino mass sum to $\sum m_{\nu} \lesssim 0.12$~eV (95\% C.L.) in a one-parameter extension to the  $\Lambda$CDM model~\cite{Aghanim:2018eyx}, about a factor of 30  stronger than is possible with present laboratory experiments~\cite{Aker:2019uuj}.  The power of cosmological probes to constrain neutrino masses derives mainly from their sensitivity to a low, sub-eV energy scale, as well as the long time scale over which non-relativistic neutrino kinematics can exert its influence on the universe's evolution.
The same feature can also be exploited to put some of the most constraining bounds on another fundamental neutrino parameter, the neutrino lifetime~$\tau_0$.

Neutrino decay is a typical phenomenon of many theories that contain non-standard neutrino interactions. Radiative decay scenarios, 
which count at least one photon in the final state,
can be constrained using CMB spectral distortions~\cite{Mirizzi:2007jd,Aalberts:2018obr} and the 21~cm hydrogen line~\cite{Chianese:2018luo}. Harder to probe however are scenarios of \textit{invisible} neutrino decays, as the decay products are now a lower-mass neutrino plus another light particle that often interacts only with the neutrino sector. One such example are Marjoron models, where a spontaneously broken global $U(1)_{B-L}$ symmetry leads to a new Goldstone boson that couples to neutrinos~\cite{Gelmini:1980re,Chikashige:1980ui,Schechter:1981cv}. 
Another possibility arises within gauged $L_i-L_j$ models~\cite{Dror:2020fbh}, where a new vector boson now plays the role of the light particle. 
Invisible neutrino decays have also been proposed as an avenue to relax cosmological neutrino mass bounds~\cite{Escudero:2020ped}. 

Non-cosmological constraints on the neutrino lifetime based upon invisible decays are generally very weak, mainly because all directly detectable neutrinos are ultra-relativistic, so that their decay rate in the laboratory frame is strongly Lorentz-suppressed. Analyses of the  neutrino disappearance rate in JUNO and KamLAND+JUNO give respectively a lower limit of  \(\tau_0/m_{\nu} > 7.5 \times 10^{-11} (\text{s/eV})\)~\cite{Porto-Silva:2020gma} for a normal mass hierarchy and  \(\tau_0/m_{\nu} > 1.1 \times 10^{-9} (\text{s/eV})\)~\cite{Porto-Silva:2020gma} for an inverted mass hierarchy.   The future experiment DUNE is expected to yield  \(\tau_0/m_\nu > (1.95 \to  2.6) \times 10^{-10} (\text{s/eV})\)~\cite{Coloma:2017zpg}.

Remarkably, lower limits on the neutrino lifetime from measurements of the CMB  anisotropies have been claimed to lie in the ballpark $\tau_0 \gtrsim \mathcal{O}(10^{10})  \, \text{s} \, (m_{\nu}/50 \, \text{meV})^3$~\cite{Escudero:2019gfk,Basboll:2008fx,Hannestad:2005ex,Archidiacono:2013dua}, i.e., some twenty orders of magnitude stronger than the aforementioned laboratory bounds.   These limits follow from a commonly-used argument that for sub-eV neutrino masses and  sufficiently large coupling, neutrino decay and its inverse process will create a tightly-coupled and locally isotropic relativistic fluid of the mother and daughter particles around the CMB formation epoch.  The resulting loss of anisotropic stress in such a scenario is incompatible with CMB observations, which can in turn be translated into a bound on the neutrino lifetime.  The same  line of argument has also been invoked to constrain neutrino self-interactions~\cite{Lancaster:2017ksf,Oldengott:2017fhy,Kreisch:2019yzn,Barenboim:2019tux}, which has gained some recent interest  in relation to the  Hubble tension.

In this construct, the crucial ingredient is the rate at which anisotropic stress is lost due to the decay and its inverse process in the relativistic limit, 
and, as a means to constrain the neutrino lifetime, how this loss rate relates to fundamental parameters of the underlying theory.   So far 
this rate has always been estimated based on heuristic arguments.  In fact, it has never been demonstrated rigorously that the exponential damping of anisotropic stress assumed in the analyses of~\cite{Escudero:2019gfk,Basboll:2008fx,Hannestad:2005ex,Archidiacono:2013dua}
is even a phenomenology of relativistic neutrino decay or its most constraining observable  in the cosmological context.  
Likewise, for neutrino decay in the non-relativistic limit, we have noticed a number of questionable approximations in the existing literature that deserve  closer scrutiny.

In this work, we address these issues using a first-principles approach, where we systematically fold in the effects of invisible neutrino decay into the framework of linear cosmological perturbation theory via a collisional integral in the Boltzmann equations for the decaying neutrino and its decay products.  We derive effective equations of motion for the decay system at both the homogeneous  and the inhomogeneous level compatible for use in a linear cosmological Boltzmann code such as \CLASS{}~\cite{Blas:2011rf}.

With this system of equations in hand, we perform a critical survey of recent works on cosmological  invisible neutrino decay~\cite{Escudero:2019gfk,Basboll:2008fx,Hannestad:2005ex,Archidiacono:2013dua,Kaplinghat:1999xy,Chacko:2019nej,Chacko:2020hmh} in both the relativistic and non-relativistic limits, 
 and point out several in our view ill-justified approximations previously applied to reduce the complexity of the numerical problem.  Two particularly relevant results are: (i)~in the non-relativistic limit,  we find some simplified equations of motion in the existing literature to formally violate momentum conservation as well as  gauge invariance,  and (ii)~in the relativistic limit,  an exponential damping of the anisotropic stress at a rate $\Gamma_{\rm T}  \sim (1/\tau_0)(m_\nu/E_\nu)^3$
is not a consistent phenomenology of the system; rather, the damping rate is a significantly smaller $\sim (1/\tau_0)(m_\nu/E_\nu)^5$.
While the former finding is likely to have only a small impact on the cosmology of invisible neutrino decay in the non-relativistic limit, the latter suggests a new and  considerably relaxed lower bound on the neutrino lifetime of  $\tau_0 \gtrsim (4 \times 10^5 \to 4 \times 10^6)\, \text{s} \, (m_{\nu}/50 \, \text{meV})^5$.

The rest of the paper is organised  as follows. We describe in section~\ref{sec:Model} our model of invisible neutrino decay. Section \ref{sec:PreviousWorks} summaries existing works on invisible neutrino decay in the cosmological context
and outlines our points of critique. In section~\ref{sec:OurFormalism}, we present for the first time the background Boltzmann equations and first-order Boltzmann hierarchies in the presence of invisible neutrino decay for the mother and daughter particles.  Numerical solutions to the background equations are shown in section~\ref{sec:numerical}.   We discuss and contrast our results with existing ones in section~\ref{sec:Comparison}, and, where applicable, derive new constraints on~$\tau_0$.  We conclude in 
 section~\ref{sec:Conclusion}.  Technical details of our calculations are reported in four appendices.

%%%%%%%%%%
%%%%%%%%%%%
%%%%%%%%%%%%%%%%%

\section{The physical system}
\label{sec:Model}

We study non-standard neutrino interactions described by the effective Lagrangian
\begin{equation}
\mathcal{L}_{\text{int}} = \mathfrak{g}_{ij} \bar{\nu}_i \nu_j \phi.
\label{eq:Lagrangian}
\end{equation}
Here, $\phi$ is a new, light scalar particle which  for our purposes shall be assumed to be {\it massless}; the indices $i,j$ label mass eigenstates; and $\mathfrak{g}_{ij}$ is the coupling matrix which, for simplicity, we assume to be universal, i.e.,
\begin{equation}
\mathfrak{g}_{ij}= \mathfrak{g} 
\begin{pmatrix}
1 & 1 & 1 \\
1 & 1 & 1 \\
1 & 1 & 1 
\end{pmatrix}.
\label{eq:coupling}
\end{equation} 
Kinematics permitting, interactions enabled by the Lagrangian~\eqref{eq:Lagrangian}  include neutrino decay $\nu_H \leftrightarrow \nu_l + \phi$ and  and its inverse process, where the subscripts ``$H$'' and ``$l$'' stand respectively for heavy and light, as well as the $2 \leftrightarrow 2$ processes $\nu \nu \leftrightarrow \nu \nu$,  $\nu \phi \leftrightarrow \nu \phi$, and $\nu \nu \leftrightarrow \phi \phi$.%

\begin{figure}
	\centering
	\includegraphics[width=\textwidth]{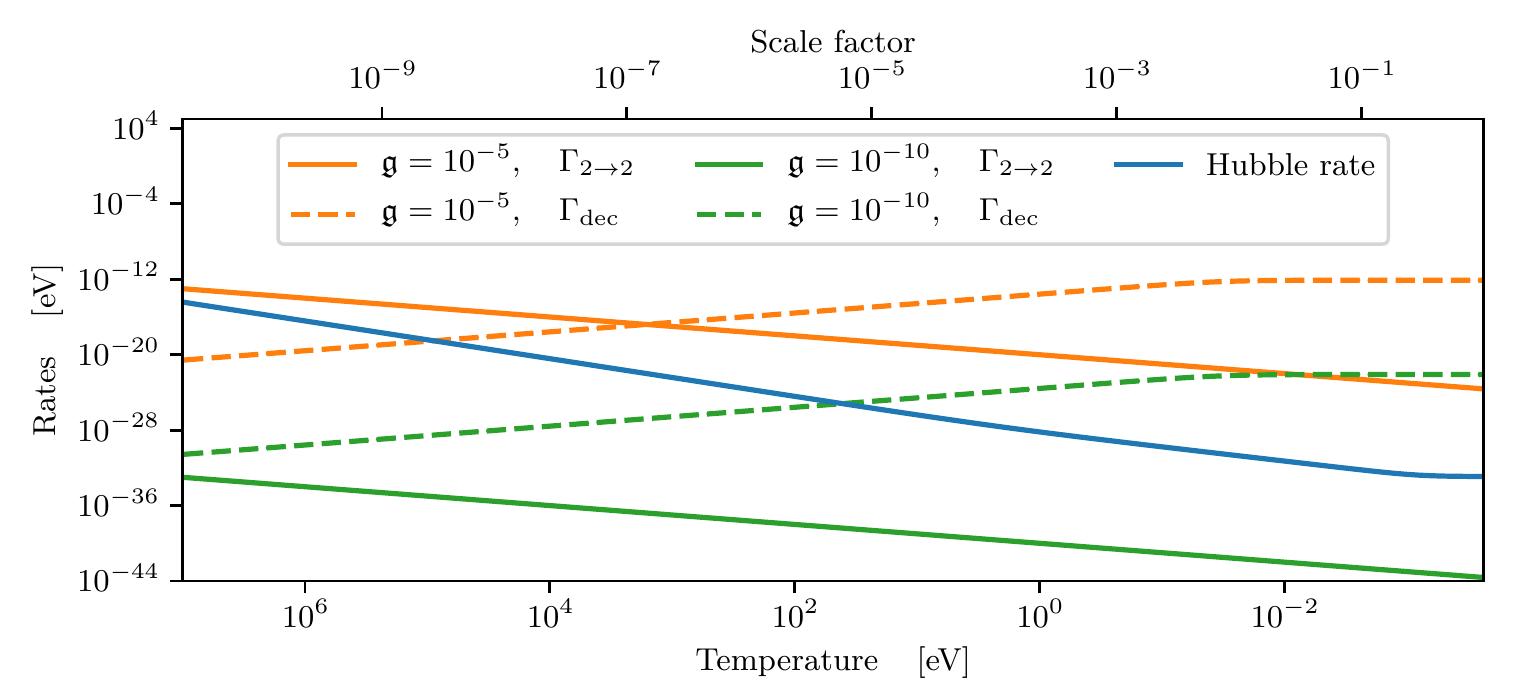}
	\caption{Schematic plot of the various relevant rates as functions of the universe's temperature~$T$. The solid blue line shows the Hubble expansion rate $H$, assuming a spatially flat universe with matter density $\Omega_{\text{m}}=0.321$ and hence matter-radiation equality at $z\simeq 3000$; the additional solid lines denote the $2\rightarrow2$ scattering rate $\Gamma_{2\rightarrow 2}$ as given by equation~\eqref{eq:2to2}  for coupling strengths $\mathfrak{g} = 10^{-5}$  (orange) and $10^{-10}$ (green); and the dashed lines depict the Majorana neutrino decay rate $\Gamma_{\text{dec}}$ of equation~(\ref{eq:decayrate}) assuming $m_{\nu}=0.1$ eV, evaluated at a typical energy $E_{\nu}=(m_{\nu}^2+(3T)^2)^{1/2}$ again for  $\mathfrak{g} = 10^{-5}$ (orange) and $10^{-10}$ (green).	}
	\label{fig:Rates}
\end{figure}

We work with the assumption that $\phi$ has no interaction other than those described by the Lagrangian~\eqref{eq:Lagrangian} and coupling matrix~\eqref{eq:coupling}, so that to order $\mathfrak{g}^4$ production of a $\phi$ population in the early universe can proceed only via neutrino annihilation into $\phi$ pairs and/or neutrino decay.
The former, $2 \rightarrow 2$ process has an interaction rate per particle, $\Gamma_{2 \rightarrow 2}$, scaling with the particle population's temperature $T$ as
\begin{equation}
\label{eq:2to2}
\Gamma_{2 \rightarrow 2} \sim \mathfrak{g}^4 T,
\end{equation}
assuming  $T$ exceeds the particle and mediator masses.
Relative to the Hubble expansion rate $H$ ($\sim T^2/m_{\rm planck}$ during radiation domination), figure~\ref{fig:Rates} shows that these $2 \to 2$ processes are typically out of equilibrium (i.e., $\Gamma_{2 \to 2} \ll H$) at high temperatures but, depending on the value of~$\mathfrak{g}$, may reach equilibrium (i.e., $\Gamma_{2 \to 2} \gg H$) at a later time as the universe cools --- we shall label this second state of affairs ``recoupling''.
Once recoupling takes place, the production of $\phi$ particles ensues, which, subject to the time of recoupling, 
may or may not impact on the background expansion rate (e.g.,~\cite{Escudero:2019gfk}; see also section~\ref{sec:largecoupling}).
Equilibration of the elastic scattering processes $\nu \phi \leftrightarrow \nu \phi$ and $\nu \nu \leftrightarrow \nu \nu$, however, always impact non-trivially on the spatial fluctuations of the density fields~\cite{Hannestad:2005ex,Basboll:2008fx,Cyr-Racine:2013jua,Oldengott:2014qra,Forastieri:2015paa,Oldengott:2017fhy,Lancaster:2017ksf,Archidiacono:2013dua,Forastieri:2017oma,Kreisch:2019yzn,Forastieri:2019cuf}.

%%%%%%%%%%%%%%
%%%%%%%%%%%%%
%%%%%%%%%%%%%
%%%%%%%%%%%%%%

On the other hand, the $\nu_H$ decay rate is given by 
\begin{equation}
\begin{aligned}
\label{eq:decayrate}
\Gamma_{\text{dec}} & = \frac{1}{\gamma \tau_0}\\
& \simeq  
\begin{dcases}
\frac{\mathfrak{g}^2}{4 \pi} m_{\nu H} \left( \frac{m_{\nu H}}{E_{\nu H}} \right), & {\rm Majorana} \\
\frac{\mathfrak{g}^2}{16 \pi} m_{\nu H} \left( \frac{m_{\nu H}}{E_{\nu H}} \right), & {\rm Dirac} \\
\end{dcases}
\end{aligned}
\end{equation}
where $\tau_0 \equiv 1/\Gamma_{\rm dec}^0$ is the (rest-frame) neutrino lifetime, $\gamma= (E_{\nu H}/m_{\nu H})$  the Lorentz boost factor,  $m_{\nu H}$ the $\nu_H$ mass and $E_{\nu H} \simeq (m_{\nu H}^2+p^2)^{1/2}$ its energy, and 
we have assumed at the second (approximate) equality  that the daughter neutrino $\nu_l$ is massless as well. Figure~\ref{fig:Rates} shows $\Gamma_{{\rm dec}}$ evaluated at a typical momentum $p \sim 3 T$ for $m_{\nu H} = 0.1$~eV.  As with the $2 \to 2$ processes, neutrino decay and its inverse process are generally unimportant at high temperatures. The decay rate does however rapidly catch up with  and can overtake  the Hubble expansion rate as the universe cools.  Depending on when and if equilibrium is established, the resulting recoupled system can again impact on the evolution of both the homogeneous background and the spatial inhomogeneities~\cite{Escudero:2019gfk,Chacko:2019nej,Chacko:2020hmh,Kaplinghat:1999xy}.

Whether it is the $2 \rightarrow 2$ processes or neutrino decay and its inverse process that are the dominant manifestation of the new interaction~\eqref{eq:Lagrangian}--\eqref{eq:coupling} depends on the coupling strength~$\mathfrak{g}$ and the neutrino masses, especially $m_{\nu H}$.  Given the general expectation of sub-eV neutrino masses, figure~\ref{fig:Rates} shows that for 
 large couplings ($\mathfrak{g} \simeq 10^{-5}$; blue lines), $\Gamma_{2\rightarrow 2}$ overtakes $H$ well before $\nu_{H}$ decay becomes relevant,  while for the smaller $\mathfrak{g} \simeq 10^{-10}$ (green lines) it is $\Gamma_{{\rm dec}}$ that overtakes $H$ first, with the $2 \rightarrow 2$ processes remaining always irrelevant.
 In this work, we shall restrict our attention to the latter, small~$\mathfrak{g}$ regime, and consider only the phenomenology of the decay $\nu_H \leftrightarrow \nu_l + \phi$ and its inverse process.

%%%%%%%%%%%%
%%%%%%%%%%%%

\section{Constraining neutrino decay with CMB and LSS: State of the art}
\label{sec:PreviousWorks}

Precision cosmological constraints on invisible neutrino decay to date have been derived broadly in two kinematic regimes: (i)~the decay becomes efficient while the bulk of the
$\nu_H$ population is ultra-relativistic, and (ii)~the decay occurs when the  $\nu_H$ population has already become non-relativistic.  Given the general expectation that neutrinos have sub-eV masses, this classification also means that relativistic decay affects primarily precision cosmological observables at ``early'' times (to be quantified), while non-relativistic decay contributes largely to ``late-time'' phenomena.  We discuss these two decay kinematic regimes below,  and also comment on the large-coupling case (i.e., $2 \to 2$ scattering) at the end of the section.

%%%%%%%%%%%%
%%%%%%%%%%%%

\subsection{Relativistic decay}
\label{sec:reldecay}

If $\nu_H$ decays while ultra-relativistic, once a significant $\phi$ population has been generated, inverse decay must also proceed at a similar rate.  Then, constraints on neutrino decay in the relativistic regime follow essentially from free-streaming arguments, which we outline below. 

Standard-model neutrinos decouple at $T \sim 1$~MeV, well before the scales probed by CMB and LSS measurements enter the horizon.%
\footnote{As a rule of thumb, the smallest scale probed by CMB and linear LSS measurements can be taken to be $k \sim 0.1$~Mpc, which enters the horizon at redshift $z \sim 10^{4}$ or  equivalently, $T\sim 1$~eV.} 
An immediate corollary is that standard-model neutrinos are free-streaming particles, as far as the said measurements are concerned.  On subhorizon scales, 
  free-streaming transfers power from the monopole (density) and dipole (velocity divergence) anisotropies of the neutrino fluid to higher multipole moments (e.g., anisotropic stress), thereby reducing the neutrino contribution to the gravitational potentials relative to the no free-streaming scenario.  In terms of observables, this effect is most prominent during and shortly after radiation domination and hence manifests itself predominantly in the CMB primary anisotropies as a suppression of power at multipoles $\ell \gtrsim 200$ in the CMB temperature power spectrum.

The transfer of neutrino monopole and dipole power to higher multipoles can be inhibited, however, by way of collisional processes --- such as relativistic $\nu_H$ decay and especially its accompanying inverse decay --- which generally damp out anisotropic stress and higher multipole moments in a fluid.  If these collisions should occur efficiently for some duration of time {\it after} the smallest probeable scale enters the horizon and {\it before} recombination ($z = z^*\simeq 1100$ or, equivalently, $T = T^*\sim 0.2$~eV), then one should expect to see an enhancement of the CMB temperature power at $\ell \gtrsim 200$ --- the exact $\ell$ range affected depends on the decay model --- relative to the standard-model case.

%%%%%%%%%%%%%%
%%%%%%%%%%%%%

\subsubsection{Transport rate}

The first CMB constraint on neutrino decay~\cite{Hannestad:2005ex} had been derived by simply demanding that  neutrinos must remain free-streaming up to the time of recombination at $z = z^*$.  
To connect this physics statement to the parameters of the Lagrangian and hence the neutrino lifetime, the authors of~\cite{Hannestad:2005ex} formally required that the rate at which the neutrino fluid isotropises --- the ``transport rate'' $\Gamma_{\rm T}$ --- to be at most as large as the Hubble expansion rate at recombination, i.e.,  $\Gamma_{\rm T} (z^*) \lesssim H(z^*)$.  

The {\it key point} of the argument of~\cite{Hannestad:2005ex}  is that the transport rate $\Gamma_{\rm T}$ is {\it not} the neutrino decay rate, but relates to the rest-frame decay rate via
\begin{equation}
\Gamma_{\mathrm{T}}   \simeq \Gamma_{\rm dec}^{0}  \left(\frac{m_{\nu H}}{E_{\nu H}} \right)^3.\\
\label{eq:transport_rate}
\end{equation}
Here, one power of Lorentz boost factor $E_{\nu H}/m_{\nu H}$ accounts as usual for time dilation. The other two powers account for near-collinearity of the decay products (because the mother neutrino decays while ultra-relativistic), such that some $\sim (E_{\nu H}/m_{\nu H})^2$ number of decay and inverse decay events are required to effect the transport of an initial momentum  to a transverse direction.%
\footnote{A similar argument for the transport rate~\eqref{eq:transport_rate} was used in \cite{Chacko:2003dt} to investigate the cosmological consequences of the process $\nu \nu \leftrightarrow \phi$, which assumes $m_{\phi}>m_{\nu}$.}
Then, by demanding $\Gamma_{\rm T} (z^*) \lesssim H(z^*)$, the authors of~\cite{Hannestad:2005ex}  found  a lower bound of
\begin{equation}
\tau_0 \gtrsim 2 \times 10^{10} \, \mathrm{s} \left( \frac{m_{\nu H}}{50 \,\mathrm{meV}} \right)^3
\label{eq:lifetime_bound}
\end{equation} 
on the neutrino lifetime.

\paragraph{Refinement 1} The same transport argument was applied in~\cite{Basboll:2008fx,Archidiacono:2013dua} with a small twist.  In these works, the combined neutrino--scalar sector is modelled initially as a single, fully free-streaming fluid, but   the free-streaming behaviour is instantaneously switched off (i.e., all $\ell \geq 2$ multipole moments  are set to zero instantaneously) at some redshift $z_{\rm fs}$.  A Markov Chain Monte Carlo (MCMC) analysis --- of the WMAP-5 data in~\cite{Basboll:2008fx} and Planck 2013 in~\cite{Archidiacono:2013dua} --- is then performed to find a lower bound on the redshift $z_{\rm fs}$, which is then translated into a lower bound on the neutrino life time by evaluating the transport rate~(\ref{eq:transport_rate}) at $z_{\rm fs}$.

In effect, this approach to constraining the neutrino lifetime amounts to demanding that $\Gamma_{\mathrm{T}}(z_{\rm fs}) \lesssim H(z_{\rm fs})$, which is generally a weaker requirement than the original $\Gamma_{\mathrm{T}}(z^*) \lesssim H(z^*)$ requirement of~\cite{Hannestad:2005ex} if $z_{\rm fs} > z^*$.  In other words, the CMB anisotropies power spectrum can tolerate some amount of non-free-streaming in the neutrino sector.  Indeed, references~\cite{Basboll:2008fx} and \cite{Archidiacono:2013dua} found respectively $z_{\rm fs} \lesssim 1500$~(95\% C.L.) and $z_{\rm fs} \lesssim 1965$~(95\% C.L.), leading to  neutrino lifetime constraints that are weaker 
 than previous constraints by a marginal factor of 2  in~\cite{Basboll:2008fx}, $\tau_0 \gtrsim 1 \times 10^{10} \, \mathrm{s} \left( m_{\nu H}/50 \,\mathrm{meV} \right)^3$~(95\% C.L.), and by an order of magnitude in~\cite{Archidiacono:2013dua}, $\tau_0 \gtrsim 1.24 \times 10^9 \, \mathrm{s} \left(m_{\nu H}/50 \,\mathrm{meV} \right)^3$~(95\% C.L.).

\paragraph{Refinement 2}
Yet another variant of the transport rate argument was adopted in~\cite{Escudero:2019gfk}, wherein the transport rate $\Gamma_{\rm T}$ is incorporated directly into the combined  neutrino--scalar Boltzmann hierarchy in the form of a damping term for multipoles $\ell \geq 2$, i.e.,
\begin{equation}
 \frac{\partial {\mathcal{F}}_{\ell}}{\partial \tau} = \cdots- a \Gamma_{\mathrm{T}} \mathcal{F}_{\ell},
 \label{eq:exponentialsuppression}
\end{equation}
where $\tau$ is the conformal time (to be properly defined in equation~(\ref{eq:synchronous})), and ${\cal F}_{\ell}$ is the $\ell$th multipole moment of the combined neutrino--scalar fluid.
We shall discuss this approach in more detail in section~\ref{sec:ComparisonLorentz}.    But note at this point that, in the absence of other contributions, equation~\eqref{eq:exponentialsuppression} is solved by ${\cal F}_{\ell} (a) = {\cal F}_{\ell}(a_{\rm ini}) \exp\left[-(1/n) (\Gamma_{\rm T}/H) \right]$, for $\Gamma_{\rm T}/H  \propto a^n$.  For the case at hand, $n=5$ during radiation domination and $n=9/2$ during matter domination.
Thus, while this modelling does amount nominally to switching off the combined fluid's free-streaming behaviour in a more gradual manner, in practice the transition introduced by the exponential damping factor is very sharp and approximates a step-function.  Indeed, testing against the Planck 2018 data, the authors of~\cite{Escudero:2019gfk} found  $\tau_0 \gtrsim (0.3 \to 1.2) \times 10^9 \, \mathrm{s} \left(m_{\nu H}/50 \,\mathrm{meV} \right)^3$, comparable to the result of~\cite{Archidiacono:2013dua} derived under the instantaneous-switching assumption.

%%%%%%%%%%
%%%%%%%%%%%

\subsubsection{Critique}
\label{sec:critique}

Clearly, the central and most critical element in all of the neutrino lifetime constraints discussed above is the transport rate~\eqref{eq:transport_rate} advocated by~\cite{Hannestad:2005ex}.  As such, the form of $\Gamma_{\rm T}$ demands closer scrutiny.  Already in~\cite{Basboll:2009qz} it has been argued that 
 an extra, additive  term proportional to $(m_{\nu H}/E_{\nu H})^2$ should appear in equation~\eqref{eq:transport_rate} to  account for a thermal  neutrino background and that this additional term would dominate and eventuate in neutrino lifetime bounds roughly three orders of magnitude stronger than those derived from the $(m_{\nu H}/E_{\nu H})^3$ term alone. We shall discuss  in section \ref{sec:ComparisonLorentz} our take on this transport rate issue and its implementations in~\cite{Hannestad:2005ex,Basboll:2008fx,Archidiacono:2013dua,Escudero:2019gfk}.
Let us however comment on two other aspects of these works below.

Firstly, it must be borne in mind that using the free-streaming argument as a means to constrain the neutrino lifetime hinges on the inverse decay process being effective throughout the CMB epoch.  If, say, for kinematic reasons inverse decay should be suppressed and only $\nu_H$ decay remains operative, then the system must reach a $\nu_l$--$\phi$-only end-state that is {\it fully free-streaming}, irrespective of the size of the coupling~$\mathfrak{g}$.  As such, no lifetime bound can be placed on $\nu_H$ on the basis of anisotropic stress loss, simply because such a loss ceases to be a phenomenology of the system. (Recall that the (re)generation of anisotropic stress in a fluid only requires that the fluid be free-streaming in an anisotropic  gravitational field.)

An immediate corollary is that  all neutrino lifetime bounds quoted thus far carry the implicit assumption that the decaying neutrino is ultra-relativistic throughout the CMB epoch.
  In other words, absent a more general modelling of neutrino decay in precision cosmology, these bounds apply {\it a priori} only to neutrino masses satisfying $m_{\nu H} \ll T^*  \sim 0.2$~eV.
Together with the minimum mass condition $m_{\nu H} \geq \sqrt{\Delta m_{\rm sun}^2}$, where $\Delta m_{\rm sun}^2$ denotes the solar squared mass splitting, 
current cosmological constraints on the neutrino lifetime lie in the range of $\tau_0=\mathcal{O}(10^7 \to10^{10})$~s.

Secondly, a common approximation in the  MCMC analyses of~\cite{Basboll:2008fx,Archidiacono:2013dua,Escudero:2019gfk,Basboll:2009qz} is to combine all of $\nu_H, \nu_l$ and $\phi$ into a single, massless fluid, thereby neglecting the evolution of the individual species at both the homogeneous and the inhomogeneous level.  This is technically a valid approximation, as long as the assumption of an ultra-relativistic $\nu_H$ population applies across the cosmological epoch of interest; indeed, as we shall show  in section~\ref{sec:numerical}, even if we were to push $m_{\nu H}$ to the maximum allowed by conservative cosmological mass bounds ($\sum m_\nu \lesssim 0.3$~eV),%
\footnote{It has yet to be investigated how cosmological neutrino mass bounds will change in the presence of relativistic neutrino decay such as that discussed in this work.}
the $\nu_H$ population is largely ultra-relativistic across the time frame in which the primary CMB anisotropies are formed, and decays away  only after recombination.

However, this reasoning ignores the fact that the CMB anisotropies are sensitive also to physics {\it after} the recombination era, through such secondary effects as the integrated Sachs--Wolfe (ISW) effect and weak gravitational lensing.   As already discussed above, once the $\nu_H$ population has become fully non-relativistic and completely decayed away, the remaining $\nu_l$--$\phi$ system must revert to a fully free-streaming one independently of the coupling~$\mathfrak{g}$, because of the kinematic suppression of inverse decay.  Thus, the single-fluid approach of~\cite{Basboll:2008fx,Archidiacono:2013dua,Escudero:2019gfk,Basboll:2009qz} --- which effectively assumes free-streaming to be lost forever once recoupling is established --- is at least to some extent unable to correctly describe the complete dynamics of CMB anisotropy formation.
 This alone gives motivation enough to scrutinise the relativistic decay scenario in some detail.

%%%%%%%%%%%
%%%%%%%%%%%

\subsection{Non-relativistic decay}

Free-streaming constraints from the CMB primary anisotropies that form the basis of the neutrino lifetime bound~(\ref{eq:lifetime_bound})
can be largely circumvented if recoupling is established only {\it after} recombination.
 In such a scenario, the $\nu_H$ population is likely non-relativistic, so that inverse decay is kinematically suppressed. 
 Then, the main effect of the decay process consists in transferring energy from the matter sector to the radiation sector (assuming massless decay products), which is observable in the CMB anisotropies as an enhanced late ISW effect~\cite{Lopez:1998jt,Kaplinghat:1999xy} and a suppressed weak gravitational lensing signal~\cite{Chacko:2019nej} (if the decay happens at $z \gtrsim 3 \to 4$), as well as in the large-scale structure matter distribution as an across-the-board suppression in the present-day matter power spectrum.

The CMB weak gravitational lensing signal is especially interesting in that, on the one hand, the current generation of cosmological neutrino mass bounds owes their restrictiveness primarily to this signal~\cite{RoyChoudhury:2019hls}.  On the other hand, such a reliance on the lensing signal also provides a means for us to exploit  non-relativistic neutrino decay to evade, or at least relax, cosmological neutrino mass bounds~\cite{Chacko:2019nej}.  The argument of~\cite{Chacko:2019nej} goes as follows: in order to compensate for a suppressed CMB lensing signal in neutrino decay scenarios, the inferred total matter density $\omega_{m,0}$ must be larger than in the standard-model case.  This automatically implies that, for the same neutrino fraction $f_\nu \equiv \omega_{\nu,0}/\omega_{m,0}$ --- the main parameter controlling the small-scale LSS power suppression due to neutrino masses, the actual neutrino energy density $\omega_{\nu,0} = \sum m_\nu/(94~{\rm eV})$ and hence the neutrino mass sum $\sum m_\nu$ in decay scenarios can be larger than is allowed in the standard, no-decay case.  Future high-redshift LSS surveys will be able to break this degeneracy~\cite{Chacko:2020hmh}.

%%%%%%%%
%%%%%%%%%%

\subsubsection{Modelling non-relativistic decay}
\label{sec:discrepancy}

Because, unlike relativistic decay, decay in the non-relativistic limit modifies the distribution of matter and radiation in the universe in an out-of-equilibrium fashion,  it is generally not appropriate to treat the mother and daughter particles as a single fluid, and some manner of (i)~energy transfer at the homogeneous background level and (ii)~transfer of the spatial fluctuations must be included in the modelling of the system.  Where inverse decay can be neglected and the daughter particles are massless, this is in principle a straightforward exercise and one that has been implemented previously in~\cite{Chacko:2019nej,Chacko:2020hmh,Kaplinghat:1999xy} in the form of (i)  two coupled background Boltzmann equations for the decaying neutrino and the decay products, and (ii) two corresponding Boltzmann hierarchies for the spatial fluctuations.  

Crucially, however, any modelling of the non-relativistic neutrino decay scenario must converge formally to the decaying cold dark matter (CDM) paradigm in the limit of an extremely heavy and hence cold $\nu_H$ population.  In this regard, we observe that the Boltzmann hierarchies presented in~\cite{Chacko:2019nej,Chacko:2020hmh}  for non-relativistic neutrino decay appear in the said limit to diverge from the decaying CDM results of~\cite{Audren:2014bca}.  Notably, as we shall show in section~\ref{sec:Comparison_Non-rel},  some terms that are required to preserve momentum conservation and gauge invariance are missing from the former works' daughter hierarchy at multipole $\ell = 1$ (and also beyond).  We argue that these terms should be reinstated for a consistent description of the system.

%%%%%%%%
%%%%%%%%

\subsection{Large coupling and $2 \to 2$ scattering}
\label{sec:largecoupling}

To complete our review on the phenomenology of and current constraints on the interaction Lagrangian~(\ref{eq:Lagrangian})  and coupling~\eqref{eq:coupling}, we briefly comment on the case of $2 \to 2$ scattering.

As already discussed in section~\ref{sec:Model}, $2 \rightarrow 2$ processes are the dominant phenomenology of the Lagrangian~(\ref{eq:Lagrangian})  and coupling~\eqref{eq:coupling} for large coupling constants~$\mathfrak{g}$.
If $\mathfrak{g}$ is so  large that  recoupling occurs before neutrino decoupling ($T \sim 1$~MeV), a thermal population of $\phi$ particles can be produced at the expense of the photon bath, consequently raising the number of (non-photon) relativistic degrees of freedom --- or the  effective number of neutrinos,  $N_{\mathrm{eff}}$ --- from its standard-model value $N_{\rm eff}^{\rm SM}=3.045$~\cite{deSalas:2016ztq}.%
\footnote{If $\phi$ production takes place after neutrino decoupling, it proceeds at the expense of the already-decoupled neutrinos, in which case $N_{\rm eff}$ does not deviate from its value attained at neutrino decoupling.}
Current constraints on $\mathfrak{g}$ in the large-coupling regime follow from null measurements of $\Delta N_{\rm eff} \equiv N_{\rm eff}- N_{\rm eff}^{\rm SM}>0$ in the observed primordial helium-4 and deuterium abundances.   These in turn set a lower neutrino lifetime bound of \(\tau_0 \gtrsim 10^{-3} \text{ s}\)~\cite{Escudero:2019gfk}.

%%%%%%%%%%

%%%%%%%%
%%%%%%%%%

\section{Our formalism}
\label{sec:OurFormalism}

The discussion of section~\ref{sec:PreviousWorks} reveals that current modellings of the cosmological effects of neutrino decay in both the relativistic and non-relativistic regimes  all leave something to be desired.  There is furthermore no fundamental reason why neutrino decay phenomenology needs to be discussed and modelled exclusively in two kinematic extremes.  Indeed, this is an especially pertinent issue in the relativistic decay scenario in that, given our current knowledge of neutrino masses, any initially ultra-relativistic decaying neutrino population must eventually transition to a non-relativistic one as the universe cools.  This simple observation alone calls for a kinematically unified approach.  

We present in this section such a unified approach, and derive from first principles the relevant Boltzmann equations to track the time evolution of the $\{\nu_H,\nu_l, \phi\}$ system at both the homogeneous (background) and the inhomogeneous (first-order) level.
Throughout the work we use the notation of~\cite{Ma:1995ey,Oldengott:2014qra}.

%%%%%%%%

\subsection{Preliminaries}

Our starting point is the Boltzmann equation for the phase space density of the $i$th particle species, $f_i(\mathbf{x},\mathbf{P},x^0)$, given in 
relativistic notation by
\begin{equation}
P^{\mu}  \frac{\partial f_i}{\partial x^{\mu}} - \Gamma^{\nu}_{\rho \sigma} P^{\rho} P^{\sigma} \frac{\partial f_i}{\partial P^{\nu}}= m_i\left(\frac{{\rm d} f_i}{{\rm d} \sigma} \right)_C,
\label{eq:Boltzmann2}
\end{equation}
where we have assumed the particle species has a mass $m_i$.
On the l.h.s., $x^\mu=(x^0,x^i)=(x^0,\mathbf{x})$ are the spacetime coordinates, $P^\mu=(P^0,P^i) = (P^0,\mathbf{P})$ the 4-momentum, $\Gamma^{\nu}_{\rho \sigma}$ the Christoffel symbols that encapsulate all gravitational physics, ${\rm d} \sigma$ is the incremental proper time along the worldline of the phase space density~$f_i$, and $f_i$ is defined such that 
\begin{equation}
f_i(\mathbf{x},\mathbf{P}, x^0) {\rm d} x^1 {\rm d} x^2 {\rm d} x^3 {\rm d}P_1 {\rm d}P_2 {\rm d}P_3 = {\rm d}N
\end{equation}
gives the number of particles ${\rm d}N$ in a differential phase space volume ${\rm d} x^1 {\rm d} x^2 {\rm d} x^3 {\rm d}P_1 {\rm d}P_2 {\rm d}P_3$, where $P_i$ is the canonical momentum conjugate to $x^i$.   On the r.h.s.\ of equation~(\ref{eq:Boltzmann2}), $m_i (\mathrm{d}f_i/\mathrm{d}\sigma)_C$ is a Lorentz-invariant collision integral that describes all scattering processes for the $i$th particle species that can be considered to happen locally at $x^\mu$; its precise form for the problem at hand will be detailed below.

We choose to work in the synchronous gauge, whose line element is given by
\begin{equation}
{\rm d}s^2 = a^2(\tau) \left[- {\rm d} \tau^2 + (\delta_{ij} + h_{ij}) {\rm d} x^i {\rm d} x^j \right],
\label{eq:synchronous}
\end{equation}
where $\tau$ is the conformal time, $a$ the scale factor, $\delta_{ij}$ a Kronecker delta, and $h_{ij}$ encodes scalar perturbations to the metric.  It is then convenient to first express  
the 4-momentum and its lower-index form as $P^\mu =(E/a,  [\delta^i_j - h^i_j/2] p^j/a)$ and $P_\mu=(-aE,  a[\delta_{ij} + h_{ij}/2] p^j)$, respectively, to linear order in $h_{ij}$, where $p^{\mu} = (E,\mathbf{p})$ is the 4-momentum in the tetrad basis, i.e., the orthonormal basis of an observer comoving with the spacetime coordinates~(\ref{eq:synchronous}).  From here we may further define a comoving momentum $\mathbf{q} \equiv a \mathbf{p}$ and a comoving energy $\epsilon \equiv (|\mathbf{q}|^2+a^2 m_i^2)^{1/2}$, in order to factor out as much as possible the effect of cosmic expansion from our final equations of motion.  We shall be using these comoving energy--momentum coordinates in the rest of the analysis.

Following usual practice, we split the phase space density $f_i(\mathbf{x},\mathbf{q},\tau)$ --- now expressed as a function of the coordinates~(\ref{eq:synchronous}) and the comoving momentum --- into (i) a homogeneous and isotropic background component and (ii) a perturbed part, i.e.,
\begin{equation}
f_i(\mathbf{x},\mathbf{q},\tau) = \bar{f}_i(|\mathbf{q}|,\tau) + F_i (\mathbf{x},\mathbf{q},\tau)
\equiv \bar{f}_i(|\mathbf{q}|,\tau) \left[1 + \Psi_i (\mathbf{x},\mathbf{q},\tau) \right].
\end{equation} 
Depending on the context, it may be simpler or physically more transparent to express the equations of motion in terms of either $F_i$ or $\Psi_i$, and we shall occasionally swap between these two notations.  Then, expanding equation~(\ref{eq:Boltzmann2}) to the  linear order in the perturbed quantities (i.e., $h_{ij}$ and $F_i$) and performing a Fourier transform $\psi(\mathbf{k}) = \mathscr{F}[\psi(\mathbf{x})]$ on all functions of the spatial coordinates~$\mathbf{x}$, we obtain for the background $\bar{f}_i$ an equation of motion
\begin{equation}
\frac{\partial \bar{f}_i}{\partial \tau}(|\mathbf{q}|,\tau)= \left( \frac{{\rm d} f_i}{{\rm d} \tau}\right)^{(0)}_C,
\label{eq:back_Boltzmann}
\end{equation}
and for the perturbed part,
\begin{equation}
\frac{\partial F_i}{\partial \tau}(\mathbf{k},\mathbf{q},\tau)+i \frac{|\mathbf{q}||\mathbf{k}|}{\epsilon} (\hat{k}\cdot \hat{q}) F_i(\mathbf{k},\mathbf{q},\tau)+\frac{\partial \bar{f}_i(|\mathbf{q}|,\tau)}{\partial \ln |\mathbf{q}|} \left[ \dot{\eta}-(\hat{k} \cdot \hat{q})^2 \frac{\dot{h}+6\dot{\eta}}{2} \right] =  \left( \frac{{\rm d} f_i}{{\rm d} \tau}\right)^{(1)}_C,
\label{eq:pert_Boltzmann}
\end{equation}
where  $\cdot \equiv \partial/\partial \tau$, $h \equiv h^i_i(\mathbf{k},\tau)$ and $6 \eta \equiv -3/(2 k^4) k^i k^j h_{ij} + 1/(2 k^2) h$  denote respectively the Fourier transforms of the trace and 
the traceless longitudinal perturbations in the space-space part of the synchronous metric~(\ref{eq:synchronous}), and    
$({\rm d}f_i/{\rm d}\tau)^{(0),(1)}_C$ on the r.h.s.\  are the coordinate- and gauge-appropriate collision integrals to zeroth and linear order,  respectively, in the perturbed quantities.  Note that equations~(\ref{eq:back_Boltzmann}) and (\ref{eq:pert_Boltzmann}) apply also to massless particles, even though we had motivated the relativistic Boltzmann equation~(\ref{eq:Boltzmann2}) by way of massive ones.

It remains to specify the forms of the collision integrals $({\rm d}f_i/{\rm d}\tau)^{(0),(1)}_C$. Firstly, we note that 
for a decay $i(p) \to j(n) + k(n')$ and its inverse decay $j(n) + k(n') \to i(p)$ process, the Lorentz-invariant collision integral for the $i$th species on the r.h.s.\ of equation~(\ref{eq:Boltzmann2}) can be written in the tetrad basis as
\begin{equation}
\begin{aligned}
m_i\left(\frac{{\rm d} f_i}{{\rm d} \sigma} \right)_C =&  \frac{1}{2} \int {\rm d} \mathbf{\Pi}_j(\mathbf{n}) \int {\rm d} \mathbf{\Pi}_k (\mathbf{n}') \, (2 \pi)^4 \, \delta^{(4)}_D(p - n - n')  \\
& \times |{\cal M}_{i\leftrightarrow j + k}|^2
\left[f_j f_k(1\pm f_i) - f_i (1\pm f_j)(1\pm f_k) \right]
\label{eq:licollisionintegral}
\end{aligned}
\end{equation}
including quantum statistics, where 
\begin{equation}
{\rm d} \mathbf{\Pi}_j (\mathbf{n}) \equiv g_j \frac{{\rm d}^3 \mathbf{n}}{(2 \pi)^3 2 E(\mathbf{n})}
\end{equation}
and so on, $g_j$ is the number of internal degrees of freedom of the $j$th particle species, $p,n$, etc.\ are physical 4-momenta, $\delta_D^{(4)}$ is the four-dimensional Dirac delta distribution, and $|{\cal M}_{i \leftrightarrow j + k}|^2$ denotes  the  Lorentz-invariant  squared  matrix  element  for  the $T$-invariant interactions $i \leftrightarrow j + k$ averaged over the spins of all particles in the initial and final states. 
Given the effective Lagrangian~(\ref{eq:Lagrangian}) and coupling matrix~(\ref{eq:coupling}), it is straightforward to establish at tree level
\begin{equation}
 |\mathcal{M}_{\nu_H \leftrightarrow \nu_l + \phi}|^2 = (2 \mathfrak{g})^2 \left( \eta_{\mu \nu} p_1^{\mu}  p_2^{\nu} +m_{\nu H} m_{\nu l} \right)
 \label{eq:matrixelement}
\end{equation}
for $\nu_H(p_1) \leftrightarrow \nu_l(p_2) + \phi(p_3)$, where the prefactor $(2 \mathfrak{g})^2$  originates from the assumption of  Majorana neutrinos which necessitates an extra factor of 2 at each vertex.%
\footnote{While we are primarily concerned with the $m_\phi=0$ case, we note that the matrix element~\eqref{eq:matrixelement} applies also to the case of a finite $m_\phi$. }

Secondly, in order to rewrite the Lorentz-invariant integral~(\ref{eq:licollisionintegral}) for use with equations~(\ref{eq:back_Boltzmann}) and~(\ref{eq:pert_Boltzmann}), we note that 
 $P^0 \partial_0 = \epsilon/a^2 \partial_\tau$ in the synchronous gauge~(\ref{eq:synchronous}).  It then follows simply that
\begin{equation}
 \left( \frac{{\rm d} f_i}{{\rm d} \tau}\right)^{(n)}_C = \frac{a^2}{\epsilon} m_i\left(\frac{{\rm d} f_i}{{\rm d} \sigma} \right)_C^{(n)},
 \label{eq:collisionmap}
\end{equation}
where $m_i({\rm d}f_i/{\rm d}\sigma)^{(n)}_C$ represents the $n$th order term in a series expansion of the Lorentz-invariant integral in the perturbed phase space density $F_i$.  The reduction of the 6-dimensional  $({\rm d}f_i/{\rm d}\tau)^{(0),(1)}_C$ integrals to one-dimensional ones is the subject of appendix~\ref{app:CollisionIntegralReduction}; we present their final, reduced forms below in sections~\ref{sec:Background} and~\ref{sec:First order perturbations}, and, in the case of the linear-order term $({\rm d}f_i/{\rm d}\tau)^{(1)}_C$, also its decomposition in terms of Legendre polynomials.

Lastly, we remark that had we used instead the Newtonian gauge, which has the line element ${\rm d}s^2 = a^2(\tau) [-(1+ 2 \psi) {\rm d}\tau^2 + (1- 2 \phi) {\rm d}x^i {\rm d}x^j]$ and hence $P^0  = (1- \psi) \epsilon/a^2 $ to linear order in $\psi$, we would have found for the first-order collision integral
the mapping
\begin{equation}
 \left( \frac{{\rm d} f_i}{{\rm d} \tau}\right)^{(1)}_{C, {\rm Newton}} = \frac{a^2}{\epsilon} \left[m_i\left(\frac{{\rm d} f_i}{{\rm d} \sigma} \right)_C^{(1)} + \psi \, m_i \left( \frac{{\rm d} f_i}{{\rm d} \sigma} \right)_C^{(0)} \right],
 \label{eq:newtonmap}
\end{equation}
i.e., relative to its synchronous gauge counterpart~(\ref{eq:collisionmap}), the linear-order Newtonian gauge mapping formally has an additional term.
Reference~\cite{Audren:2014bca} was first to highlight this extra term in the context of dark matter decay.

%%%%%%%%%%%%
%%%%%%%%%%%%%%

\subsection{Background equations}
\label{sec:Background}

Beginning with the zeroth-order Boltzmann equation~(\ref{eq:back_Boltzmann}) and the collision integral~(\ref{eq:licollisionintegral}), a set of  of coupled integro-differential equations can be established for the time evolution of the background distributions $\bar{f}_{\nu H}$, $\bar{f}_{\nu l}$, and $\bar{f}_\phi$. In their final form and using the shorthand notations $q_i \equiv |\mathbf{q}_i|$, $\epsilon_i = (q_i^2 + a^2 m_i^2)^{1/2}$, these equations of motion are: 
\begin{equation}
\begin{aligned}
\frac{\partial \bar{f}_{\nu H}(q_1)}{\partial \tau} = & \frac{\mathfrak{g}^2 a^2 (m_{\nu H} \! + \! m_{\nu l})^2}{4 \pi \, \epsilon_1 \, q_1} \! \left[ - \underbrace{q_1 \frac{(m_{\nu H}^2 \! - \! m_{\nu l}^2)}{m_{\nu H}^2} \bar{f}_{\nu H}(q_1)}_{\text{dec}} + \underbrace{\int_{q_{2-}^{(\nu H)}}^{q_{2+}^{(\nu H)}} \! \! \mathrm{d}q_2 \, \frac{q_2}{\epsilon_2} \, \bar{f}_{\nu l} (q_2) \, \bar{f}_{\phi}(\epsilon_1 - \epsilon_2)}_{\text{inv}} \right. \\
		& \hspace{3.5cm} \left. + \underbrace{\bar{f}_{\nu H}(q_1) \left( \int_{q_{2-}^{(\nu H)}}^{q_{2+}^{(\nu H)}} \mathrm{d}q_2 \, \frac{q_2}{\epsilon_2} \, \bar{f}_{\nu l}(q_2) - \int_{q_{3-}^{(\nu H)}}^{q_{3+}^{(\nu H)}} \mathrm{d}q_3 \, \bar{f}_{\phi}(q_3) \right)}_{\text{qs}} \right] ,
\label{eq:background_Boltzmann1}
\end{aligned}
\end{equation}
\begin{equation}
\begin{aligned}
\frac{\partial \bar{f}_{\nu l}(q_2)}{\partial \tau} = &  \, \frac{\mathfrak{g}^2 a^2 (m_{\nu H} + m_{\nu l})^2}{4 \pi \, \epsilon_2 \, q_2} \left[ \underbrace{\int_{q_{1-}^{(\nu l)}}^{q_{1+}^{(\nu l)}} \mathrm{d}q_1 \, \frac{q_1}{\epsilon_1} \, \bar{f}_{\nu H}(q_1) }_{\text{dec}} - \underbrace{\bar{f}_{\nu l}(q_2) \int_{q_{3-}^{(\nu l)}}^{q_{3+}^{(\nu l)}} \mathrm{d}q_3 \, \bar{f}_{\phi}(q_3)}_{\text{inv}} \right. \\
		& \hspace{3.5cm} \left. + \underbrace{ \int_{q_{1-}^{(\nu l)}}^{q_{1+}^{(\nu l)}} \mathrm{d}q_1 \, \frac{q_1}{\epsilon_1} \left( \bar{f}_{\nu H}(q_1) \, \bar{f}_{\phi}(\epsilon_1 - \epsilon_2) - \bar{f}_{\nu H}(q_1) \, \bar{f}_{\nu l}(q_2) \right) }_{\text{qs}} \right] \, ,
\label{eq:background_Boltzmann2} 
\end{aligned}
\end{equation}
\begin{equation}
\begin{aligned}
\frac{\partial \bar{f}_{\phi}(q_3)}{\partial \tau} =&  \, \frac{\mathfrak{g}^2 a^2 (m_{\nu H} + m_{\nu l})^2}{2 \pi \, q_3^2} \left[ \underbrace{\int_{q_{1-}^{(\phi)}}^{q_{1+}^{(\phi)}}  \mathrm{d}q_1 \, \frac{q_1}{\epsilon_1} \, \bar{f}_{\nu H}(q_1) }_{\text{dec}} - \underbrace{ \bar{f}_{\phi}(q_3) \int_{q_{2-}^{(\phi)}}^{q_{2+}^{(\phi)}} 
	 \mathrm{d}q_2 \, \frac{q_2}{\epsilon_2} \, \bar{f}_{\nu l}(q_2)}_{\text{inv}}  \right. \\
		& \hspace{1cm} \left. + \underbrace{ \int_{q_{1-}^{(\phi)}}^{q_{1+}^{(\phi)}} 	
			 \mathrm{d}q_1 \, \frac{q_1}{\epsilon_1} \, \left( \bar{f}_{\nu H}(q_1) \, \bar{f}_{\phi}(q_3) - \bar{f}_{\nu H}(q_1) \, \bar{f}_{\nu l} \left( \sqrt{(\epsilon_1 - q_3)^2 - a^2 m_{\nu l}^2} \right) \right)  }_{\text{qs}} \right] \, , 
\label{eq:background_Boltzmann3}
\end{aligned}
\end{equation}
where the integration limits are given by
\begin{equation}
\begin{aligned}
q_{2\pm}^{(\nu H)} & = \left| \frac{\epsilon_1(m_{\nu H}^2-m_{\nu l}^2) \pm q_1(m_{\nu H}^2+m_{\nu l}^2)}{2 m_{\nu H}^2}\right|, \\
q_{3\pm}^{(\nu H)} & =  \frac{m_{\nu H}^2-m_{\nu l}^2}{2 m_{\nu H}^2} (\epsilon_1 \pm q_1), 
\label{Equation:IntegralLimits1}
\end{aligned}
\end{equation}
\begin{equation}
\begin{aligned}
q_{1\pm}^{(\nu l)} & = \left| \frac{\epsilon_2(m_{\nu H}^2-m_{\nu l}^2) \pm q_2(m_{\nu H}^2+m_{\nu l}^2)}{2 m_{\nu l}^2}\right| 
 \underset{m_{\nu l}=0}{\to}  
\begin{dcases}
\infty \\
\left|\frac{1}{4}\frac{a^2m_{\nu H}^2}{q_2}-q_2\right| \\
\end{dcases} \, , 	 \\
q_{3,\pm}^{(\nu l)} & =  \frac{m_{\nu H}^2-m_{\nu l}^2}{2 m_{\nu l}^2} (\epsilon_2 \pm q_2) 
 \underset{m_{\nu l}=0}{\to}  
\begin{dcases}
\infty \\
 \frac{1}{4} \frac{a^2 m_{\nu H}^2}{q_2} \\
\end{dcases} \, ,
\label{Equation:IntegralLimits2} 	 
\end{aligned}
\end{equation}
\begin{equation}
\begin{aligned}
q_{1-}^{(\phi)} & = \left| \frac{a^2(m_{\nu H}^2-m_{\nu l}^2)^2-4 m_{\nu H}^2 q_3^2}{4 q_3 (m_{\nu H}^2-m_{\nu l}^2)} \right|, \qquad  q_{1+}^{(\phi)} = \infty, \\
q_{2-}^{(\phi)} & = \left| \frac{a^2(m_{\nu H}^2-m_{\nu l}^2)^2-4 m_{\nu l}^2 q_3^2}{4 q_3 (m_{\nu H}^2-m_{\nu l}^2)} \right| , \qquad \;\,  q_{2+}^{(\phi)} = \infty.
\label{Equation:IntegralLimits3}
\end{aligned}
\end{equation}
Details of the derivation can be found in appendix~\ref{app:CollisionIntegralReduction}

Note that in equations~(\ref{eq:background_Boltzmann1})--(\ref{eq:background_Boltzmann3}) we have split up the contributions to the collision integrals from $\nu_H$ decay (``dec''), inverse decay (``inv''), and the inclusion of quantum statistics (``qs''), i.e., Pauli blocking for $\nu_{H,l}$  and Bose enhancement for $\phi$.  We shall study these individual contributions numerically in section~\ref{sec:numerical}.   We remark here, however, the ``qs'' terms mix contributions from decay and inverse decay, and are hence only physically meaningful when both the ``dec'' and ``inv'' terms are present.

%%%%%%%%%%%%%%%%
%%%%%%%%%%%%%%%%

\subsection{First-order perturbation equations}
\label{sec:First order perturbations}

The first-order equation of motion~(\ref{eq:pert_Boltzmann}) for the perturbed phase space density $F_i$ can also be recast as an equation of motion for the phase space density contrast $\Psi_i (\mathbf{k},\mathbf{q}_i,\tau)= F_i(\mathbf{k},\mathbf{q}_i,\tau)/\bar{f}_i(|\mathbf{q}_i|,\tau)$:
\begin{equation}
\begin{aligned}
 \frac{\partial {\Psi}_i}{\partial \tau}(\mathbf{k},\mathbf{q}_i,\tau) + \frac{\dot{\bar{f}}_i}{\bar{f}_i} \Psi_i +i \frac{|\mathbf{q}_i||\mathbf{k}|}{\epsilon} (\hat{k}\cdot \hat{q}_i)  \Psi_i
+ \frac{\partial \ln \bar{f}_i}{\partial \ln |\mathbf{q}_i|} \left[ \dot{\eta}-(\hat{k} \cdot \hat{q}_i)^2 \frac{\dot{h}+6\dot{\eta}}{2} \right] = \frac{1}{\bar{f}_i} \left( \frac{{\rm d} f_i}{{\rm d} \tau}\right)_C^{(1)}.
\label{perturbed Liouville}
\end{aligned}
\end{equation}
This is a generalisation of equation~(40) of reference~\cite{Ma:1995ey} to include a time-dependent background distribution.
In the case of a time-independent background (e.g., standard-model neutrinos),  $\dot{\bar{f}}_i$ and hence the second term vanish.

Because the l.h.s.\ of equation~(\ref{perturbed Liouville}) depends explicitly only on $|\mathbf{q}_i|$, $|\mathbf{k}|$, and $\hat{k}\cdot \hat{q}_i$, it is standard practice to decompose $\Psi_i(\mathbf{k},\mathbf{q}_i,\tau)$ in terms of the Legendre polynomials $P_\ell(\hat{k}\cdot \hat{q}_i)$.  Here, we use the convention of~\cite{Ma:1995ey}, i.e.,
\begin{equation}
\begin{aligned}
\Psi_{i}(|\mathbf{k}|,|\mathbf{q}_i|,\hat{k}\cdot \hat{q}_i) &= \sum_{\ell=0}^{\infty}(-i)^{\ell} (2\ell+1) \Psi_{i,\ell}(|\mathbf{k}|,|\mathbf{q}_i|) P_{\ell}(\hat{k} \cdot \hat{q}_i), \\
\Psi_{i,\ell}(|\mathbf{k}|,|\mathbf{q}_i|) &= \frac{i^{\ell}}{2} \int_{-1}^{1} \mathrm{d}(\hat{k} \cdot \hat{q}_i) \, \Psi_{i}(|\mathbf{k}|,|\mathbf{q}_i|,\hat{k} \cdot \hat{q}_i) P_{\ell}(\hat{k} \cdot \hat{q}_i).
\label{Legendre_decompositio0n}
\end{aligned}
\end{equation}
Decomposing the equation of motion~(\ref{perturbed Liouville}) in a similar manner, we find an infinite hierarchy of equations of motion for the multipole moments $\Psi_{i,\ell}(|\mathbf{k}|,|\mathbf{q}_i|)$ --- the  Boltzmann hierarchy ---  of the form
\begin{equation}
\begin{aligned}
\dot{\Psi}_{i,0}(q_i) & = -\frac{q_i k}{\epsilon_i} \Psi_{i,1}(q_i)+ \frac{1}{6} \frac{\partial \ln \bar{f}_{i}}{\partial \ln q_i}\dot{h} + \mathcal{C}^{(1)}_{0}[\Psi_{i} (q_i)], \\
\dot{\Psi}_{i,1}(q_i)&= \frac{q_i k}{\epsilon_i} \left(-\frac{2}{3} \Psi_{i,2}(q_i)+\frac{1}{3}  \Psi_{i,0}(q_i) \right) +  \mathcal{C}^{(1)}_{1}[\Psi_{i} (q_i)], \\
\dot{\Psi}_{i,2}(q_i) &= \frac{q_i k}{\epsilon_i} \left(-\frac{3}{5} \Psi_{i,3}(q_i) \! + \! \frac{2}{5} \Psi_{i,1}(q_i) \right) \! - \! \frac{\partial \ln \bar{f}_{i}}{\partial \ln q_i} \! \left( \frac{2}{5} \dot{\eta} \! + \! \frac{1}{15} \dot{h} \right)  +  \mathcal{C}^{(1)}_{2}[\Psi_{i} (q_i)], \\
\dot{\Psi}_{i,\ell>2}(q_i) & =\, \frac{k}{2\ell+1} \frac{q_i}{\epsilon_i} \left[ \ell \Psi_{i,\ell-1}(q_i) - (\ell+1)\Psi_{i,\ell+1}(q_i) \right] +  \mathcal{C}^{(1)}_{\ell}[\Psi_{i} (q_i)],
\end{aligned}
\label{hierarchyPsi_nu1}
\end{equation}
where  the collision term
\begin{equation}
\mathcal{C}^{(1)}_{\ell}[\Psi_{i} (q_i)] \equiv \frac{1}{\bar{f}_i} \left( \frac{{\rm d} f_i}{{\rm d} \tau}\right)_{C,\ell}^{(1)} (q_i)- \frac{\dot{\bar{f}}_i}{\bar{f}_i} \Psi_{i,\ell} (q_i)
\end{equation}
subsumes both the order~$\ell$ Legendre moment of the first-order collision integral~(\ref{eq:collisionmap}) and the background evolution term.  Details on how to perform the Legendre decomposition of the collision integrals can be found in appendix~\ref{app:CollisionIntegralReduction}.  

For $i = \nu_H$, the order $\ell$ collision term reads 
\begin{equation}
\begin{aligned}
\mathcal{C}^{(1)}_{\ell}[\Psi_{\nu H}] \! = \frac{\mathfrak{g}^2 a^2 (m_{\nu H} + m_{\nu l})^2}{4 \pi \, \epsilon_1 \, q_1 \, \bar{f}_{\nu H}(q_1)} \left[ \underbrace{ \int_{q_{2-}^{(\nu H)}}^{q_{2+}^{(\nu H)}} \mathrm{d}q_2 \, \frac{q_2}{\epsilon_2} \, \bar{f}_{\nu l}(q_2) \, \bar{f}_{\phi}(\epsilon_1 \! - \! \epsilon_2) \, \Psi_{\nu l, \ell} (k, q_2) \, P_{\ell}\left(\cos{\alpha^*}\right) }_{\text{inv}} \right.& \\
		- \underbrace{ \Psi_{\nu H, \ell} (k, q_1) \int_{q_{2-}^{(\nu H)}}^{q_{2+}^{(\nu H)}} \mathrm{d}q_2 \, \frac{q_2}{\epsilon_2} \, \bar{f}_{\nu l}(q_2) \, \bar{f}_{\phi}(\epsilon_1 \! - \! \epsilon_2) }_{\text{inv}} & \\
		\hspace{1.8cm}+ \underbrace{ \int_{q_{3-}^{(\nu H)}}^{q_{3+}^{(\nu H)}} \mathrm{d}q_3 \, \bar{f}_{\nu l}\!\left(\!\sqrt{(\epsilon_1\!-\!q_3)^2 \! - \! a^2 m_{\nu l}^2}\right) \, \bar{f}_{\phi} (q_3) \, \Psi_{\phi, \ell} (k, q_3) \, P_{\ell}\left(\cos{\beta}^*\right) }_{\text{inv}} & \\
		+ \underbrace{ \bar{f}_{\nu H} (q_1) \int_{q_{2-}^{(\nu H)}}^{q_{2+}^{(\nu H)}} \! \mathrm{d}q_2 \, \frac{q_2}{\epsilon_2} \, \bar{f}_{\nu l} (q_2) \, \Psi_{\nu l, \ell} (k, q_2) \, P_{\ell}\left(\cos{\alpha^*}\right) }_{\text{qs}} & \\
		- \left. \underbrace{ \bar{f}_{\nu H}(q_1) \int_{q_{3-}^{(\nu H)}}^{q_{3+}^{(\nu H)}} \! \mathrm{d}q_3 \, \bar{f}_{\phi} (q_3) \, \Psi_{\phi, \ell} (k, q_3)\, P_{\ell}\left(\cos{\beta^*}\right) }_{\text{qs}} \right] & ,
\end{aligned}
\label{nuH_collision_integral}
\end{equation}
where  
\begin{equation}
\begin{aligned}
\cos{\alpha^*} =& \, \frac{2\epsilon_1 \epsilon_2 - a^2 (m_{\nu H}^2 + m_{\nu l}^2)}{2 q_1 q_2}, \\
\cos{\beta^*} =& \, \frac{2\epsilon_1 q_3 - a^2 (m_{\nu H}^2 - m_{\nu l}^2)}{2 q_1 q_3}
\end{aligned}
\label{kernel}
\end{equation} 
are the angular openings between momentum vectors allowed by energy-momentum conservation.  The integration limits $q_{2\pm}^{(\nu H)}$ and $q_{3\pm}^{(\nu H)}$ are identical to those appearing in the background $\nu_H$~collision integral~\eqref{eq:background_Boltzmann1} and given in equation~(\ref{Equation:IntegralLimits1}).  Again, we have split up the collision term into contributions from decay, inverse decay, and quantum statistics.

Immediately apparent in equation~\eqref{nuH_collision_integral} is that the collision integral $\mathcal{C}^{(1)}_\ell[\Psi_{\nu H}(q_1)]$ contains no pure decay term.  This implies that, in the limit of non-relativistic decay where inverse decay and quantum statistics can be ignored, the Boltzmann hierarchy~(\ref{hierarchyPsi_nu1}) 
 for the mother particle $i = \nu_H$ is in fact identical to the standard, free-streaming one~\cite{Kaplinghat:1999xy}.   The absence of the decay term is physically sensible and merely reflects the fact that exponential decay depletes the $\nu_H$ phase space density $f_{\nu H}(\mathbf{x},\mathbf{q},\tau)$ {\it everywhere} by exactly the same, spatially-independent factor
  \begin{equation}
\exp \left[- \int^\tau_0 {\rm d} \tau' \, \frac{\mathfrak{g}^2 a^2 (m_{\nu H} + m_{\nu l})^2  (m_{\nu H}^2 - m_{\nu l}^2)^2}{4 \pi m_{\nu H}^2\epsilon_1} \right].
\label{eq:decayfactor}
  \end{equation}
 Once this depletion has been accounted for by the evolution equation~(\ref{eq:background_Boltzmann1}) for the background distribution $\bar{f}_{\nu H}(q,\tau)$, it must disappear from the collision integral~\eqref{nuH_collision_integral} and hence the equation of motion~(\ref{hierarchyPsi_nu1}) for $\Psi_{\nu H}(\mathbf{k},\mathbf{q},\tau)= F_{\nu H}(\mathbf{k},\mathbf{q},\tau)/\bar{f}_{\nu H}(|\mathbf{q}|,\tau)$.  A corollary of this observation is that, had we chosen to work with $F_{\nu H}$ instead of $\Psi_{\nu H}$, the associated collision integral and Boltzmann hierarchy must contain a decay term.

For the daughter particles $i=\nu_l$ and $i=\phi$, the order $\ell$ collision terms are given respectively by
\begin{equation}
\begin{aligned}
\mathcal{C}^{(1)}_{\ell}[\Psi_{\nu l}(q_2)] \! = \frac{\mathfrak{g}^2 a^2 (m_{\nu H} + m_{\nu l})^2}{4 \pi \, \epsilon_2 \, q_2 \, \bar{f}_{\nu l}(q_2)} \! \left[ \underbrace{ \int_{q_{1-}^{(\nu l)}}^{q_{1+}^{(\nu l)}} \!\! \mathrm{d}q_1 \, \frac{q_1}{\epsilon_1} \, \bar{f}_{\nu H}(q_1) \left( \Psi_{\nu H, \ell}(k, q_1) P_{\ell}\left(\cos{\alpha^*}\right) \! - \! \Psi_{\nu l, \ell}(k, q_2) \right)  }_{\text{dec}} \right. & \\
		- \underbrace{ \bar{f}_{\nu l}(q_2)   \int_{q_{3-}^{(\nu l)}}^{q_{3+}^{(\nu l)}}  \mathrm{d}q_3 \, \bar{f}_{\phi}(q_3) \, \Psi_{\phi, \ell}(k, q_3) \, P_{\ell}\left(\cos{\gamma^*}\right) }_{\text{inv}} & \\
		+ \underbrace{ \int_{q_{3-}^{(\nu l)}}^{q_{3+}^{(\nu l)}}  \mathrm{d}q_3 \, \bar{f}_{\nu H}\left(\sqrt{(\epsilon_2 \! + \! q_3)^2 \! - \! a^2 m_{\nu H}^2}\right) \, \bar{f}_{\phi}(q_3) \, \Psi_{\phi, \ell}(k, q_3) \, P_{\ell}\left(\cos{\gamma^*}\right) }_{\text{qs}} & \\
		+ \underbrace{ \int_{q_{1-}^{(\nu l)}}^{q_{1+}^{(\nu l)}} \mathrm{d}q_1 \, \frac{q_1}{\epsilon_1} \, \bar{f}_{\nu H}(q_1) \, \bar{f}_{\phi}(\epsilon_1 - q_2) \, \left(\Psi_{\nu H}(k, q_1) \, P_{\ell}\left(\cos{\alpha^*}\right) - \Psi_{\nu l, \ell}(k, q_2) \right) }_{\text{qs}} & \\
		\left. - \underbrace{ \bar{f}_{\nu l}(q_2) \int_{q_{1-}^{(\nu l)}}^{q_{1+}^{(\nu l)}} \mathrm{d}q_1 \, \frac{q_1}{\epsilon_1} \, \bar{f}_{\nu H}(q_1) \, \Psi_{\nu H, \ell}(k, q_1) \, P_{\ell}\left(\cos{\alpha^*}\right) }_{\text{qs}} \right] & , 
		\label{eq:nulcol}
\end{aligned}
\end{equation}
and
\begin{equation}
\begin{aligned}
\mathcal{C}^{(1)}_{\ell}[\Psi_{\phi}(q_3)] 
		\! = \frac{\mathfrak{g}^2 a^2 (m_{\nu H} + m_{\nu l})^2}{2 \pi q_3^2 \bar{f}_{\phi}(q_3)} \left[ \underbrace{ 
			 \int_{q_{1-}^{(\phi)}}^{q_{1+}^{(\phi)}} 
			 \mathrm{d}q_1 \, \frac{q_1}{\epsilon_1} \, \bar{f}_{\nu H}(q_1) \left( \Psi_{\nu H, \ell}(k, q_1) \, P_{\ell}\left(\cos{\beta^*}\right) - \Psi_{\phi, \ell}(k, q_3) \right) }_{\text{dec}} \right. & \\
		- \underbrace{  \int_{q_{2-}^{(\phi)}}^{q_{2+}^{(\phi)}} 	
			 \mathrm{d}q_2 \, \frac{q_2}{\epsilon_2} \, \bar{f}_{\nu l}(q_2) \, \Psi_{\nu l, \ell}(k, q_2) \, \bar{f}_{\phi}(q_3) \, P_{\ell}\left(\cos{\gamma^*}\right) }_{\text{inv}} & \\
		- \underbrace{  \int_{q_{2-}^{(\phi)}}^{q_{2+}^{(\phi)}} 			
			\mathrm{d}q_2 \, \frac{q_2}{\epsilon_2} \, \bar{f}_{\nu l}(q_2) \, \Psi_{\nu l,\ell}(k, q_2) \, \bar{f}_{\nu H}\left(\sqrt{(\epsilon_2 \! + \! q_3)^2 \! - \! a^2 m_{\nu H}^2} \right) \, P_{\ell}\left(\cos{\gamma^*}\right) }_{\text{qs}} & \\
		+ \underbrace{  \int_{q_{1-}^{(\phi)}}^{q_{1+}^{(\phi)}} 
			\mathrm{d}q_1 \, \frac{q_1}{\epsilon_1} \, \bar{f}_{\nu H}(q_1) \, \bar{f}_{\nu l}\left(\sqrt{(\epsilon_1 \! - \! q_3)^2 \! - \!  a^2 m_{\nu l}^2} \right) \left(\Psi_{\phi, \ell}(k, q_3) - \Psi_{\nu H,\ell}(k, q_1) \, P_{\ell}\left(\cos{\beta^*}\right)\right) }_{\text{qs}} & \\
		\left. + \underbrace{  \int_{q_{1-}^{(\phi)}}^{q_{1+}^{(\phi)}}  \mathrm{d}q_1 \, \frac{q_1}{\epsilon_1} \, \bar{f}_{\nu H}(q_1) \, \bar{f}_{\phi}(q_3) \, \Psi_{\nu H,\ell}(k, q_1) \, P_{\ell}\left(\cos{\beta^*}\right) }_{\text{qs}} \right]  & ,
		\label{eq:phicol}
\end{aligned}
\end{equation}
 where $\cos \alpha^*$ and $\cos \beta^*$ are identically the quantities given in equation~\eqref{kernel}, and we have in addition
\begin{equation}
	\cos{\gamma^*} = \frac{2 \epsilon_2 q_3 + a^2 (m_{\nu l}^2 - m_{\nu H}^2)}{2 q_2 q_3} .
\end{equation}
As in the first-order collision integrals for $\nu_H$, the integration limits here are again the same as those found in the background collision integrals~\eqref{eq:background_Boltzmann2} and ~\eqref{eq:background_Boltzmann3} for $\nu_l$ and $\phi$; the exact expressions are given in equation~(\ref{Equation:IntegralLimits2}) and~\eqref{Equation:IntegralLimits3}.

%%%%%%%%%%%%%
%%%%%%%%%%%%

\section{Numerical implementation}
\label{sec:numerical}

Having presented the relevant equations of motion in sections~\ref{sec:Background} and~\ref{sec:First order perturbations}, we are now in a position to numerically evaluate them.

We have implemented the evolution equations~(\ref{eq:background_Boltzmann1})--(\ref{eq:background_Boltzmann3}) for the background phase space densities of the three particle species $\nu_H$, $\nu_l$, and $\phi$ 
in the linear Einstein--Boltzmann solver \CLASS{}~\cite{Blas:2011rf}. Because none of the species already implemented in \CLASS{} requires the background distribution function to be evolved numerically, the number of background equations evolved by \CLASS{} in the current implementation becomes completely dominated by the three new species. Furthermore, the system of equations may become stiff, as the interaction time-scale of some momentum bins can be vastly smaller than the age of the Universe. We therefore modify \CLASS{} to use the 
\textit{ndf15}-evolver for the background equations instead of the Runge--Kutta evolver. The number of momentum bins needed to adequately compute the number and energy densities varies significantly over the decay parameter space, and scenarios with early recoupling in particular require a fine momentum resolution. 

Implementation of the first-order perturbation equations~(\ref{hierarchyPsi_nu1}) for $\nu_H$ and $\nu_l$
could proceed in principle via modifications to existing massive neutrino Boltzmann hierarchies in \CLASS{} to incorporate the new collision integrals ${\cal C}_\ell^{(1)}[\Psi_i]$ and background terms $\partial \ln \bar{f}_i/\partial \ln q$.  The scalar~$\phi$, on the other hand, could be implemented as a non-cold dark matter (albeit massless) species with its own Boltzmann hierarchy. This implementation is however outside of the scope of the present paper, and shall be postponed, along with  an MCMC analysis, to a future publication. Nevertheless, the numerical solutions of the background distributions alone already enable us to draw important conclusions about how the first-order inhomogeneous system should behave --- to  be discussed in detail in section~\ref{sec:ComparisonLorentz}.

%%%%%%
%%%%%%

\subsection{Massless daughter neutrino}
\label{sec:Massless}

Let us begin with the simpler case of a massless $\nu_l$.  Considering current measurements of the solar and atmospheric neutrino mass splittings, $\Delta m^2_{\rm sun}$ and $\Delta m^2_{\rm atm}$, from oscillations experiments, the assumption of $m_{\nu l}=0$ is clearly not a realistic one, unless we restrict our discussions to a few select values of $m_{\nu H}$ masses, i.e., $m_{\nu, H} = \sqrt{\Delta m^2_{\rm sun}}, \sqrt{|\Delta m^2_{\rm atm}|}$, or $\sqrt{\Delta m^2_{\rm sun}+ |\Delta m^2_{\rm atm}|}$.
Nevertheless, $m_{\nu l}=0$ is the assumption adopted in the  works~\cite{Hannestad:2005ex,Basboll:2008fx,Archidiacono:2013dua,Escudero:2019gfk,Basboll:2009qz} discussed in section~\ref{sec:PreviousWorks}; we therefore devote some time to it in order  to facilitate comparisons.  Outside of our  framework,
this limit is of course also applicable to, e.g., sterile neutrino decay scenarios such as that proposed in the context of short-baseline anomalies~\cite{Denton:2018dqq,Farzan:2019yvo}, wherein the daughter active neutrino masses may be negligible relative to the sterile neutrino mass, as well as those cases in which neutrinos decay entirely into dark radiation~\cite{Chacko:2019nej}. 

%%%%%%%%%%%%
%%%%%%%%%%%
\begin{figure}
\centering
\includegraphics[width=\textwidth]{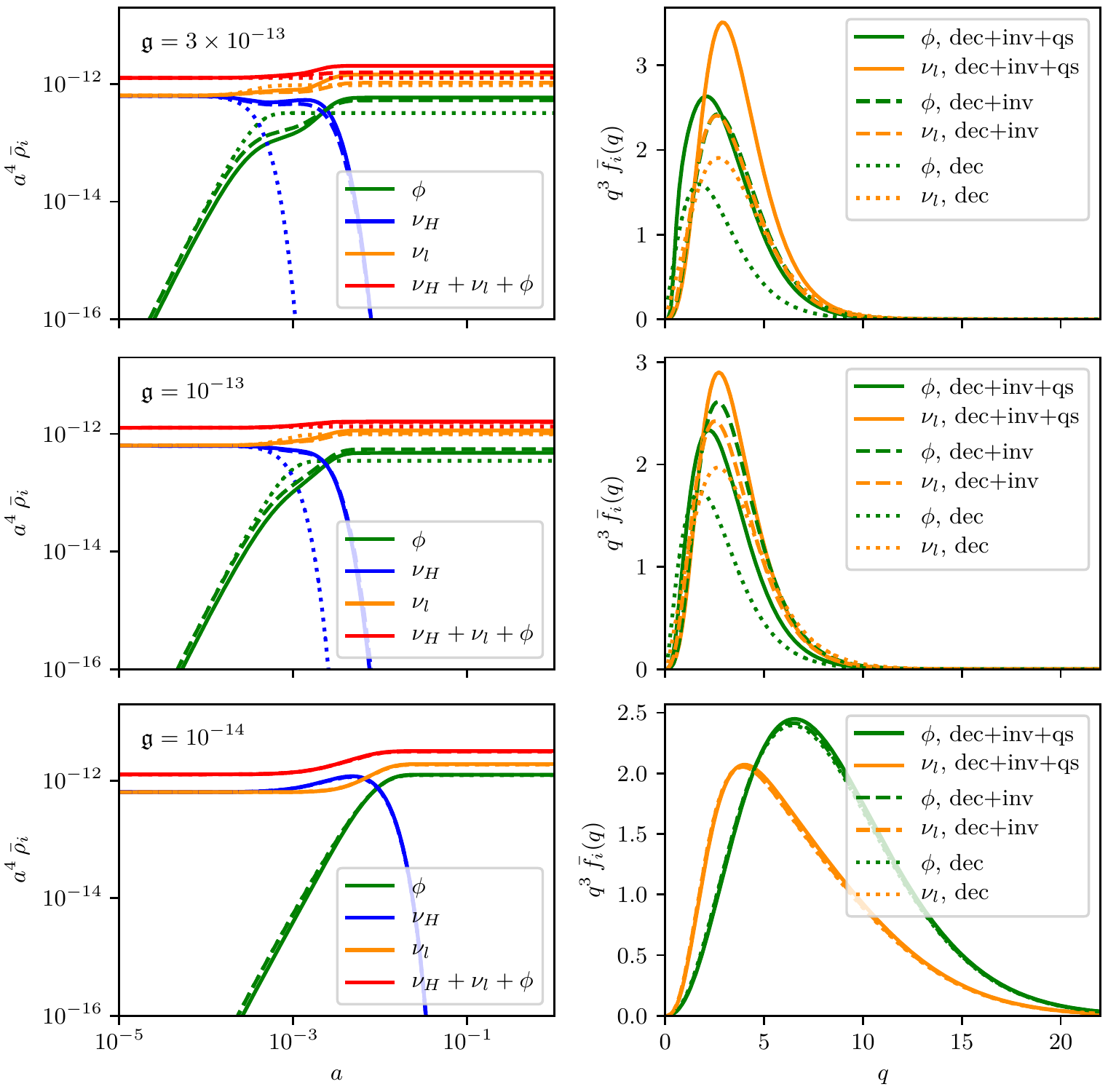}
\caption{\textit{Left:} Background energy densities of $\nu_H$ (blue), $\nu_l$ (orange), and $\phi$ (green), as well as their sum (red),
as functions of the scale factor $a$, computed under three different assumptions for the collision integrals.
Dotted lines represent computations incorporating only decay terms (``dec''), dashed lines denote contributions from decay and inverse decay (``dec+inv''), while solid lines contain all contributions (``dec+inv+qs'').
\textit{Right:} Final phase space distributions of $\nu_l$ and $\phi$. The comoving momentum $q$ is given in units of the present-day neutrino temperature $T_0$. All plots assume $m_{\nu_H}=0.3 \, \text{eV}$ and a massless $\nu_l$. }
\label{fig:energy_massless}
\end{figure}
%%%%%%%%%%%%
%%%%%%%%%%%

Figure~\ref{fig:energy_massless} shows  the evolution of the background energy densities of the various particle species for three different values of the coupling $\mathfrak{g}$ and a common $m_{\nu H}=0.3$~eV, and the associated phase space distributions of $\phi$ and $\nu_l$ at $a=1$.  
Three variants of the results are shown, representing three different combinations of the decay (``dec''), inverse decay (``inv''), and quantum statistics (``qs'') contributions to the collision integral used in the numerical solution of the background Boltzmann equations~(\ref{eq:background_Boltzmann1})--(\ref{eq:background_Boltzmann3}).
Note that a different choice of $m_{\nu H}$ would have led to significant quantitative changes to the evolution of the system; the qualitative conclusions of this section, however, remain valid.

%%%%%%%
%%%%%%%

\paragraph{Relativistic decay}

The limit of relativistic $\nu_H$ decay is represented in figure~\ref{fig:energy_massless} by the two larger coupling values of $\mathfrak{g}$ depicted in the top and middle rows.  Here, we see that neglecting inverse decay results in a significantly earlier disappearance of $\nu_H$ (blue lines).  Indeed, while in a decay-only calculation the disappearance of $\nu_H$  must be controlled by $\Gamma_{{\rm dec}}^0$ alone,
comparing the top left and middle left panels, we see that the total depletion of $\nu_H$ in the full dec+inv+qs calculation  is instead controlled by $m_{\nu H}$, or, equivalently, by the 
transition of the $\nu_H$ population from ultra-relativistic to non-relativistic, which shuts off inverse decay as it becomes increasingly kinematically unviable.

\begin{figure}
	\centering
	\includegraphics[width=\textwidth]{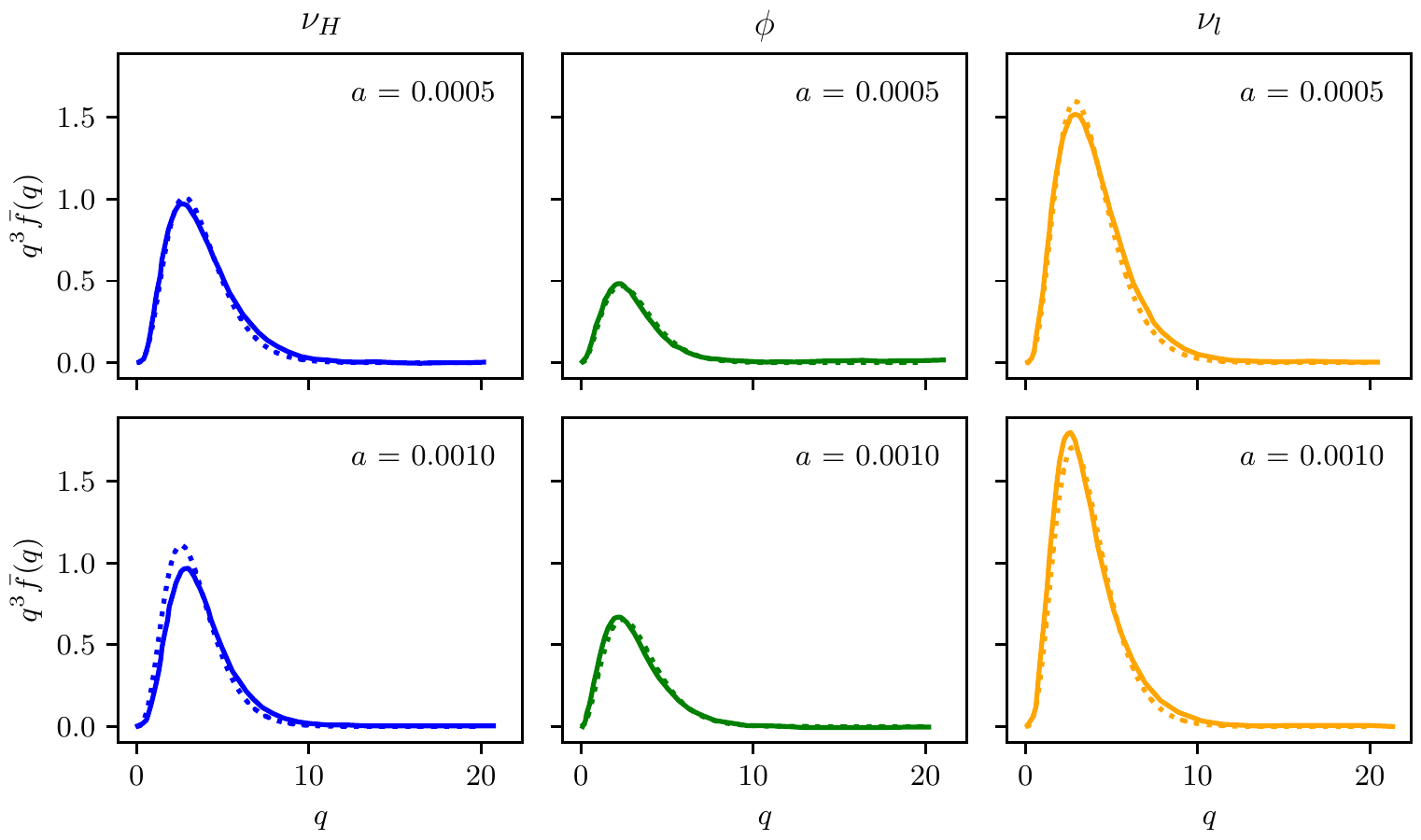}
	\caption{Background phase space distributions of, from left to right,  $\nu_H$, $\phi$, and $\nu_l$ at scale factors $a=0.0005$ (top) and $a=0.001$ (bottom).   The comoving momentum $q$ is given in units of the present-day neutrino temperature $T_0$.   All plots assume $m_{\nu H}=0.3$~eV, $\mathfrak{g}=3 \times 10^{-13}$, and massless decay products. 
		Solid lines denote the actual numerical output of \CLASS{}.  Dotted lines represent thermal distributions, $q^3 \bar{f}_{i,{\rm Th}} (q)= q^3/(\exp[(\epsilon_i(q)-\mu_i)/T_i] \pm 1)$, 
		 with temperatures $T_{\nu H, \phi, \nu l}/T_0= \{0.88, 0.8, 0.9\} $ and chemical potentials $\mu_{\nu H, \phi, \nu l}= \{0.15, -0.33, 0.51\}$ at $a=0.0005$, and  $T_{\nu H, \phi, \nu l}/T_0 = \{0.8, 0.84, 0.85\}$ and $\mu_{\nu H, \phi, \nu l} = \{0.46, -0.2,0.7\}$ at $a=0.001$.
		  	\label{fig:equil}}
\end{figure}

Another notable feature, especially in the top left panel of figure~\ref{fig:energy_massless}, is the appearance of an intermediate, quasi-steady state between the initial onset of decay and the final depletion of the $\nu_H$ population, where the energy densities of all species are temporarily almost constant or change at a much slower rate.  This is clearly a manifestation of the decay and inverse decay processes attaining a quasi-equilibrium on a time scale $\sim 1/\Gamma_{{\rm dec}}$, in the sense that at any instant the phase space densities $\bar{f}_i(|{\bf q}_i|)$ are essentially thermal, with an approximately common temperature and chemical potentials close to satisfying $\mu_{\nu H} = \mu_{\nu l} + \mu_\phi$;  figure~\ref{fig:equil} demonstrates this quasi-equilibrium.
 Because the final depletion of the $\nu_H$ population is determined by~$m_{\nu H}$ alone while the steady-state/quasi-equilibrium regime is triggered by $\Gamma_{\rm dec}$,  for the same  mass $m_{\nu H}$ we generally expect the duration of the steady state to be shorter for smaller couplings~$\mathfrak{g}$.

%%%%%%%%%%%%
%%%%%%%%%%%
%%%%%%%%%%%%
%%%%%%%%%%%

Furthermore, because inverse decay replenishes the $\nu_H$ population and holds off its disappearance until $\nu_H$ has transitioned to a non-relativistic species, the effect of the relativistic-to-non-relativistic transition is reflected in the total energy density of the system. Indeed, in the top left and middle left panels of figure~\ref{fig:energy_massless}, we see that the total energy of the system (solid red line) decreases momentarily as $\propto a^{-3}$, rather than $\propto a^{-4}$. In contrast, if we were to neglect inverse decay, the $\nu_H$ population would have been negligible before the mass could become relevant. This explains why the total energy density computed with decay-only (red dotted line) essentially follows the $\propto a^{-4}$ trajectory at all times. We therefore conclude that (i) a realistic treatment of relativistic decay must also account for the background effects of inverse decay, and (ii) any analysis that treats the neutrino--scalar system as a single massless fluid must invariably miss these effects.

Quantum statistics, on the other hand, do not change the evolution of the total energy density of the system, and play but a minor role in determining the onset of $\nu_H$ disappearance and of $\nu_l$ and $\phi$ production.  They do however alter the partition of the decay energy between the daughter $\nu_l$ and the $\phi$ sectors, an effect most easily discernible in the upper right panel of figure~\ref{fig:energy_massless}. Recall that, without quantum statistics,  the $\nu_l$ and $\phi$ background Boltzmann equations~(\ref{eq:background_Boltzmann2}) and (\ref{eq:background_Boltzmann3}) differ only by an overall factor of $2$ and their initial conditions. As such, the final $\nu_l$ and $\phi$ phase space distributions (dashed lines) are rather similar.  However, once Pauli blocking  and Bose enhancements have been accounted for, the difference between the two phase space distributions (solid lines) becomes noticeably larger.

%%%%%%%%%%
%%%%%%%%%

\paragraph{Non-relativistic decay}  For the smaller coupling value $\mathfrak{g}=10^{-14}$, decay happens only when the bulk of the $\nu_H$ population has become non-relativistic.  Expectedly, inverse decay is kinematically suppressed and quantum statistics likewise turn out to be negligible, as is evident in the bottom left panel of figure~\ref{fig:energy_massless}.
We can therefore conclude that, in the non-relativistic decay scenario, the collision part of the background Boltzmann equations \eqref{eq:background_Boltzmann1}--\eqref{eq:background_Boltzmann3} is indeed well approximated by the decay term alone.

Observe also in the lower right plot of figure \ref{fig:energy_massless} that non-relativistic $\nu_H$ decay tends to populate the high-momentum tail of the daughter particles' phase space distributions --- for reference, a thermal distribution peaks at $q \sim 2 \to3$.  The reason is that while the decay turns $m_{\nu H}$ into kinetic energy for the daughter particles, there is no particle scattering within the $\nu_l$ and $\phi$ populations to redistribute this energy to lower momenta (recall that inverse decay is now kinematically suppressed).  Consequently,  for a given~$m_{\nu H}$, the smaller the coupling~$\mathfrak{g}$, the higher the momentum of the tail section that the decay tends to populate.

%%%%%%%%%%%
%%%%%%%%%%%%

\subsection{Realistic neutrino mass ordering}
\label{sec:Massive}

Consider now a realistic mass spectrum for the three active neutrino species, whose mass eigenstates are denoted $\nu_1,\nu_2$ and $\nu_3$.  Two orderings of their masses are currently allowed by oscillations experiments: (i)~the normal hierarchy (NH), in which $m_1<m_2<m_3$ by standard convention, and (ii)~the inverted hierarchy (IH), with
 $m_3<m_1<m_2$. In both cases, the squared mass splittings obey~\cite{Tanabashi:2018oca}
\begin{equation}
\begin{aligned}
\Delta m_{21}^2 & \equiv m_2^2-m_1^2 = (7.53 \pm 0.018) \times 10^{-5} \,  \text{eV}^2, \\
\Delta m_{32}^2 & \equiv  m_3^2 - \frac{m_1^2+m_2^2}{2} =
\begin{cases}
(2.444 \pm 0.034) \times 10^{-3} \, \text{eV}^2 &(\text{NH})\\
(-2.55 \pm 0.04) \times 10^{-3} \, \text{eV}^2 \hspace{1cm} &(\text{IH}) 
\end{cases}.
\end{aligned}
\label{eq:mass_splittings}
\end{equation}
Correspondingly, the rest-frame decay rate of neutrino species~$i$ to a lighter neutrino species~$j$ can now be written as
\begin{equation}
\Gamma_{i \rightarrow j}^0= \frac{1}{\tau_0}= \frac{\mathfrak{g}^2}{4 \pi} \frac{(m_i^2-m_j^2)(m_i+m_j)^2}{m_i^3},
\label{eq:RestDecayRate}
\end{equation}
where we note that the expression differs from the much simpler equation~(\ref{eq:decayrate}) because of phase space blocking from a nonzero daughter neutrino mass.

%%%%%%%%%
%%%%%%%%%

\subsubsection{Simplification from a three-state to a two-state system}
\label{sec:twostate}

In general,  three disparate neutrino mass eigenstates and a nonzero transition probability between all possible pairs mean that all three neutrino species and the common $\phi$ particle must now partake in two different decay/inverse decay processes each. This necessitates that we compound the set of background Boltzmann equations~\eqref{eq:background_Boltzmann1}--\eqref{eq:background_Boltzmann3} {\it and} the first-order Boltzmann hierarchy~(\ref{hierarchyPsi_nu1})
each with a second set of collision integrals.
However, because of an inherent hierarchy in the neutrino mass splittings~(\ref{eq:mass_splittings}), under certain conditions it  is quite sufficient to treat $\nu_1$ and $\nu_2$ as degenerate species, and hence effectively reduce the three-state system to a two-state one.  

%%%%%%%%%
%%%%%%%%
\begin{figure}
\includegraphics[width=\textwidth]{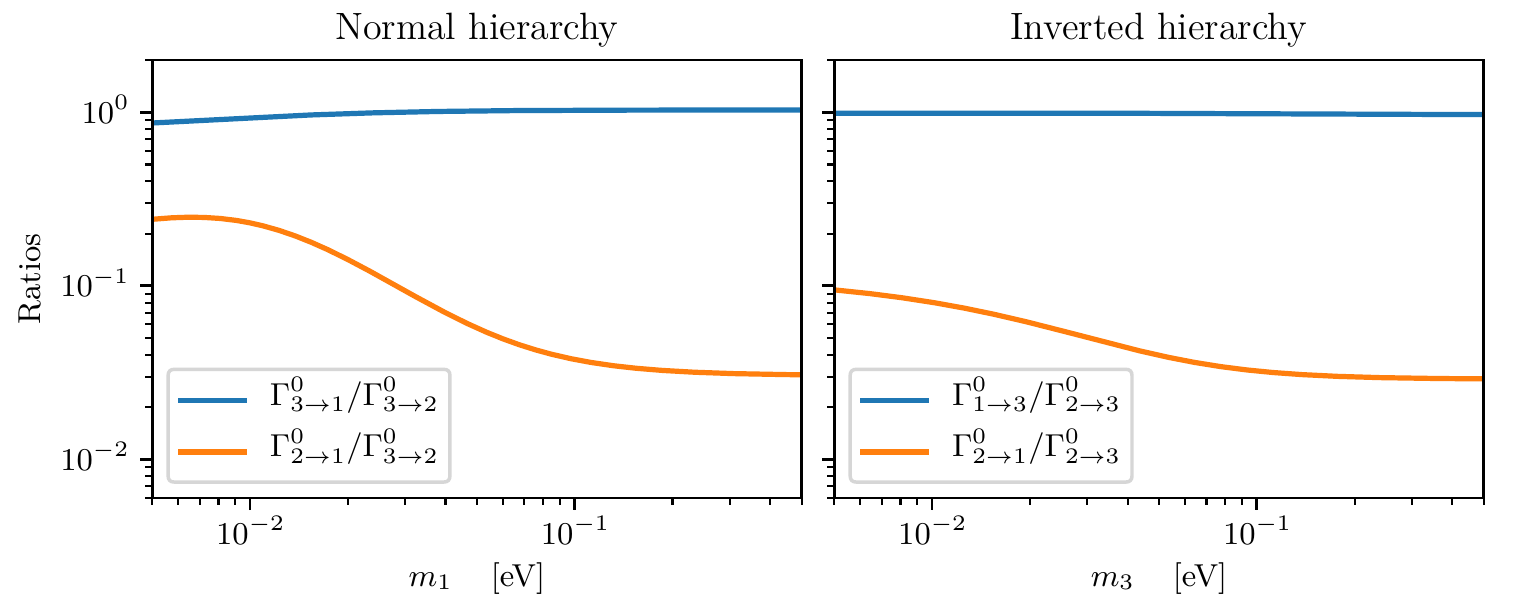}
\caption{Ratios of the rest-frame decay rates for the normal hierarchy (left) and the inverted hierarchy (right) as functions of the lightest neutrino mass.}
\label{fig:Massive_approx}
\end{figure}
%%%%%%%%
%%%%%%%%%

In the case of the normal hierarchy, the two lighter states $\nu_1$ and $\nu_2$ can be considered effectively degenerate in our Boltzmann framework if the conditions
\begin{equation}
\begin{aligned}
\Gamma^0_{3 \rightarrow 1} &\simeq \Gamma^0_{3 \rightarrow 2},\\
 \Gamma^0_{2 \rightarrow 1} &\ll \Gamma^0_{3 \rightarrow 2}
 \label{eq:nhcond}
 \end{aligned}
 \end{equation}
 are satisfied; the left panel of figure~\ref{fig:Massive_approx} shows the ratios $\Gamma^0_{3 \rightarrow 1} / \Gamma^0_{3 \rightarrow 2}$ and $\Gamma^0_{2 \rightarrow 1}/ \Gamma^0_{3 \rightarrow 2}$ as a function of the lightest mass $m_{1}$ .
 The analogous conditions for the inverted hierarchy are
 \begin{equation}
 \begin{aligned}
 \Gamma^0_{1 \rightarrow 3} & \simeq \Gamma^0_{2 \rightarrow 3}, \\
 \Gamma^0_{2 \rightarrow 1} & \ll \Gamma^0_{2 \rightarrow 3},
 \label{eq:ihcond}
 \end{aligned}
 \end{equation}
 where $\nu_1$ and $\nu_2$ are now the two heavier states; the ratios  $\Gamma^0_{1 \rightarrow 3} /\Gamma^0_{2 \rightarrow 3}$ and $\Gamma^0_{2 \rightarrow 1} /\Gamma^0_{2 \rightarrow 3}$ are displayed on the right panel of figure~\ref{fig:Massive_approx} as a function of $m_3$.
 
To deem the condition~(\ref{eq:nhcond}) or (\ref{eq:ihcond}) satisfied, we might demand that  $\Gamma^0_{2 \rightarrow 1}/\Gamma^0_{3 \rightarrow 2} < 0.1$ and
 $\Gamma^0_{3 \rightarrow 1} / \Gamma^0_{3 \rightarrow 2}> 0.9$ in the normal hierarchy, and  
$\Gamma^0_{2 \rightarrow 1}/\Gamma^0_{2 \rightarrow 3}< 0.1$ and $\Gamma^0_{1 \rightarrow 3} /\Gamma^0_{2 \rightarrow 3}> 0.9$ in the inverted case. 
Then, by these criteria, figure~\ref{fig:Massive_approx} suggests that the inverted hierarchy can, across the whole $m_3$ range, be well described by the degenerate-$\nu_{1,2}$ approximation.  On the other hand, the approximation may break down for masses $m_1 \lesssim 0.03$ eV in the normal hierarchy, where  $\Gamma^0_{2 \rightarrow 1}/\Gamma^0_{3 \rightarrow 2}$ in this $m_1$ region rises to above $0.25$.  We shall therefore restrict our attention to 
$m_1 \gtrsim 0.03$ eV in our analysis of the normal hierarchy to follow.

At the practical level, the background Boltzmann equations~\eqref{eq:background_Boltzmann1}--\eqref{eq:background_Boltzmann3} and the first-order Boltzmann hierarchy~(\ref{hierarchyPsi_nu1}) can be modified to describe a three-state system in the two-state, degenerate-$\nu_{1,2}$ limit as follows:
\begin{itemize}
\item For the normal hierarchy,  multiply by 2 all collision terms in the $\phi$ and $\nu_H$ Boltzmann equations, as well as all momentum-integrated quantities (e.g., energy density) of~$\nu_l$.

\item For the inverted hierarchy, same procedure as for the normal hierarchy, but with the interchange $\nu_l \leftrightarrow \nu_H$.
\end{itemize}
The heavy and light neutrino masses are always related in this limit by
\begin{equation}
m_{\nu H}= \sqrt{m_{\nu l}^2+ |\Delta m_{32}^2|},
\label{eq:mnuH}
\end{equation} 
irrespective of the neutrino mass ordering.

%%%%%%%%%%
%%%%%%%%%%%

\subsubsection{Decays in the two-state approximation}

%%%%%%%%%%%%%%%
\begin{figure}
\centering
\includegraphics[width=\textwidth]{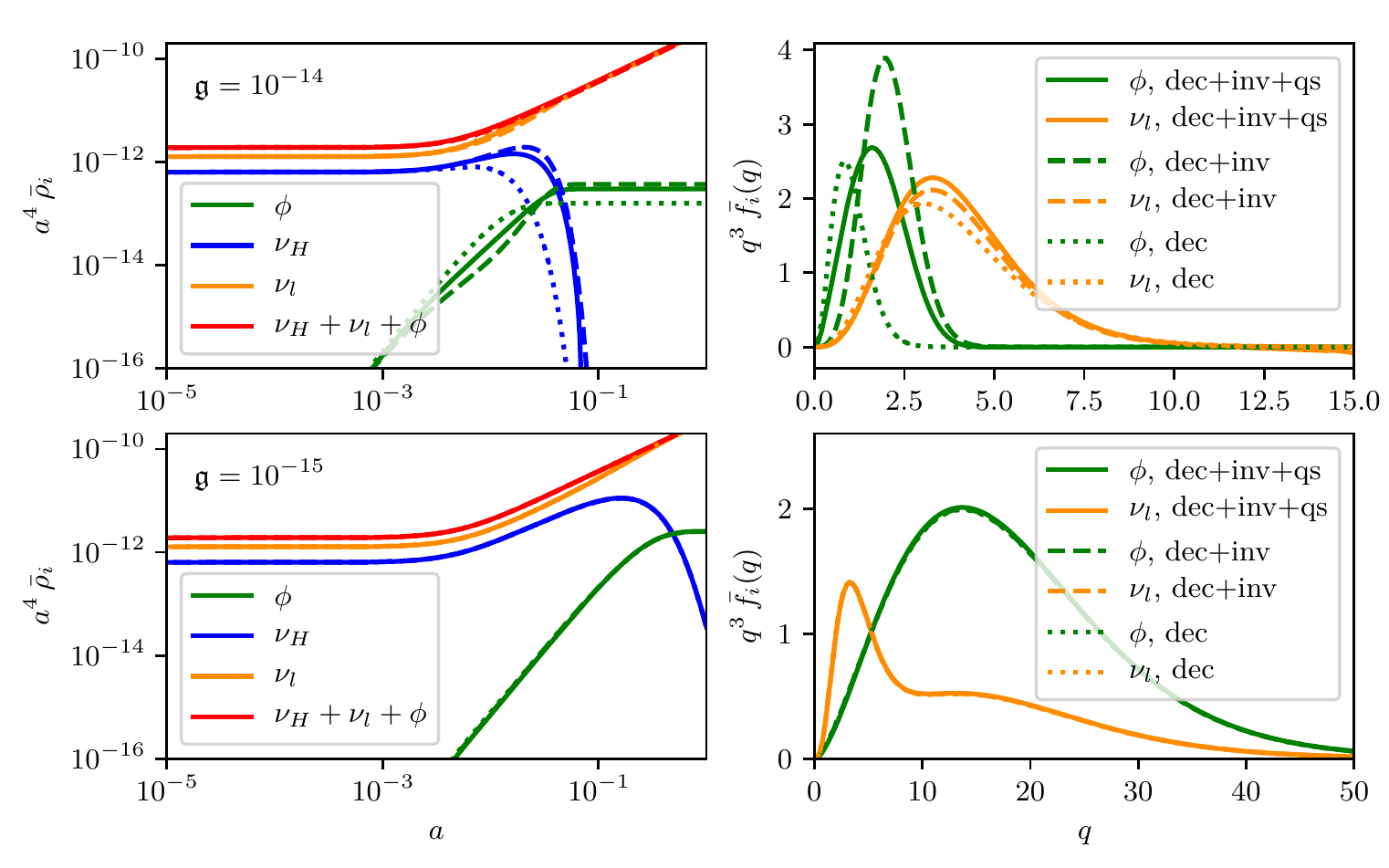}
\caption{Same as figure~\ref{fig:energy_massless}, but for a three-state system in the two-state limit (see section~\ref{sec:twostate}), assuming $m_{\nu l}=0.1$ eV, a normal mass ordering, and $m_{\nu H}$ satisfying equation~(\ref{eq:mnuH}). }
\label{fig:energy_massive}
\end{figure}
%%%%%%%%%%%%%%

Similar to figure~\ref{fig:energy_massless}, figure~\ref{fig:energy_massive} shows the evolution of the various energy densities (left) and final phase space distributions (right) for the normal hierarchy in the two-state 
approximation,
where we have assumed the  couplings $\mathfrak{g}=10^{-14}$ (top) and $\mathfrak{g}=10^{-15}$ (bottom), and a common light neutrino mass~$m_{\nu l}=0.1$~eV (and hence $m_{\nu H} \simeq 0.11$~eV according to equation~(\ref{eq:mnuH})).

As before, inverse decay and  quantum statistics impact significantly only on the case of a large coupling~$\mathfrak{g}$ ($=10^{-14}$), wherein decay begins while the bulk of the $\nu_H$ species is still ultra-relativistic.  For late decays ($\mathfrak{g}=10^{-15}$) that occur when the $\nu_H$ population is already non-relativistic, we again find that using the decay-only collision term alone suffices the capture the full (background) dynamics of the system.  Note that, like the massless $\nu_l$ case studied in section~\ref{sec:Massless},
the late decay scenario is again accompanied by an over-population of the high-momentum tail of the daughter neutrino distribution, although, for the same $m_{\nu H}$, not to the same extreme because a portion of the energy released in the decay is reserved for a finite $m_{\nu l}$.  A large population of massive $\nu_l$ in the high energy tail 
delays the onset of the species' transition to a non-relativistic gas, and this fact has previously been exploited to relax cosmological bounds on the neutrino mass sum~\cite{Oldengott:2019lke}.

\begin{figure}[t]
\centering
\includegraphics[width=\textwidth]{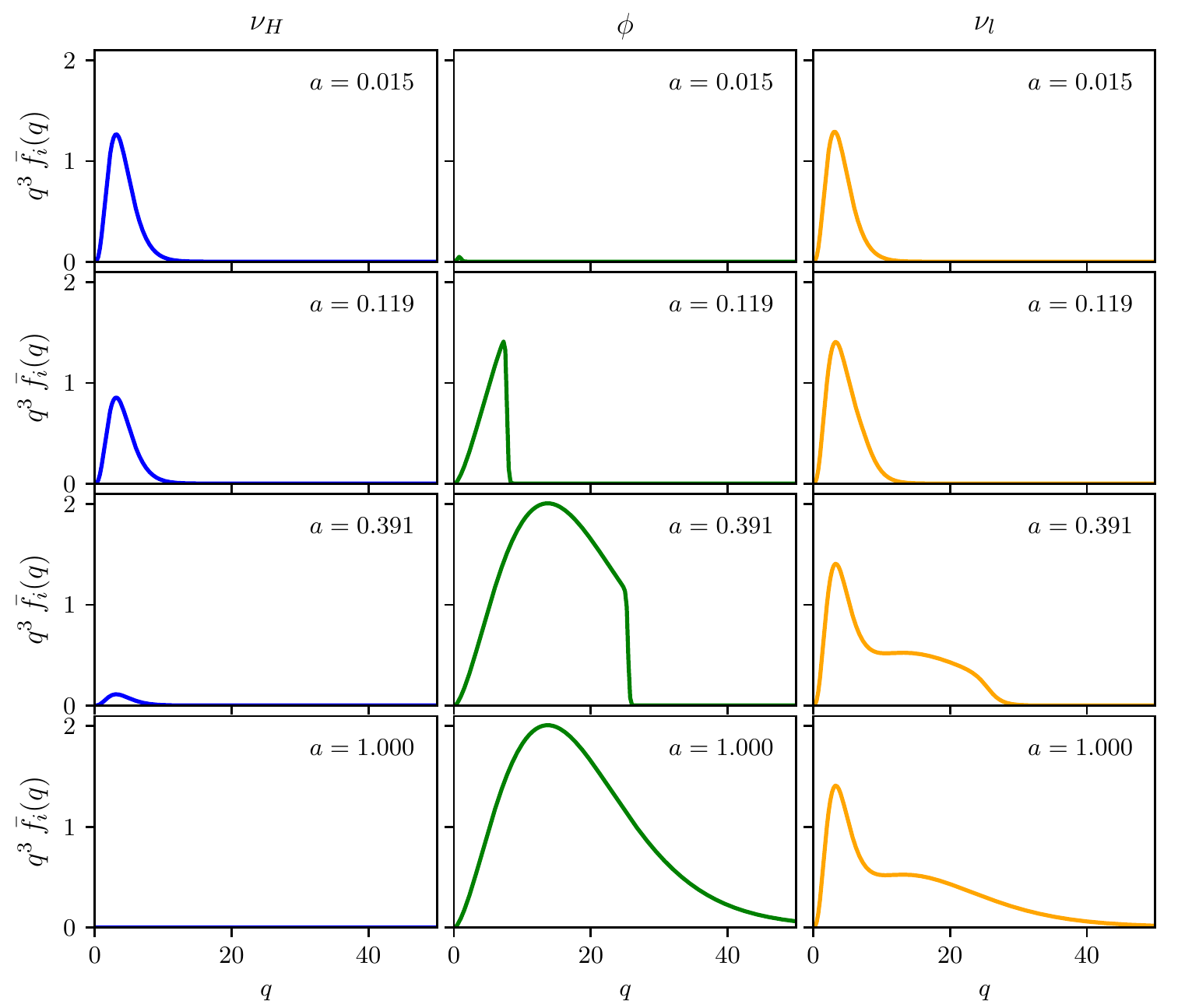}
\caption{Background phase space distributions of, from left to right,  $\nu_H, \phi$, and $\nu_l$ at scale factors, from top to bottom, $a=0.015, 0.119, 0.391,1$.   The comoving momentum $q$ is given in units of the present-day neutrino temperature $T_0$.   All plots assume $m_{\nu l}=0.1$ eV, $\mathfrak{g}=10^{-15}$, and a normal neutrino mass hierarchy.}
\label{fig:evolution}
\end{figure}

Let us also scrutinise the evolution of the phase space distributions in the non-relativistic decay limit in some detail.  Figure~\ref{fig:evolution} shows $q^3 \bar{f}_i(q)$ for $i,=\nu_H, \phi, \nu_l$ at scale factors~$a=0.015, 0.119, 0.391, 1$, assuming the parameter combination  $\mathfrak{g}=10^{-15}$, $m_{\nu l}=0.1$~eV, and a normal mass hierarchy.
 At intermediate times ($a=0.119$), the  $\phi$ distribution develops a sharp edge that progresses to larger (comoving) momenta with time.  The $\nu_l$ distribution, on the contrary, is always smooth.  These contrasting outcomes can be understood as follows.
 
 %%%%%%%%%%
 %%%%%%%%%%%

 In the rest frame of the decaying $\nu_H$, the decay products $\phi$ and $\nu_l$ are emitted in opposite directions but with a common absolute comoving momentum 
 \begin{equation}
 q'=a \frac{m_{\nu H}^2-m_{\nu l}^2}{2 m_{\nu H}}=a \frac{\Delta m^2_{32}}{2 m_{\nu H}},
 \label{eq:qprime}
 \end{equation}
which increases with the scale factor~$a$ --- the main reason why the edge in the $\phi$ distribution moves right with time.  Boosting to the cosmic frame, this comoving momentum transforms according to 
\begin{equation}
\begin{aligned}
q_{\phi} &= q' \, \gamma_{\nu H} \left( 1-v_{\nu H} \cos \theta' \right) \simeq q' \, \left(1-v_{\nu H} \cos \theta' \right), \\
q_{\nu l} & = q'\, \left[ \sin^2 \theta' + \gamma_{\nu H}^2 \left(\cos \theta' + v_{\nu H}  \frac{\epsilon'_{\nu l}}{q'_{\nu l}} \right)^2 \right]^{1/2}
\simeq q'\, \left(1 + v_{\nu H}  \frac{\epsilon'_{\nu l}}{q'_{\nu l}} \cos \theta' \right),
\label{eq:qphiqnul}
\end{aligned}
\end{equation}
where $\gamma_{\nu H} \equiv \epsilon_{\nu H}/a m_{\nu H}$ is the Lorentz factor, $v_{\nu H}$ the velocity of $\nu_H$, $\cos \theta'$  the rest-frame emission angle of $\phi$ relative to the boost direction, and  we have expanded both expressions to linear order in $v_{\nu H}$ at the approximate equality.

If all emission angles $\cos \theta' \in [-1,1]$ are equally probably, then we see immediately in equation~\eqref{eq:qphiqnul} that, in the limit of  non-relativistic $\nu_H$ decay ($v_{\nu H} \ll 1$), 
$\phi$ can be emitted only within an extremely narrow comoving momentum range in the cosmic frame:
\begin{equation}
q_\phi  \simeq  q' \pm a\frac{m^2_{\nu H}- m^2_{\nu l}}{2 m_{\nu H}}  v_{\nu H}.
\end{equation}
In contrast, the comoving momentum interval in which $\nu_l$ can be emitted in the same $v_{\nu H} \ll 1$ limit  is enhanced by a factor of $\epsilon'_{\nu l}/q'_{\nu l}=(m_{\nu H}^2+m_{\nu l}^2)/(m_{\nu H}^2-m_{\nu l}^2)$, i.e., 
\begin{equation}
q_{\nu l}   \simeq  q'  \pm a\frac{m_{\nu H}^2 + m_{\nu l}^2}{2 m_{\nu H}} v_{\nu H}.
\end{equation}
This explains why the $\phi$ distribution in figure~\ref{fig:evolution} comes with a sharp edge, while the $\nu_l$ distribution is relatively smooth at all times.
We remark in passing that the same sharp feature appears in fact also in the scenario of non-relativistic $\nu_H$ decay into massless daughters studied in section~\ref{sec:Massless}.  However, because in that case the mothers neutrino's mass is entirely converted into the daughter particle's momenta, the sharp edges in the $\phi$ and $\nu_l$ distributions appear at extremely large momenta, making them difficult to illustrate graphically.

Lastly, note that while we have presented results only for the normal hierarchy, the qualitative features of decay in the inverted hierarchy are in fact very similar to those shown in figures~\ref{fig:energy_massive} and~\ref{fig:evolution}.  The main quantitative differences trace their origins to the various factors of 2 inserted at different, hierarchy-dependent  points in the Boltzmann equations in the two-state approximation:  For the same $\mathfrak{g}$ and $m_{\nu H}$, 
 decay is delayed in the inverted hierarchy relative to the normal hierarchy, because $\nu_H$ has only half the number of decay channels in IH compared with NH.  The same delay also causes the $\nu_l$ and $\phi$ phase space distributions to be populated at even higher momenta in IH than those shown in figure~\ref{fig:energy_massive} for NH.

%%%%%%%%%%%
%%%%%%%%%%%

\section{Comparison with other works}
\label{sec:Comparison}

In light of our numerical results for the background phase space distributions and energy densities presented  in section~\ref{sec:numerical},  we are now in a position to re-examine several of references~\cite{Hannestad:2005ex,Basboll:2008fx,Archidiacono:2013dua,Escudero:2019gfk,Basboll:2009qz,Kaplinghat:1999xy, Chacko:2019nej,Chacko:2020hmh}'s claims discussed earlier in section~\ref{sec:PreviousWorks}. 

%%%%%%%
%%%%%%%

\subsection{Relativistic decay}
\label{sec:ComparisonLorentz}

As explained in section~\ref{sec:reldecay}, the crucial quantity in the relativistic $\nu_H$ decay scenario is the rate at which the combined neutrino--scalar fluid isotropises.  This rate is assumed in~\cite{Hannestad:2005ex,Basboll:2008fx,Archidiacono:2013dua,Escudero:2019gfk} to be the transport rate given in equation~\eqref{eq:transport_rate}, which has the peculiar feature that it is the rest-frame $\nu_H$ decay rate $\Gamma_{\rm dec}^0$ scaled {\it not} only by one inverse Lorentz factor $(m_{\nu H}/E_{\nu H})$, but by  $(m_{\nu H}/E_{\nu H})^3$,
and the isotropisation is implemented either as an exponential or a step-function damping of the combined fluid's anisotropic stress.
Our task in this section, therefore, is to see if such exponential damping at a rate scaling as $(m_{\nu H}/E_{\nu H})^3$ arises naturally in our first-order Boltzmann hierarchy~\eqref{hierarchyPsi_nu1} and associated collision integrals.
Of particular interest is the \(\ell=2\) multipole moment, because this is the lowest-order Legendre moment that represents anisotropy in the system.

%%%%%%%%%%%%
%%%%%%%%%%%%%

\subsubsection{Effective collision integral for the anisotropic stress at leading order}
\label{sec:effectivecollision}

The CMB observables are sensitive to the shear in the metric perturbations sourced by anisotropic stresses in various energy density components.  This is most easily seen in the conformal Newtonian gauge,
\begin{equation}
	k^2 (\phi - \psi) = 12 \pi a^2  G (\bar{\rho} + \bar{P}) \sigma \equiv 12 \pi G a^2 \, \Pi \, ,
\end{equation}
where we shall use the term total anisotropic stress when referring to the symbol~$\Pi$.  The contribution of the combined neutrino--scalar fluid to $\Pi$, $\Pi_{\nu \phi}$, is given in the $m_{\nu l} = m_{\phi} = 0$ limit by~\cite{Ma:1995ey}%
\footnote{The normalisation of our expression differs at face value from the definition of~\cite{Ma:1995ey},  because the latter chose to absorb the factor $g_i/(2 \pi)^3$ into their definition of the phase space density~$f$.}
\footnote{The quantity ${\cal F}_2$ used in reference~\cite{Escudero:2019gfk}
	(see also equation~\eqref{eq:exponentialsuppression})
	 is related to $\Pi_{\nu \phi}$ via $\Pi_{\nu \phi} = (\bar{\rho}_{\nu \phi} + \bar{P}_{\nu \phi}) {\cal F}_2 /2$.\label{footnote:F}}
\begin{equation}
	\begin{aligned}
		a^4 \Pi_{\nu\phi} \equiv \frac{1}{3 \pi^2} \left[ g_{\nu H} \int \mathrm{d}q_1 \, q_1^2 \, \frac{q_1^2}{\epsilon_1} \, F_{\nu H, 2}(q_1) + g_{\nu l} \int \mathrm{d}q_2 \, q_2^3 \, F_{\nu l, 2}(q_2) + g_{\phi} \int \mathrm{d}q_3 \, q_3^3 \, F_{\phi, 2}(q_3) \right] \, ,
		\label{eq:a4pidef}
	\end{aligned}
\end{equation}
which has an effective collision integral   
\begin{equation}
	\begin{aligned}
	&\left(\frac{\mathrm{d} (a^4\Pi_{\nu\phi})}{\mathrm{d}\tau}\right)_C  = \\
	& \qquad \frac{1}{3 \pi^2}
		\left[	g_{\nu H} \int \mathrm{d}q_1 \, q_1^2 \, \frac{q_1^2}{\epsilon_1} \left(\frac{\mathrm{d} f_{\nu H}}{\mathrm{d}\tau}\right)_{C,  2}^{(1)} + g_{\nu l} \int \mathrm{d}q_2 \, q_2^3 \, \left(\frac{\mathrm{d} f_{\nu l}}{\mathrm{d}\tau}\right)_{C,  2}^{(1)} + g_{\phi} \int \mathrm{d}q_3 \, q_3^3 \left(\frac{\mathrm{d}f_{\phi}}{\mathrm{d}\tau}\right)_{C,  2}^{(1)}  \right]
	\label{eq:dpidt}
	\end{aligned}
\end{equation}
constructed from the  $\ell=2$ collision terms for the individual particle species presented in section~\ref{sec:First order perturbations} and appendix~\ref{app:CollisionIntegralReduction}.

To extract physically meaningful conclusions out of equation~\eqref{eq:dpidt}, 
let us first examine the scattering kernels, i.e., the Legendre polynomials $P_\ell(x)$ that go into the first-order collision integrals at $\ell=0,1,2$. In the relativistic limit, the bulk of the $\nu_H$ and $\nu_l$ populations have energies much larger than $m_{\nu H}$.  This prompts us to expand $P_\ell(x)$ in the small quantities $a m_{\nu H} / \epsilon_1$ and $a m_{\nu H} / \epsilon_2$.   Expanding to $\mathcal{O}(a^2 m_{\nu H}^2 / \epsilon_{1,2}^2)$ yields the relation
\begin{equation}
	P_2(x) = 3 P_1(x) - 2 P_0(x) + \mathcal{O}(a^4 m_{\nu H}^4 / \epsilon_i^4) \, , 
	\label{eq:maplegendre}
\end{equation}
which holds for all of $x = \cos{\alpha^*}, \cos{\beta^*},  \cos{\gamma^*}$.
Then, to $\mathcal{O}(a^2 m_{\nu H}^2 / \epsilon_{1,2}^2)$, we can recast the $\ell = 2$ collision integrals as
\begin{equation}
	\left(\frac{\mathrm{d}f_i}{\mathrm{d}\tau}\right)_{C, \ell = 2}^{(1)} \simeq - 2 \left(\frac{\mathrm{d}f_i}{\mathrm{d}\tau}\right)^{(1)}_{C, \ell = 0} (F_{j,0} \to F_{j,2}) +
	3 \left(\frac{\mathrm{d}f_i}{\mathrm{d}\tau}\right)_{C, \ell = 1}^{(1)}(F_{j,1} \to F_{j,2})  \, ,
	\label{eq:replaced}
\end{equation}
where the notation $F_{j,0} \to F_{j, 2}$ indicates that all multipole moments $F_{j,0}$ in the original $\left(\mathrm{d}f_i/\mathrm{d}\tau\right)_{C, \ell = 0}^{(1)}$ integrands
 need to be replaced with their $F_{j,2}$ equivalent and so on.
Physically, this replacement means that to $\mathcal{O}(a^2m_{\nu H}^2 / \epsilon_{1,2}^2)$, the evolution of $F_{i, 2}$ resembles an admixture of the monopole and the dipole behaviours.  Thus, in the same way that $F_{i, 0}$ and $F_{i, 1}$ are losslessly exchanged between species by the decay and inverse decay processes --- thereby leading to energy and momentum conservation in the neutrino--scalar system --- we expect the leading-order behaviour of $F_{i,2}$ to be a lossless inter-species transfer as well.

To estimate the rate at which $\Pi_{\nu \phi}$ evolves, we substitute equation~\eqref{eq:replaced} into the effective collision integral~\eqref{eq:dpidt} and likewise expand it out in powers of $a m_{\nu H} / \epsilon_{1,2}$.  The leading-order (LO) result is
\begin{equation}
\begin{aligned}
	&  \left(\frac{\mathrm{d}(a^4 \Pi_{\nu \phi})}{\mathrm{d}\tau}\right)_C^{\rm LO} \simeq  \\
	 & \qquad \frac{4 a^2 m_{\nu H}^2}{\langle\epsilon_1^2\rangle} \, \frac{1}{3 \pi^2} \int \mathrm{d}q_1 \, \langle\epsilon_1^2 \rangle \, \frac{q_1^2}{\epsilon_1} \left[\left(\frac{\mathrm{d}f_{\nu H}}{\mathrm{d}\tau}\right)^{(1)}_{C, 0} (F_{j,0} \to F_{j,2}) - \frac{3}{4}  \left(\frac{\mathrm{d}f_{\nu H}}{\mathrm{d}\tau}\right)_{C, 1}^{(1)} (F_{j,1} \to F_{j,2}) \right],
	\label{Eq:SigmaEvo}
	\end{aligned}
\end{equation}
where we have inserted a characteristic squared energy~$\langle \epsilon_1^2 \rangle \sim (5 \to 10) \, T_{\nu H}^2$ to preserve the dimension of the momentum-integral.  Observe that the rate of change of $\Pi_{\nu \phi}$ depends at leading order on the rate of change of the  $\nu_H$ population's $\ell = 2$ perturbations alone, but suppressed by  inverse Lorentz factor squared, $a^2 m_{\nu H}^2 / \langle \epsilon_1^2 \rangle$.
The dependence on the  $\ell=2$ collision integrals of $\nu_l$ and $\phi$ has vanished by the conservation equations~\eqref{eq:energy_conservation} and \eqref{eq:momentum_conservation}, because to $\mathcal{O}(a^2m_{\nu H}^2 / \epsilon_{1,2}^2)$ the inter-species exchange of $F_{i,2}$ is lossless.  

Thus, equation~\eqref{Eq:SigmaEvo} appears consistent with the heuristic intuition that, at leading order, the time scale on which the  anisotropic stress of the  neutrino--scalar system evolves is loosely some fundamental time scale of the system multiplied by a factor $\langle \epsilon_1^2 \rangle/a^2 m_{\nu H}^2$.   If that fundamental time scale is taken to be $1/\Gamma_{{\rm dec}}$, then equation~\eqref{Eq:SigmaEvo} implies a rate of change of $a^4 \Pi_{\nu \phi}$ proportional to $(a m_{\nu H}/\epsilon_1)^3 \Gamma_{{\rm dec}}^0$, like the transport rate~$\Gamma_{\rm T}$ of equation~\eqref{eq:transport_rate}.

%%%%%%%%%%
%%%%%%%%%%

\subsubsection{Solving the leading-order effective equation of motion}

Our rate estimates so far appear to be consistent with assumptions in the existing literature.
Let us now address a crucial question: does the leading-order effective collision integral~\eqref{Eq:SigmaEvo} necessarily imply an {\it exponential damping} of the anisotropic stress at a rate  proportional to $(a m_{\nu H}/\epsilon_1)^3 \Gamma_{{\rm dec}}^0$?  That is, can equation~\eqref{Eq:SigmaEvo}  be equivalently recast into  a relaxation form
\begin{equation}
\left(\frac{\mathrm{d}(a^4 \Pi_{\nu \phi})}{\mathrm{d}\tau}\right)_C^{\rm LO} \simeq - a \Gamma_{\rm T} (a^4 \Pi_{\nu \phi}),
\label{eq:expform}
\end{equation}
with a damping or relaxation rate~$\Gamma_{\rm T}$ equal to the transport rate of equation~\eqref{eq:transport_rate}?
 To answer these questions, we must examine $\left(\mathrm{d}f_{\nu H}/\mathrm{d}\tau\right)^{(1)}_{C,2}$ and hence the evolution of $F_{\nu H, 2}$.  
 
Recall from the discussions above that to 
$\mathcal{O}(a^2m_{\nu H}^2 / \epsilon_{1,2}^2)$,  $F_{i, 2}$ evolves like an admixture of $F_{i, 0}$ and $F_{i,1}$.  In fact, if we are concerned only with gross behaviours, then $P_2(x) \simeq P_1(x) \simeq P_0(x)$ to zeroth order in $a m_{\nu H} / \epsilon_{1,2}$, and $F_{i,0}$, $F_{i,1}$, and $F_{i,2}$ all evolve alike. 
This observation allows us to further simplify equation~\eqref{Eq:SigmaEvo} to
\begin{equation}
\left(\frac{\mathrm{d}(a^4\Pi_{\nu \phi})}{\mathrm{d}\tau}\right)_C^{\rm LO} \simeq  \frac{a^2 m_{\nu H}^2}{\langle \epsilon_1^2 \rangle} \,  \frac{1}{3 \pi^2}  \int \mathrm{d}q_1 \, \langle \epsilon_1^2 \rangle \, \frac{q_1^2}{\epsilon_1} \left(\frac{\mathrm{d}f_{\nu H}}{\mathrm{d}\tau}\right)^{(1)}_{C,\ell= 0}\!\! (F_{j,0} \!\to\! F_{j,2}), 
\label{Eq:SigmaEvo2}
\end{equation}
with the next-order correction entering at  $\mathcal{O}(a^4 m_{\nu H}^4 / \langle \epsilon_{1,2}^4 \rangle)$.

Our next task is to evaluate $\left(\mathrm{d}f_{\nu H}/\mathrm{d}\tau\right)^{(1)}_{C,0}$.  However, since our  modified version of \CLASS{} does not include first-order collision terms, we need to estimate it by other means. 
To this end, we observe firstly that $\left(\mathrm{d}f_{\nu H}/\mathrm{d}\tau\right)^{(1)}_{C,0}$
shares the same structure and scattering kernels with
the background collision integral $\left(\mathrm{d}f_{\nu H}/\mathrm{d}\tau\right)^{(0)}_{C}$.  This must be so, because the monopole perturbations to the phase space densities are but the isotropic component. 
 Secondly, collision integrals are always local in space,
in the sense that the Boltzmann collision operator does not contain spatial derivatives.
This means that in the absence of non-local effects caused by free-streaming and gravitational clustering (which are described by spatial derivatives), the   ``background'' Boltzmann equations~\eqref{eq:background_Boltzmann1}--\eqref{eq:background_Boltzmann3} must also apply locally at ${\bf x}$, provided we formally replace all occurrences of the background variables~$\bar{f}_i(q_i,\tau)$ with their local counterparts~$f_i({\bf x},q_i,\tau)$.  Then, establishing $\left(\mathrm{d}f_{\nu H}/\mathrm{d}\tau\right)^{(1)}_{C,0}$ and hence the evolution of $F_{\nu H,0}$ --- and $F_{\nu H,2}$ via the $F_{j,0} \to F_{j,2}$ mapping --- {\it under collisions alone} becomes simply a matter of running our modified version of \CLASS{} twice with two perturbatively different CMB temperatures (to represent spatial fluctuations) but otherwise the same cosmological parameters, and taking the differences of the two sets of outcomes.

%%%%%%%%%%%%%%%
\begin{figure}[t]
	\centering
	\includegraphics[width=1\textwidth]{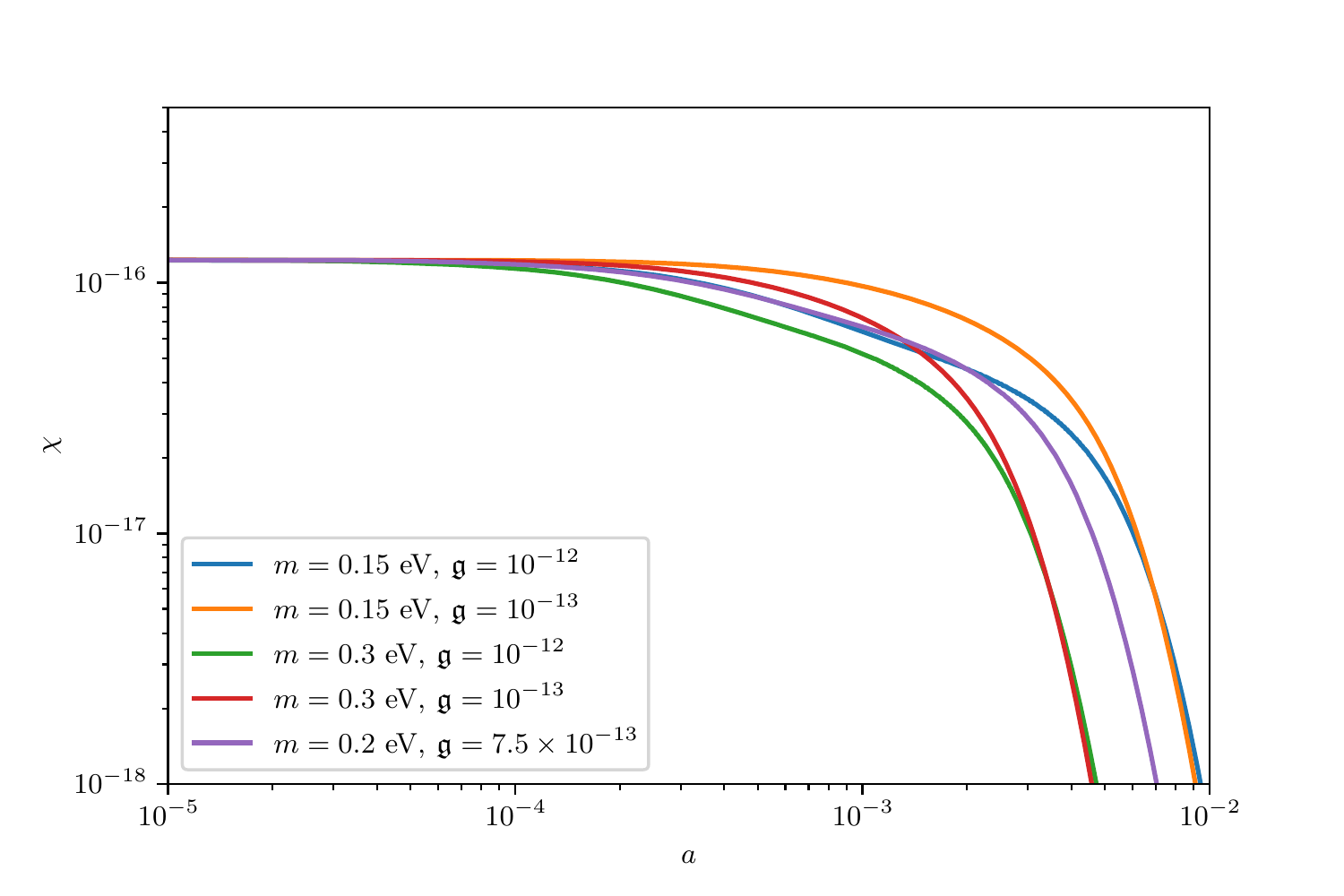}
	\caption{The function $\chi(a)$, defined in equation~\eqref{eq:chi}, for various combinations of $\nu_H$ masses and couplings.  The assumption of $m_{\nu l} = m_{\phi}=0$ is implicit.  Note  that $\chi$ has no physical interpretation;  it is merely a computational device we use to estimate the change in the total anisotropic stress according to the leading-order effective collision integral~\eqref{Eq:SigmaEvo2}.
	\label{fig:chi}}
\end{figure}
%%%%%%%%%%%%%%

Choosing for reasons of numerical stability a 1\% difference in the CMB temperature between runs, figure~\ref{fig:chi} shows our estimates of the quantity 
\begin{equation}
\chi(a) \equiv \frac{1}{3 \pi^2} \int {\rm d} q_1 \, \langle \epsilon_1^2 \rangle\, \frac{q_1^2}{\epsilon_1} F_{\nu H, 2} (q_1, a)
\label{eq:chi}
\end{equation}
for a range of $m_{\nu H}$ and coupling values.
 Note that because this estimate represents a linear perturbation, the normalisation is arbitrary.  
Like the background energy densities shown in figure~\ref{fig:energy_massless}, the function $\chi$ shows a steady decline after the onset of decay and suffers a complete depletion in the $m_{\nu H} \gtrsim T_{\nu H}$ regime.  One should bear in mind however that, while a useful computational device in conjunction with equation~\eqref{Eq:SigmaEvo2} for estimating the leading-order change in $a^4 \Pi_{\nu \phi}$,  the function~$\chi$ does not in fact correspond to any physical quantity.

\begin{figure}[t]
	\centering
	\includegraphics[width=1\textwidth]{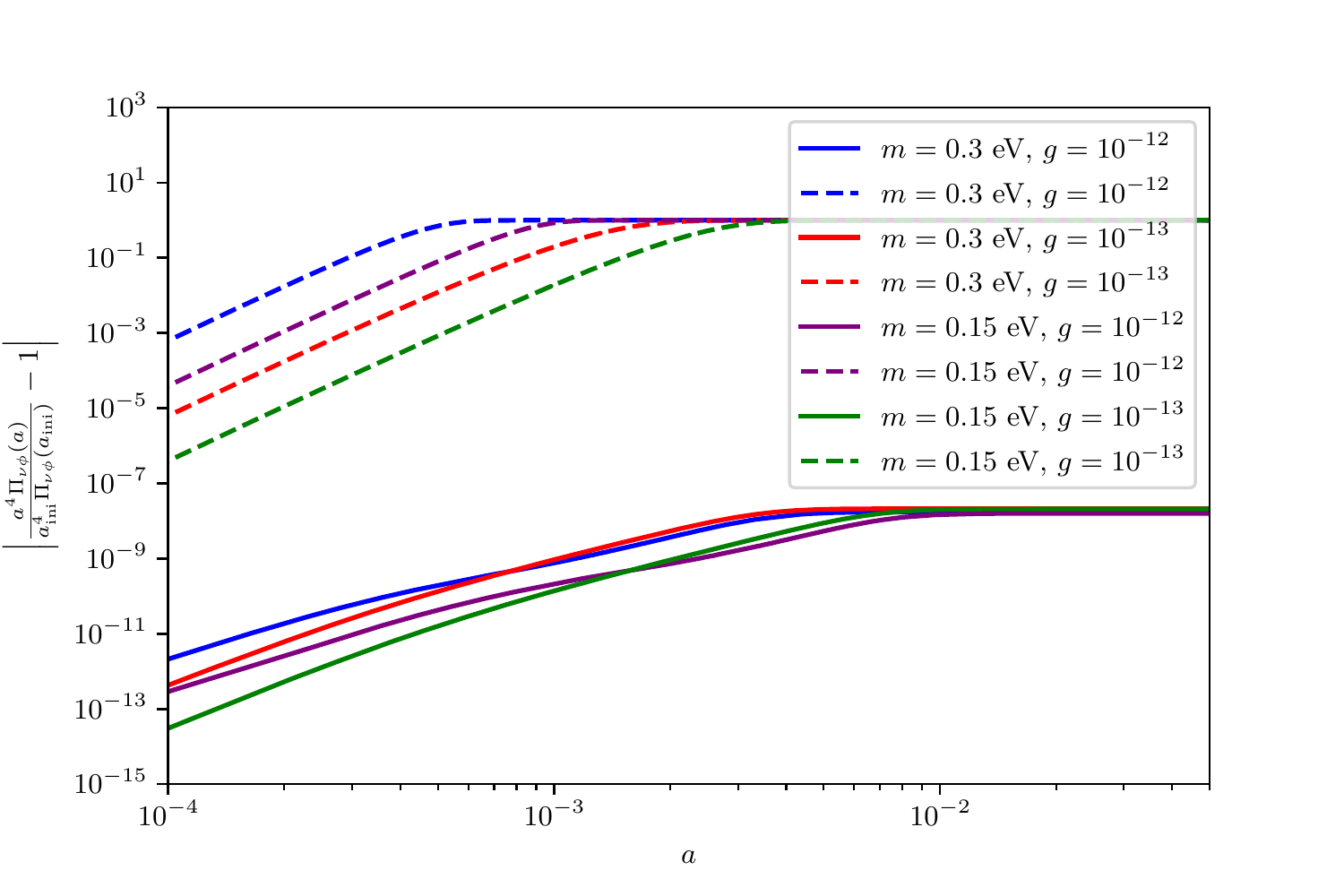}
	\caption{Leading-order fractional change in the total anisotropic stress $a^4 \Pi_{\nu \phi}$ of the combined neutrino--scalar system (solid lines), as estimated from equation~\eqref{eq:a4Pi} together with the result of figure~\ref{fig:chi}, for various combinations of $\nu_H$ masses and couplings.  The assumption of $m_{\nu l} = m_{\phi}=0$ is implicit.  For comparison we plot for each parameter combination also the corresponding exponential loss at the transport rate $\Gamma_{\rm T} =  (m_{\nu H}/E_{\nu H})^3 \, \Gamma_{\rm dec}^0$ (dashed lines). \label{fig:Sigma}}
\end{figure}

Using $\chi(a)$ as an external source on the r.h.s. of equation~\eqref{Eq:SigmaEvo2}, we can now write down an approximate analytical solution for the total anisotropic stress~$\Pi_{\nu \phi} (a)$ of the form  
\begin{equation}
a^4 \Pi_{\nu \phi}^{\rm LO} (a)  \simeq m_{\nu H}^2 \left[ \frac{a^2 \chi(a)}{ \langle \epsilon_1^2(a) \rangle} - \frac{a_{\rm in}^2\chi(a_{\rm in})}{ \langle \epsilon_1^2(a_{\rm in})\rangle} - 2 \int_{\ln a_{\rm in}}^{\ln a}  {\rm d} \ln a' \, \frac{a'^2 \chi(a')}{ \langle \epsilon_1^2(a') \rangle} \right] + a_{\rm in}^4 \Pi_{\nu \phi} (a_{\rm in})
\label{eq:a4Pi}
\end{equation}
at leading-order in $a m_{\nu H}/\epsilon_1$.   Figure~\ref{fig:Sigma} shows this solution as a fractional loss of the initial anisotropic stress,
$a^4 \Pi_{\nu \phi}^{\rm LO}(a)/a_{\rm in}^4 \Pi_{\nu \phi} (a_{\rm in}) -1$, for various decay parameter combinations.  Evidently, contrary to the assumptions of~\cite{Escudero:2019gfk,Basboll:2008fx,Hannestad:2005ex,Archidiacono:2013dua},
our first-principle derivation demonstrates that the leading-order effective collision integral~\eqref{Eq:SigmaEvo2} {\it does not cause an exponential damping} of the total anisotropic stress   in the limit of relativistic $\nu_H$ decay.  Instead, the actual loss of $a^4 \Pi_{\nu \phi}$ at this order in $a m_{\nu H}/\epsilon_1$ is very small if not altogether vanishing, and switches off completely in the limit the decay and inverse decay processes  attain an exact equilibrium, as we shall show below in section~\ref{sec:separable}

%%%%%%%%%%%%%%%
%%%%%%%%%%%%%%

The result of figure~\ref{fig:Sigma} also means that equation~\eqref{eq:expform}, with a transport rate given  by $\Gamma_{\rm T} = (a m_{\nu H}/\epsilon_1)^2 \, \Gamma_{\rm dec} = (a m_{\nu H}/\epsilon_1)^3 \, \Gamma_{\rm dec}^0$, is {\it not} a correct model to describe  the actual behaviour of the total anisotropic stress of the neutrino--scalar system.   Consequently, cosmological bounds on invisible neutrino decay and hence the neutrino lifetime based upon the transport rate of equation~\eqref{eq:transport_rate} --- either from solving equation~\eqref{eq:expform} (or, equivalently, equation~\eqref{eq:exponentialsuppression}) or its step-function variant discussed in section~\ref{sec:reldecay} --- are invalid.

%%%%%%%%%%%%%%
%%%%%%%%%%%%%%%

\subsubsection{Next-order loss rate and implications for neutrino lifetime bounds}
\label{sec:separable}

Having demonstrated quite generally that there is virtually no loss of anisotropic stress at the leading-order rate $\Gamma_{\rm T} = (a m_{\nu H}/\epsilon_1)^2 \, \Gamma_{\rm dec} = (a m_{\nu H}/\epsilon_1)^3 \, \Gamma_{\rm dec}^0$, let us now return to the original (and exact) collision integral~\eqref{eq:dpidt} for $a^4 \Pi_{\nu \phi}$, and extract from it the lowest-order non-vanishing contribution.
To make the calculation tractable, we make three simplifying assumptions: 
\begin{enumerate}
\item We assume equilibrium Maxwell--Boltzmann statistics, meaning that the background phase space density of each particle species can be described at any instant by $\bar{f}_i(q_i) = e^{-(\epsilon_i-\mu_i)/T_0}$, with a common temperature $T_0$ and chemical potentials satisfying $\mu_{\nu H} = \mu_{\nu l} + \mu_\phi$.  Physically, this means we take the equilibration time scale $1/\Gamma_{\rm dec}$ to be much shorter than the time scale of  anisotropic stress loss we are interested to compute.
As demonstrated in section~\ref{sec:Massless} and  figure~\ref{fig:equil}, the equilibrium conditions are reasonably well satisfied during the steady-state/quasi-equilibrium regime of relativistic decay.  
\item We take the Legendre moments $F_{i,\ell}(k,q_i)$ to be separable functions of $k$ and $q_i$.  This is a useful trick in near-equilibrium systems, when the momentum-dependence of the background phase space densities is largely constant in time.  For ultra-relativistic particle species, this separable ansatz takes the form~\cite{Oldengott:2017fhy}
\begin{equation}
F_{i,\ell} (k, q_i)  \simeq - \frac{1}{4} \frac{{\rm d} \bar{f}_i}{{\rm d} \ln q_i}\, {\cal F}_{i,\ell} (k),
\label{eq:separable}
\end{equation}
where ${\cal F}_{i,2}$ is the anisotropic stress contrast related to the actual anisotropic stress $\Pi_i$ in the species $i$ via $\Pi_i = (\bar{\rho}_i + \bar{P}_i) {\cal F}_{i,2}/2$.  See also footnote~\ref{footnote:F}.

\item In standard cosmology, anisotropic stress is induced by gravity after a $k$-mode enters the horizon.  Gravity induces the same perturbation contrast ${\cal F}_{i,\ell}$ in all particle species at the same point in space, i.e., 
\begin{equation}
 {\cal F}_{\nu H,\ell} (k) \simeq  {\cal F}_{ \nu l,\ell} (k)  \simeq  {\cal F}_{\phi,\ell} (k) \equiv  {\cal F}_{\ell} (k).
 \label{eq:sameperturbation}
\end{equation}
This constitutes our third approximation.
\end{enumerate}

Implementing these approximations into equations~\eqref{eq:a4pidef} and~\eqref{eq:dpidt} is a fairly straightforward exercise. The detailed calculation can be found in appendix~\ref{app:separable}.  Here, we merely quote the result:
\begin{equation}
\left(\frac{\mathrm{d}(a^4\Pi_{\nu \phi})}{\mathrm{d}\tau}\right)_C =
a \tilde{\Gamma}_{{\rm dec}} T_0^{-4}  \left\{
\int  \mathrm{d} q_1 \, q_1^3  e^{-\epsilon_1/T_0}
\left[ \frac{2\epsilon_1}{q_1} - \frac{q_1}{ \epsilon_1}- \frac{q_1^3}{\epsilon_1^3} -  \frac{3 a^2 m_{\nu H}^2}{\epsilon_1 q_1} 
\right] \right\}
(a^4 \Pi_{\nu \phi}),
\label{eq:loss1}
\end{equation}
with
\begin{equation}
\begin{aligned}
\tilde{\Gamma}_{\rm dec}& \equiv \Gamma_{{\rm dec}}^0 \frac{a m_{\nu H}}{24 T_0} \frac{\bar{n}_{\nu H}}{\sum_i \bar{n}_i} \left[\frac{1}{2}
\left(\frac{a m_{\nu H}}{T_0} \right)^2 K_2 \left(\frac{a m_{\nu H}}{T_0} \right)
\right]^{-1}\\
& \sim \Gamma_{{\rm dec}} \, ; \hspace{0.5cm} \text{for } \, \, a m_{\nu H}/T_0 \ll 1 \, , 
\label{eq:gammatilde}
\end{aligned}
\end{equation}
where $K_2(x)$ is a modified Bessel function of the second kind, and $\tilde{\Gamma}_{\rm dec}$ is, in the limit of a relativistic $\nu_H$ population, comparable to the (boosted) decay rate of equation~\eqref{eq:decayrate}. 
Expanding the term in the square brackets  in $a m_{\nu H}/\epsilon_1$, we find
\begin{equation}
\begin{aligned}
\left(\frac{\mathrm{d}(a^4\Pi_{\nu \phi})}{\mathrm{d}\tau}\right)_C & \simeq  -
a \tilde{\Gamma}_{{\rm dec}} T_0^{-4} 
 \left\{
\int  \mathrm{d} q_1 \, q_1^3 \,  e^{-\epsilon_1/T_0} \frac{a^4 m_{\nu H}^4}{\epsilon_1^4} \right\}
(a^4 \Pi_{\nu \phi})\\
& = - a \tilde{\Gamma}_{{\rm dec}} \left(\frac{a m_{\nu H} }{T_0} \right)^4   {\mathscr F}\left(\frac{a m_{\nu H}}{T}\right) \, (a^4 \Pi_{\nu \phi}) ,
\label{eq:newtransporteqn}
\end{aligned}
\end{equation}
where the dimensionless function~${\mathscr F}(x)$ is given by
\begin{equation}
{\mathscr F}(x) = 	\frac{1}{2}e^{-x} \left[-1 + x + e^x x^2 {\rm Ei}(-x)\right] + \Gamma(0,x),
\end{equation}
with Ei and $\Gamma$ representing respectively an exponential integral and an incomplete gamma function.  For $x \sim 10^{-10} \to 0.1$, 
${\mathscr F}(x)$ evaluates to  $\sim 20 \to 1$.  

Equation~\eqref{eq:newtransporteqn} tells us that, in the limit that decay and inverse decay occur at 
equilibrium rates, the leading-order anisotropic stress loss associated with the transport rate $(a m_{\nu H}/\epsilon_1)^3 \, \Gamma_{\rm dec}^0$ is {\it exactly vanishing}.  The first non-vanishing contribution is in fact proportional to $(a m_{\nu H}/\epsilon_1)^5 \, \Gamma_{\rm dec}^0$, i.e., the rest-frame decay rate suppressed by {\it five powers} of the inverse Lorentz factor.  Most interestingly, this  contribution is always negative and of the relaxation form
\begin{equation}
\begin{aligned}
\left(\frac{\mathrm{d}(a^4\Pi_{\nu \phi})}{\mathrm{d}\tau}\right)_C  \sim - a \, \Gamma_{{\rm dec}}^0 \left( \frac{m_{\nu H}}{E_{\nu H}}\right)^5
 (a^4 \Pi_{\nu \phi}), 
 \label{eq:power5}
\end{aligned}
\end{equation}
indicating an exponential damping of $a^4\Pi_{\nu \phi}$.

 We therefore conclude that the free-streaming argument outlined in section~\ref{sec:reldecay} commonly used to constrain relativistic invisible neutrino decay in the cosmological context is phenomenologically correct.  The anisotropic stress damping rate or transport rate, however, needs to be updated to that given in  equation~\eqref{eq:power5}.   
  Because for the same $\tau_0$ and $m_{\nu H}$ the revised rate is significantly smaller than the old transport rate~\eqref{eq:transport_rate}, assuming the loss of anisotropic stress and its effect on the CMB primary anisotropies to be the dominant phenomenology,
  we expect the lower limit on $\tau_0$ to be generally relaxed relative to the old constraint $\tau_0^{\rm old}$.
 Demanding that  $\Gamma_{{\rm dec}}^0 \left( m_{\nu H}/E_{\nu H}\right)^5 (a_{\rm fs})=H(a_{\rm fs})$ and $z_{\rm fs } \lesssim 1965$~\cite{Archidiacono:2013dua},  we find a new limit
 \begin{equation}
 \begin{aligned}
 \tau_0 & \gtrsim \tau_0^{\rm old} \, a_{\rm fs}^2 \left(\frac{50\, {\rm meV}}{y T_0} \right)^2 \left(\frac{m_{\nu H}}{50~\, {\rm meV}}
 \right)^5\\
 & \simeq (4\times 10^5 \to 4 \times 10^{6})\, {\rm s} \left(\frac{m_{\nu H}}{50\, {\rm meV}}
 \right)^5,
 \label{eq:newbound}
\end{aligned}
 \end{equation}
depending on what exactly we assume for $y = 1\to 3$ when evaluating $\epsilon_1 \simeq  y T_0$.

%%%%%%%%%%%%
\begin{figure}
	\centering
	\includegraphics[width=0.7\textwidth]{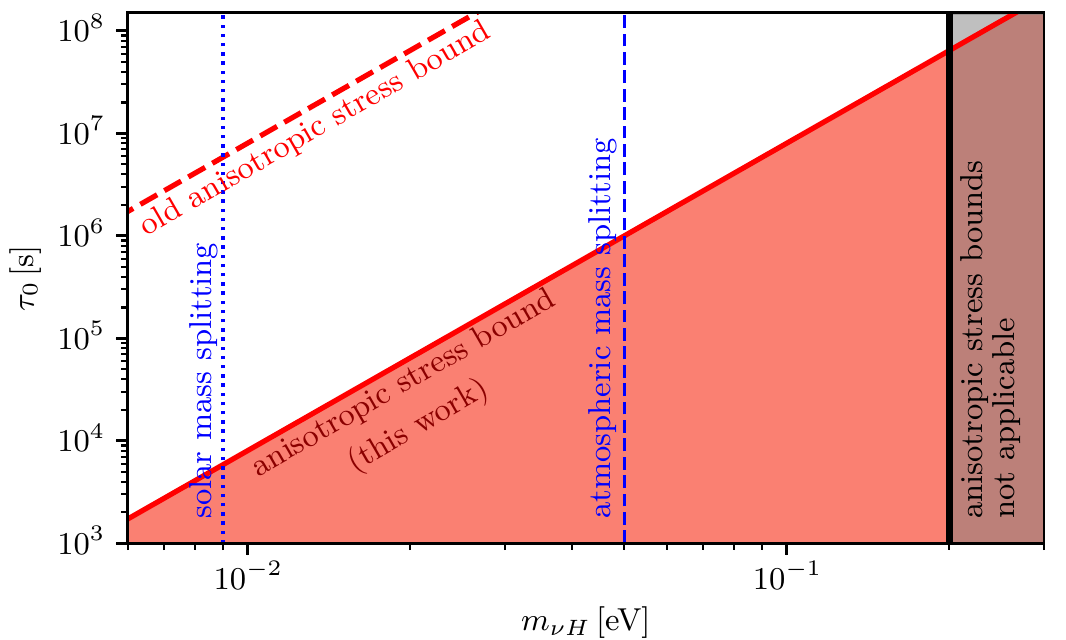}
	\caption{New constraint on the neutrino lifetime~$\tau_0$ in relation to the mother neutrino mass~$m_{\nu H}$ as per equation~\eqref{eq:newbound}, alongside the old constraint of references~\cite{Archidiacono:2013dua,Escudero:2019gfk}.
		Both constraints follow from the same anisotropic stress argument of section~\ref{sec:reldecay}, but have been derived using different damping rates.
		Note that the validity of these bounds are {\it  a priori} restricted to values of $m_{\nu H}$ smaller than the recombination temperature $T^* \sim 0.2$~eV.  The two sets of blue vertical dotted lines denote the lower mass limits set respectively by the solar and atmospheric squared mass splittings, $\sqrt{\Delta m^2_{\rm sun}}$ and solar $\sqrt{\Delta m^2_{\rm atm}}$. Note that the limits on $\tau_0$ have been derived assuming a massless daughter neutrino, which is strictly a correct assumption only if $m_{\nu H}$ does not exceed  $\sim \sqrt{\Delta m^2_{\rm atm}}$.   See also footnote~\ref{foot}. 
	\label{fig:transportrate_bounds2}}
\end{figure}
%%%%%%%%%s

Figure~\ref{fig:transportrate_bounds2} shows this new limit alongside the old one in  the $(m_{\nu H},\tau_0)$-plane.%
\footnote{For historical reasons we have presented the limit on $\tau_0$ in figure~\ref{fig:transportrate_bounds2} for a range of $m_{\nu H}$ up to $m_{\nu H} = 0.2$~eV, even though the limit is strictly valid only for a mother neutrino mass $m_{\nu H}$ smaller than the atmospheric mass splitting $\sqrt{\Delta m_{\rm atm}^2}$; for $m_{\nu H}$ values greater than  $\sqrt{\Delta m_{\rm atm}^2}$, the assumption of a massless daughter neutrino is not in fact compatible with neutrino oscillations parameters established by experiments.  Determining the exact impact of a massive daughter neutrino on the $\tau_0$ bound is however beyond the scope of the present work and will be deferred to a future publication.\label{foot}} 
 Again, we remind the reader that anisotropic stress bounds on the neutrino lifetime are only meaningful if $\nu_H$ should remain ultra-relativistic throughout the CMB epoch, a regime we have demarcated with a vertical black line in figure~\ref{fig:transportrate_bounds2}, representing $m_{\nu H} = T^* \sim 0.2$~eV, where $T^*$ is the recombination temperature.
Besides the physical reasons already discussed in section~\ref{sec:critique}, such a restriction is also necessary from the point of view of technicality: the derivation of the new transport rate~\eqref{eq:newtransporteqn} explicitly assumes $m_{\nu H} \ll E_{\nu H}$; any lifetime bound that follows from it must also be subject to the same kinematic constraints.

%%%%%%%
%%%%%%%%

\subsubsection{Final remarks}

While we have arrived at our new neutrino lifetime constraints~\eqref{eq:newbound} assuming for simplicity isotropic decay described by 
a Yukawa interaction~\eqref{eq:Lagrangian}, 
the general reasoning presented in all of section~\ref{sec:ComparisonLorentz} in fact generalises to other coupling structures which may give rise to more complex angular dependences.  The key argument of section~\ref{sec:effectivecollision}, for example,  i.e., the approximate mapping of the Legendre polynomials in equation~\eqref{eq:maplegendre}, follows simply from kinematics, and applies to all relativistic $1 \to 2$ decays.

Likewise, the structure of the integrand of equation~\eqref{eq:loss1} is purely a consequence of kinematics, and can be easily adjusted to accommodate a more complex decay matrix element via a common,  $\cos \alpha$-dependent multiplicative factor prior to integration over $q_2$ in equations~\eqref{eq:term1} and~\eqref{eq:term2a}.   Because any new appearances of the angle $\cos \alpha$ 
must be evaluated at its kinematically-determined value  $\cos \alpha^*$ --- which tends to unity in the limit of relativistic decay --- such a common multiplication can only affect higher-order terms in the subsequent small-$m_{\nu H}$ expansion; 
the leading-order term given in equation~\eqref{eq:newtransporteqn} is unaffected by new angular dependences brought on by a departure from the isotropic-decay assumption.
Consequently, the damping rate of the anisotropic stress is always suppressed by five powers of the inverse Lorentz factor in the limit of relativistic decay, irrespective of the exact Lagrangian that effects the interaction.

Our results also reveal that the original argument of near-collinearity of the decay products  used in~\cite{Chacko:2003dt,Hannestad:2005ex,Basboll:2008fx,Archidiacono:2013dua,Escudero:2019gfk} to establish the anisotropic stress damping rate ultimately tells only half of the story, as in its present form the argument is incompatible with  detailed balance.  Indeed, if transporting momentum by an angle $\pi/2$ using $\sim (E_{\nu H}/m_{\nu H})^2$ number of decay and inverse decay events was a sufficient condition to wipe out an isolated system's quadrupole anisotropy, then logically a similar number of events would flip the momentum direction by~$\pi$ and wipe out the system's dipole perturbation too.  This conclusion is clearly wrong, as any net momentum in an isolated system must be conserved by collisions.  

Rather, the argument needs to be extended to take into account  that only the portion of momentum associated with the transverse directions of the two decay products is transportable; the longitudinal component is conserved.  Because  the transportable component is suppressed by  $m_{\nu H}/E_{\nu H}$ relative to the conserved one --- which accounts for one new power of the inverse Lorentz factor in the correct damping rate,  it takes $\sim E_{\nu H}/m_{\nu H}$  events to make the transportable part comparable to the conserved part in one event --- which accounts for the second new power of $\gamma^{-1}$.  The near-collinearity argument follows on from this point.

To conclude this section, we comment on the analysis  of~\cite{Basboll:2009qz}, which found an additional, fast contribution to the transport rate that is suppressed  only by one power of the  inverse Lorentz factor $m_{\nu H}/E_{\nu  H}$ relative to the fundamental rate of the system.  This contribution arises purportedly from including a thermal background in the rate estimation. However, it remains  unclear to us how a single power of $m_{\nu H}/E_{\nu  H}$ can pop out of the estimation, since any small-$m_{\nu H}$ expansion of the collision integrals must produce a series in even powers of $m_{\nu H}$.  
Furthermore, reference~\cite{Basboll:2009qz} drew their conclusions exclusively from estimates of the {\it initial} rates of the system, i.e., rates computed by substituting in an initial value whenever a quantitative estimate of a {\it dynamical} function is called for.  Determining how a system should evolve on the basis of the initial conditions alone is a very misleading exercise,
as these initial rates quantify merely the evolution of transients. 
  Indeed, while the method of~\cite{Basboll:2009qz} does yield a  seemingly physically consistent negative rate of change for the anisotropic stress  for the initial conditions at hand,  one can easily engineer a different set of initial conditions that would cause the initial rate to flip sign.
Such estimates therefore must be avoided.

%%%%%%%%%
%%%%%%%%%

\subsection{Non-relativistic decay}
\label{sec:Comparison_Non-rel}

In section~\ref{sec:discrepancy} we argued that in the limit of an extremely large $\nu_H$ mass, the Boltzmann hierarchies in the neutrino decay scenario must converge to those of a decaying CDM.  In this connection, we noted that there appears to be a discrepancy between the daughter neutrino's Boltzmann hierarchy presented in~\cite{Kaplinghat:1999xy, Chacko:2019nej,Chacko:2020hmh} for the neutrino decay case 
and the analogous equations for CDM decay derived in~\cite{Audren:2014bca}.  We elaborate on  this discrepancy below.

Let us first examine the Boltzmann hierarchies for the mother particle $\nu_H$ and the decay products $\nu_l+\phi$, in the limit (i)~the $\nu_H$ population  approaches a fully cold fluid (i.e., $a m_{\nu H} \gg q_{\nu H}$ and $\epsilon_{\nu H} \simeq a m_{\nu H}$), (ii)~$\nu_l$ and $\phi$ are exactly massless, and (iii)  only the decay term enters the collision integrals. Then, 
defining the density contrast~$\delta_{i}$ and velocity divergence~$\theta_i$ of the $i$th fluid to be~\cite{Ma:1995ey}
\begin{equation}
\begin{aligned}
\delta_{i} & \equiv   \frac{(2 \pi^2 a^4)^{-1} g_i  \int \mathrm{d}q \, q^2 \epsilon_i \, F_{i, 0}}{\bar{\rho}_{i} }=\frac{\int \mathrm{d}q \, q^2 \epsilon_i \,  F_{i, 0}}{\int \mathrm{d}q \, q^2 \epsilon_i \, \bar{f}_{i} (q)}, \\
\theta_{i} & \equiv \frac{(2 \pi^2 a^4)^{-1} g_i \, k \int \mathrm{d}q \, {  q^3 }  \,  F_{i, 1}}{\bar{\rho}_{i} + \bar{P}_{i} }
=  \frac{k \, \int \mathrm{d}q \, { q^3 }  \, F_{i, 1}}{\int \mathrm{d}q \, q^2 \left(\epsilon_i + \frac{q^2}{3 \epsilon_i}\right)\,\bar{f}_{i} (q)}, 
\label{eq:deltatheta}
\end{aligned}
\end{equation}
it follows straightforwardly from our arguments immediately before equation~(\ref{eq:decayfactor}) that the familiar equations of motions for $\delta_{\nu H}$ and $\theta_{\nu H}$ describing a {\it  non-interacting} CDM, namely,
\begin{align}
\dot{\delta}_{\nu H} & = - \left ( \theta_{\nu H} + \frac{\dot{h}}{2} \right),
\label{eq:deltaH} \\
\dot{\theta}_{\nu H} & = -\frac{\dot{a}}{a} \, \theta_{\nu H},
\label{eq:thetaH}
\end{align}
must also be applicable to a fully {\it cold and decaying} $\nu_H$ population.  The same equations of motion have also been obtained previously in~\cite{Audren:2014bca} for the CDM decay scenario.%
\footnote{It is common practice to define the synchronous coordinates as the rest frame of the CDM fluid~\cite{Ma:1995ey}, in which case the velocity divergence is by definition zero.}

On the other hand, in order to match previous results for the decay products, we combine the $\nu_l$ and $\phi$ populations into a single massless fluid $\nu_l + \phi$, with a combined  density contrast and velocity divergence defined following~(\ref{eq:deltatheta}) by
\begin{equation}
\begin{aligned}
\delta_{\nu l+\phi} & \equiv  \frac{(2 \pi^2 a^4)^{-1} \int \mathrm{d}q \, q^3 \,\left( g_{\nu l} F_{\nu l, 0} + g_{\phi} F_{\phi, 0} \right)}{\bar{\rho}_{\nu l + \phi}} 
= \frac{\int \mathrm{d}q \, q^3 \,\left(2F_{\nu l, 0} + F_{\phi, 0} \right)}{\int \mathrm{d}q \, q^3\, \left(2\bar{f}_{\nu l} + \bar{f}_\phi \right)}, \\
\theta_{\nu l + \phi} & \equiv  \frac{(2 \pi^2 a^4)^{-1} k\, \int \mathrm{d}q \, q^3  \, \left( g_{\nu l} F_{\nu l, 1} + g_{\phi} F_{\phi, 1} \right)}{(4/3)  \bar{\rho}_{\nu l + \phi} }
= \frac{3k}{4}  \frac{\int \mathrm{d}q \, q^3  \, \left(2 F_{\nu l, 1} +  F_{\phi, 1} \right)}{\int \mathrm{d}q \, q^3 \,\left(2\bar{f}_{\nu l}  + \bar{f}_\phi  \right)},
\label{eq:deltanulphi}
\end{aligned}
\end{equation}
where $\bar{\rho}_{\nu l+\phi} \equiv \bar{\rho}_{\nu l}+ \bar{\rho}_\phi$.
Note that these definitions are only possible because we have taken the limit $\epsilon_{\nu l} = \epsilon_{\phi} = q$; they do not apply if either $\nu_l$ or $\phi$ is massive, nor is it generally feasible to derive a formally closed set of equations of motion for a mixed  fluid. 

Differentiating the expressions~\eqref{eq:deltanulphi} with respect to conformal time gives
\begin{eqnarray}
\dot{\delta}_{\nu l +\phi} & =&  \frac{\int \mathrm{d}q \, q^3 \left(2\dot{F}_{\nu l, 0}+ \dot{F}_{\phi, 0} \right)}{\int \mathrm{d}q \, q^3 \left(2\bar{f}_{\nu_l} +  \bar{f}_\phi\right)} - \frac{\int \mathrm{d}q \, q^3 \left(2\dot{\bar{f}}_{\nu l}+\dot{\bar{f}}_{\phi} \right)}{\int \mathrm{d}q \, q^3 \left(2\bar{f}_{\nu_l} +  \bar{f}_{\phi} \right)}  \delta_{\nu l + \phi}, \label{eq:delta_2}
 \\
 \dot{\theta}_{\nu l+\phi} & =& \frac{3k}{4}  \frac{\int \mathrm{d}q \, q^3 \left(2 \dot{F}_{\nu l, 1}+ \dot{F}_{\phi, 1}\right)}{\int \mathrm{d}q \, q^3 \left(2\bar{f}_{\nu_l}+  \bar{f}_{\phi} \right)} - \frac{\int \mathrm{d}q \, q^3 \left(2\dot{\bar{f}}_{\nu l}+\dot{\bar{f}}_{\phi} \right)}{\int \mathrm{d}q \, q^3 \left(2 \bar{f}_{\nu_l}+ \bar{f}_{\phi}\right)}  \theta_{\nu l + \phi}, \label{eq:theta_2}
\end{eqnarray}
where $\dot{\bar{f}}_i$ and $\dot{F}_{i,\ell}$ can be read off the background Boltzmann equations~\eqref{eq:background_Boltzmann1}--\eqref{eq:background_Boltzmann3}, the first-order Boltzmann hierarchy~(\ref{hierarchyPsi_nu1}), and their corresponding collision terms  in the appropriate limits.  The details of this calculation can be found in appendix~\ref{app:nonrel}; we show here only the outcome:
\begin{align}
\dot{\delta}_{\nu l+\phi} &= -\frac{4}{3} \left( \theta_{\nu l+\phi} + \frac{\dot{h}}{2} \right) - \frac{a}{ \tau_0} \frac{\bar{\rho}_{\nu H}}{\bar{\rho}_{\nu l+\phi}} \left( \delta_{\nu l+\phi}-\delta_{\nu H} \right),
\label{eq:delta_1} \\ 
\dot{\theta}_{\nu l+\phi} &= k^2 \left(\frac{1}{4} \delta_{\nu l+\phi} - \sigma_{\nu l+\phi} \right) - \frac{a}{ \tau_0} \frac{3 \bar{\rho}_{\nu H}}{4 \bar{\rho}_{\nu l+\phi}} \left( \frac{4}{3} \theta_{\nu l+\phi} - \theta_{\nu H}  \right),
\label{eq:theta_1}
\end{align}
where 
\begin{equation}
\sigma_{\nu l + \phi} \equiv  \frac{(3 \pi^2 a^4)^{-1} \int \mathrm{d}q \, q^3  \, \left(2 F_{\nu l, 2} + F_{\phi, 2} \right)}{(4/3) \bar{\rho}_{\nu l + \phi} }
= \frac{1}{2} \, \frac{\int \mathrm{d}q \, q^3  \, \left(2 F_{\nu l, 2} + F_{\phi, 2} \right)}{\int \mathrm{d}q \, q^3 \,\left(2 \bar{f}_{\nu l}  +  \bar{f}_\phi  \right)}
\end{equation}
is the anisotropic stress of the $\nu_l+\phi$ fluid.  

As with equations~(\ref{eq:deltaH}) and (\ref{eq:thetaH}), equations~(\ref{eq:delta_1}) and~(\ref{eq:theta_1}) have also previously appeared in~\cite{Audren:2014bca} in the context of CDM decay.  At first glance, the totality of these four equations
appears to violate energy--momentum conservation.  The key point to note in this regard, however,  is that the conserved quantities of the system are  {\it not} in fact sums of the {\it relative} perturbations \(\sum_i \delta_{i}\) and \(\sum_i \theta_{i}\), but rather sums of the {\it absolute} perturbations $\sum_i \bar{\rho}_{i} \delta_{i}$ and $\sum_i  \left(\bar{\rho}_i+\bar{P}_i  \right) \theta_i$. 
  We refer the interested reader to appendix~\ref{appendix:conservation_laws} for an explicit demonstration of energy--momentum conservation.

Turning our attention now to the non-relativistic neutrino decay studies of~\cite{Kaplinghat:1999xy, Chacko:2019nej,Chacko:2020hmh}, we note that they also use equations~(\ref{eq:deltaH}) and (\ref{eq:thetaH}) to describe the decaying particle,%
\footnote{This sameness may not be immediately apparent in the notation of~\cite{Chacko:2019nej}, where the Boltzmann hierarchies are always written in terms of the {\it absolute} perturbation to the phase space distributions, $F_{i,\ell}$,  instead of the {\it relative} perturbation or {\it phase space density contrast}~$\Psi_{i,\ell}$. See discussion immediately after equation~(\ref{eq:decayfactor}).  We note that publicly available Boltzmann solvers such as \CLASS{}~\cite{Blas:2011rf} use~$\Psi_{i,\ell}$ in their implementation of the standard neutrino Boltzmann hierarchy.}
and equation~(\ref{eq:delta_1}) for the decay products.  Their equivalent of equation~(\ref{eq:theta_1}), however, misses the last term proportional to $\theta_{\nu H}$, an omission that, following the logic above, must immediately violate  momentum conservation. Gauge invariance is likewise broken, since by effectively setting $\theta_{\nu H}=0$ (and hence equal to 
the cold dark matter divergence $\theta_{\rm CDM}$ in the synchronous gauge), the equation of motion cannot be consistently transformed to a different gauge without additional manual adjustments. As an example, consider transforming to a ``neutrino synchronous gauge'', wherein $\theta_{\nu H}$ is fixed at zero. If the same equations of motion are to be recovered, then the CDM velocity divergence $\theta_{\rm{CDM}}$ must be manually set to zero. There is in general no justification for such an adjustment, as $\theta_{\rm CDM}$ is {\it a priori} unknown.

The origin of the missing term can in each case be traced to an incorrect assumption in the derivation pipeline:
\begin{itemize}
\item References~\cite{Chacko:2019nej,Chacko:2020hmh} assume the emission direction of the decay products to be uncorrelated with the direction of the mother particle's momentum.
Isotropic decay in the {\it rest frame} of $\nu_H$  is in itself a reasonable simplification of the problem, and one that we also adopt in our analysis.  The assumption of~\cite{Chacko:2019nej,Chacko:2020hmh}, on the other hand, amounts effectively to assuming the decay to be isotropic in the {\it cosmic frame} and ignoring explicitly that the decaying neutrino has a non-vanishing bulk velocity divergence  
that can be transferred via decay to the daughter species.

\item In the case of~\cite{Kaplinghat:1999xy}, we believe the error originates from their discarding all terms in the $\nu_l$ and $\phi$ collision integrands proportional $q_1/\epsilon_1$ and its higher powers. This step effectively renders the first-order $\ell \geq 1$  collision terms   zero, which at $\ell=1$ has the same final outcome as omitting the $\theta_{\nu H}$ term in 
equation~\eqref{eq:theta_1}.  See appendix~\ref{app:nonrel}.
Contrary to the common practice of discarding $(q_1/\epsilon_1)^2$ terms and higher for a cold fluid (because these terms encode the fluid's velocity dispersion), $q_1/\epsilon_1$ terms represent the fluid's velocity divergence, which is {\it always} tracked in large-scale structure studies no matter the coldness of the fluid, and hence must never be relinquished.

\end{itemize}
We have not assessed the errors induced by the missing term in the predictions of cosmological observables.  In all likelihood its absence  incurs only a very small numerical shift in the predicted values, because (i)~cosmological observations already constrain the neutrino mass fraction $f_\nu$ to less than 1\%, and (ii)~non-relativistic neutrino decay kicks in primarily during matter domination, when the energy density of the relativistic  $\nu_l+\phi$  fluid is largely inconsequential. Even so, it is our view that whatever equations of motion one uses to model a system, they should at  least respect conservation laws where such laws are known to exist within the adopted theoretical framework.  We defer a numerical investigation --- using the correct equations of motion --- of the non-relativistic neutrino decay scenario of~\cite{Chacko:2019nej,Chacko:2020hmh}  to a future publication.

%%%%%%%%%%
%%%%%%%%%%%
 
\section{Conclusions}
\label{sec:Conclusion}

It has long been argued that cosmological observations provide the tightest constraints on invisible neutrino decay and hence the neutrino lifetime~\cite{Escudero:2019gfk,Basboll:2008fx,Hannestad:2005ex,Archidiacono:2013dua}.  We have revisited  this topic in this work, by way of a first-principles approach to understanding the CMB and LSS phenomenology of invisible neutrino decay.   Our decay model consists of a mother neutrino $\nu_H$ disintegrating into a daughter neutrino $\nu_l$ and a scalar particle $\phi$ via a Yukawa interaction.  Both $\nu_H$ and $\nu_l$ are standard-model mass eigenstates, and, as in the existing literature,  we take $\phi$ to have no other interactions.

Assuming a perturbed Friedmann--Lema\^{\i}tre--Robertson--Walker  universe, 
we have derived from first principles the complete set of Boltzmann equations,  at both the spatially homogeneous and the first-order inhomogeneous levels,  for the phase space densities of $\nu_H$, $\nu_l$, and $\phi$ in the presence of the decay and its inverse process.
Our base calculation presented in appendix~\ref{app:CollisionIntegralReduction} is completely general, and  applies to any particle mass spectrum satisfying $m_{\nu H} > m_{\nu l}, m_\phi$.  Our main presentation of this system of equations in sections~\ref{sec:Background} and \ref{sec:First order perturbations}, however, focuses on the case of $m_\phi=0$, and, occasionally in subsequent analyses,  we set $m_{\nu l}=0$ as well to match assumptions in the existing literature.

We have implemented the set of background Boltzmann equations~\eqref{eq:background_Boltzmann1}--\eqref{eq:background_Boltzmann3} into the linear cosmological Boltzmann code \CLASS{}, and used their numerical solutions to identify some time scales of the system as functions of the decay parameters (i.e., masses and coupling).
 Implementation of the corresponding first-order Boltzmann hierarchies~\eqref{hierarchyPsi_nu1} and associated collision terms~\eqref{nuH_collision_integral}, \eqref{eq:nulcol}, and \eqref{eq:phicol}, however,  proves to be highly non-trivial, and we defer this exercise to a later publication.  Nonetheless,  even at the level of the equations, it is clear that neutrino decay and its inverse process must introduce similar time scales in both the spatially homogeneous and inhomogeneous systems.  This realisation turns out to be extremely useful when it comes to establishing the behaviours of the spatial perturbations.

With our system of Boltzmann equations in hand together with numerical solutions of the background quantities,
we have performed a critical survey of recent works on cosmological  invisible neutrino decay~\cite{Escudero:2019gfk,Basboll:2008fx,Hannestad:2005ex,Archidiacono:2013dua,Kaplinghat:1999xy,Chacko:2019nej,Chacko:2020hmh}, in both limits of decay while $\nu_H$ is ultra-relativistic and non-relativistic, 
and assessed the validity of the approximations used in these works to simplify the numerical problem.   Our two main findings in this regard are:
\begin{enumerate}
\item In the case of non-relativistic $\nu_H$ decay,  a series of in our view ill-justified simplifying steps has led to effective equations of motion in~\cite {Kaplinghat:1999xy,Chacko:2019nej,Chacko:2020hmh} that formally violate momentum conservation and gauge invariance at the equation level.

\item In the ultra-relativistic limit, we find virtually no loss in the anisotropic stress of the combined neutrino--scalar system associated with the rate $\Gamma_{\rm T} = \Gamma_{\rm dec} (m_{\nu H}/E_{\nu H})^2= \Gamma_{\rm dec}^0 (m_{\nu H}/E_{\nu H})^3$, 
a model commonly used to derive cosmological bounds on the neutrino lifetime~\cite{Escudero:2019gfk,Basboll:2008fx,Hannestad:2005ex,Archidiacono:2013dua}.  Rather, anisotropic stress is exponentially damped at a much slower rate $\sim \Gamma_{\rm dec}^0 (m_{\nu H}/E_{\nu H})^5$, i.e., the rest-frame decay rate scaled by five powers of the inverse Lorentz factor.
\end{enumerate}
Both findings are model-independent. The second,  in particular,   implies  a substantial weakening of current cosmological limits on the rest-frame neutrino lifetime from $\tau_0^{\rm old} \gtrsim 1.2 \times 10^9\, {\rm s}\, (m_{\nu H}/50\, {\rm meV})^3$~\cite{Escudero:2019gfk,Archidiacono:2013dua} to $\tau_0 \gtrsim (4 \times 10^5 \to 4 \times 10^6)\, {\rm s}\, (m_{\nu H}/50\, {\rm meV})^5$, depending on modelling details. Though we anticipate the impact to be small, the precise implications of the first finding will be investigated by way of a full implementation of the first-order inhomogeneous Boltzmann hierarchies in \CLASS{} together with a Markov Chain Monte Carlo analysis  in a subsequent publication.

%%%%%%%%%%
%%%%%%%%%%

\acknowledgments

JZC acknowledges support from an Australian Government Research Training Program Scholarship.
Y$^3$W is supported in part by the Australian Government through the Australian Research Council's  Future Fellowship (project FT180100031) funding scheme. GB and IMO acknowledge support from FPA2017-845438  and the Generalitat Valenciana under grant   PROMETEOII/2017/033. T.T. was supported by a research grant (29337) from VILLUM FONDEN.

%%%%%%%%%%%%%%%%%%%%%%
%%%%%%%%%%%%%%%%%%%%%%

\appendix

\newpage
\section{Collision integral reduction}
\label{app:CollisionIntegralReduction}

We derive in this appendix the background Boltzmann equations~\eqref{eq:background_Boltzmann1}-(\ref{eq:background_Boltzmann3}), as well as the Boltzmann hierarchies for the first order perturbations presented in section~\ref{sec:First order perturbations}, in the presence of neutrino decay and inverse decay.  To keep the derivation general we retain the possibility of a finite scalar mass~$m_{\phi}$ for most parts of the calculation, and only present the $m_{\phi}=0$ result explicitly in those cases where taking that said limit requires some care.

\subsection{Zeroth-order collision integrals}
\label{sec:C0_nuH}

Beginning with equations~(\ref{eq:licollisionintegral}) and (\ref{eq:collisionmap}) and the matrix element~(\ref{eq:matrixelement}), 
the collision integrals for the background evolution of $\nu_H$, $\nu_l$ and $\phi$ can be written as
\begin{eqnarray}
\left(\frac{{\rm d} f_{\nu H} }{{\rm d} \tau}\right)^{(0)}_C (\mathbf{q}_1)
&= & \, \frac{4 \mathfrak{g}^2}{\epsilon_1} \int \frac{\mathrm{d}^3 \mathbf{q}_2}{(2\pi)^3 2 \epsilon_2} \int \frac{\mathrm{d}^3 \mathbf{q}_3}{(2\pi)^3 2 \epsilon_3} (2\pi)^4 \delta^{(4)}_D(q_1 - q_2 - q_3) \nonumber \\ 
		&& \qquad \quad \times (\eta_{\mu \nu} q_1^{\mu} q_2^{\nu} + a^2 m_{\nu H} m_{\nu l})  \Lambda(|\mathbf{q}_1|,|\mathbf{q}_2|,|\mathbf{q}_3|), 
\label{Eq:Master} \\
\left(\frac{{\rm d} f_{\nu l} }{{\rm d} \tau}\right)^{(0)}_C (\mathbf{q}_2) &= & \, -\frac{4 \mathfrak{g}^2}{\epsilon_2} \int \frac{\mathrm{d}^3\mathbf{q}_1}{(2\pi)^3 2 \epsilon_1} \int \frac{\mathrm{d}^3\mathbf{q}_3}{(2\pi)^3 2 \epsilon_3} (2\pi)^4 \delta^{(4)}_D(q_1 - q_2 - q_3)  \nonumber \\
&& \qquad \quad \times (\eta_{\mu \nu} q_1^{\mu} q_2^{\nu} + a^2 m_{\nu H} m_{\nu l}) \Lambda(|\mathbf{q}_1|,|\mathbf{q}_2|,|\mathbf{q}_3|),
\label{Eq:Master2} \\
\left(\frac{{\rm d} f_{\phi} }{{\rm d} \tau}\right)^{(0)}_C (\mathbf{q}_3)& = & \, -\frac{8 \mathfrak{g}^2}{\epsilon_3} \int \frac{\mathrm{d}^3 \mathbf{q}_1}{(2\pi)^3 2 \epsilon_1} \int \frac{\mathrm{d}^3 \mathbf{q}_2}{(2\pi)^3 2 \epsilon_2} (2\pi)^4 \delta^{(4)}_D(q_1 - q_2 - q_3) \nonumber  \\
&& \qquad \quad \times \left(\eta_{\mu \nu} q_1^{\mu} q_2^{\nu} + a^2 m_{\nu H} m_{\nu l} \right) \Lambda(|\mathbf{q}_1|,|\mathbf{q}_2|,|\mathbf{q}_3|), 
\label{Eq:Master3} 
\end{eqnarray}
where 
\begin{equation}
 \Lambda(|\mathbf{q}_1|,|\mathbf{q}_2|,|\mathbf{q}_3|) =   \bar{f}_{\nu l}(|\mathbf{q}_2|) \bar{f}_{\phi}(|\mathbf{q}_3|) - \bar{f}_{\nu H}(|\mathbf{q}_1|) + \bar{f}_{\nu H}(|\mathbf{q}_1|) \bar{f}_{\nu l}(|\mathbf{q}_2|) - \bar{f}_{\nu H}(|\mathbf{q}_1|) \bar{f}_{\phi}(|\mathbf{q}_3|),\\
 \label{eq:lambdafactor}
\end{equation}
is a common phase space distribution factor that depends only on the momentum magnitudes because of our assumption of homogeneity and isotropy.
Our goal is to reduce these nominally 6-dimensional integrals to one-dimensional ones in $|\mathbf{q}_1|$, \(|\mathbf{q}_2|\), or \( |\mathbf{q}_3|\), by first integrating out the angular dependences.  For a phase space distribution factor~$\Lambda$ that contains no angular dependence, this is a straightforward exercise and many recipes exist for the purpose.  Here, we follow the procedure of~\cite{Oldengott:2014qra}, as it is applicable also to those cases wherein the angular dependence of $\Lambda$ may be non-trivial.

%%%%%%%%%%%

\subsubsection{Eliminating one momentum-integral}

First of all, in order to consistently evaluate  the 4-dimensional Dirac delta distribution, we make use of the relation
\begin{equation}
	\begin{aligned}
		& \int \mathrm{d}^4q \, \delta_D(q^2 - a^2 m^2) \, \Theta(\epsilon) \\
		& \qquad =  \int \mathrm{d}\epsilon \, \mathrm{d}^3 \mathbf{q} \, \delta_D(\epsilon^2 - |\mathbf{q}|^2 - a^2 m^2) \, \Theta(\epsilon) \\
		& \qquad  =  \int \mathrm{d}\epsilon \, \mathrm{d}^3\mathbf{q} \, \frac{1}{2\epsilon} \left( \delta_D(\epsilon - \sqrt{|\mathbf{q}|^2 + a^2 m^2}) + \delta_D(\epsilon + \sqrt{|\mathbf{q}|^2 + a^2 m^2}) \right) \, \Theta(\epsilon)  \\
	& \qquad 	=  \int \frac{\mathrm{d}^3 \mathbf{q}}{2\epsilon}.
	\end{aligned}
	\label{eq:delta_relation1}
\end{equation}
Applying in particular the second and fourth lines to the integrals (i)~over $\mathbf{q}_3$ in equations~(\ref{Eq:Master})-(\ref{Eq:Master2}), and (ii) over~$\mathbf{q}_2$ in equation~(\ref{Eq:Master3}),  we obtain
\begin{eqnarray}
\left(\frac{{\rm d} f_{\nu H} }{{\rm d} \tau}\right)^{(0)}_C (\mathbf{q}_1)
		&=& \, \frac{\mathfrak{g}^2}{2\pi^2 \epsilon_1} \int \frac{\mathrm{d}^3 \mathbf{q}_2}{\epsilon_2} \, \delta_D\left( (\epsilon_1 - \epsilon_2)^2 - |\mathbf{q}_1 - \mathbf{q}_2|^2 - a^2 m_{\phi}^2 \right) \, \Theta(\epsilon_1 - \epsilon_2) \nonumber \\
		&& \qquad \qquad \times (q_{1\mu} q_2^{\mu} + a^2 m_{\nu H} m_{\nu l})  \Lambda(|\mathbf{q}_1|,|\mathbf{q}_2|,|\mathbf{q}_3|),
	\label{C_nuH_step1} \\
	\left(\frac{{\rm d} f_{\nu l} }{{\rm d} \tau}\right)^{(0)}_C (\mathbf{q}_2)
	&=& \,- \frac{\mathfrak{g}^2}{2\pi^2 \epsilon_2} \int \frac{\mathrm{d}^3 \mathbf{q}_1}{\epsilon_1} \, \delta_D\left( (\epsilon_1 - \epsilon_2)^2 - |\mathbf{q}_1 - \mathbf{q}_2|^2 - a^2 m_{\phi}^2 \right) \, \Theta(\epsilon_1 - \epsilon_2) \nonumber \\
	&& \qquad \qquad \times (q_{1\mu} q_2^{\mu} + a^2 m_{\nu H} m_{\nu l})  \Lambda(|\mathbf{q}_1|,|\mathbf{q}_2|,|\mathbf{q}_3|),
	\label{C_nul_step1}\\
		\left(\frac{{\rm d} f_{\phi} }{{\rm d} \tau}\right)^{(0)}_C (\mathbf{q}_3)
	&=& \, \frac{\mathfrak{g}^2}{\pi^2 \epsilon_3} \int \frac{\mathrm{d}^3 \mathbf{q}_1}{\epsilon_1} \, \delta_D\left( (\epsilon_1 - \epsilon_3)^2 - |\mathbf{q}_1 - \mathbf{q}_3|^2 - a^2 m_{\nu l}^2 \right) \, \Theta(\epsilon_1 - \epsilon_3) \nonumber \\
	&& \qquad \qquad \times \left[q_{1\mu} q_3^{\mu} - a^2 m_{\nu H}(m_{\nu H}+ m_{\nu l}) \right]  \Lambda(|\mathbf{q}_1|,|\mathbf{q}_2|,|\mathbf{q}_3|),
	\label{C_phi_step1}
\end{eqnarray}
where it is understood $|\mathbf{q}_3| = [(\epsilon_1-\epsilon_2)^2 - m_\phi^2]^{1/2}$ and $|\mathbf{q}_2| = [(\epsilon_1-\epsilon_3)^2 - m_{\nu l}^2]^{1/2}$
 following  from energy conservation.

%%%%%%%

\subsubsection{Angular integration}
\label{sec:angular}

In their current forms, the integrands~(\ref{C_nuH_step1}) and (\ref{C_nul_step1}) depend angularly only on $\mathbf{q}_1 \cdot \mathbf{q}_2$, while integrand~(\ref{C_phi_step1}) is a function of $\mathbf{q}_1 \cdot \mathbf{q}_3$.  It is therefore convenient to align the fixed, external momentum vector in the $z$-direction while parameterising the dummy vector (i.e., the one integrated over) in terms of its magnitude and angle subtended with the $z$-axis.  In the case of the $\nu_H$ collision integral~(\ref{C_nuH_step1}), for example, this means
\begin{equation}
	\begin{aligned}
		\mathbf{q}_1 =& \,|\mathbf{q}_1| \, (0,0,1) \, , \\
		\mathbf{q}_2 =& \,|\mathbf{q}_2| \, (0,\sin\alpha,\cos\alpha) \, ,
	\end{aligned}
\end{equation}
so that  the integration variables become \( \int \mathrm{d}^3 \mathbf{q}_2  = 2\pi \int_0^\infty  |\mathbf{q_2 }|^2 \,  \mathrm{d} |\mathbf{q}_2| \, \int_{-1}^{1} \mathrm{d}\cos{\alpha}\).   Then, equations~(\ref{C_nuH_step1})-(\ref{C_phi_step1}) simplify to
\begin{eqnarray}
\left(\frac{{\rm d} f_{\nu H} }{{\rm d} \tau}\right)^{(0)}_C (\mathbf{q}_1)&= & \frac{\mathfrak{g}^2}{\pi \epsilon_1} \int \mathrm{d} |\mathbf{q}_2| \, \mathrm{d}\cos{\alpha} \, \Theta(\epsilon_1 - \epsilon_2)  \frac{|\mathbf{q}_2|^2}{\epsilon_2} \, K_\alpha (|\mathbf{q}_1|,|\mathbf{q}_2|,\cos \alpha) \, \Lambda, 
\label{eq:C0_nuH_dcosa}		\\
\left(\frac{{\rm d} f_{\nu l} }{{\rm d} \tau}\right)^{(0)}_C (\mathbf{q}_2)&= & - \frac{\mathfrak{g}^2}{\pi \epsilon_2} \int \mathrm{d} |\mathbf{q}_1| \, \mathrm{d}\cos{\alpha} \,  \Theta(\epsilon_1 - \epsilon_2)  \frac{|\mathbf{q}_1|^2}{\epsilon_1}\, K_\alpha (|\mathbf{q}_1|,|\mathbf{q}_2|,\cos \alpha) \, \Lambda, 
\label{eq:C0_nul_dcosa}	\\
\left(\frac{{\rm d} f_{\phi} }{{\rm d} \tau}\right)^{(0)}_C (\mathbf{q}_3)&= & \frac{2 \mathfrak{g}^2}{\pi \epsilon_3} \int \mathrm{d} |\mathbf{q}_1| \, \mathrm{d}\cos{\beta} \, \Theta(\epsilon_1 - \epsilon_3)  \frac{|\mathbf{q}_1|^2}{\epsilon_1}\, K_\beta (|\mathbf{q}_1|,|\mathbf{q}_3|,\cos \beta) \,  \Lambda,
\label{eq:C0_phi_dcosa}	
\end{eqnarray}
where 
\begin{equation}
\begin{aligned}
K_\alpha (|\mathbf{q}_1|,|\mathbf{q}_2|,\cos \alpha) \equiv & \, \delta_D\left( g(\cos \alpha) \right)  \left(\epsilon_1 \epsilon_2 - |\mathbf{q}_1||\mathbf{q}_2|\cos{\alpha} + a^2 m_{\nu H} m_{\nu l} \right),  \\
K_\beta (|\mathbf{q}_1|,|\mathbf{q}_3|,\cos \beta) \equiv & \, \delta_D\left( h(\cos \beta) \right) \left(\epsilon_1 \epsilon_3 - |\mathbf{q}_1||\mathbf{q}_3|\cos{\beta}- a^2 m_{\nu H} (m_{\nu H}+ m_{\nu l}) \right)  \label{eq:kernels1}
\end{aligned}
\end{equation}
are the scattering kernels with $\cos \alpha \equiv \hat{q}_1 \cdot \hat{q}_2$ and $\cos \beta \equiv \hat{q}_1 \cdot \hat{q}_3$,
 and the functions
\begin{equation}
\begin{aligned}
g(\cos \alpha) & \equiv \, 2|\mathbf{q}_1||\mathbf{q}_2|\cos{\alpha} - 2 \epsilon_1 \epsilon_2 + a^2 (m_{\nu H}^2 + m_{\nu l}^2 - m_{\phi}^2), \\
h(\cos \beta) & \equiv \, 2|\mathbf{q}_1||\mathbf{q}_3|\cos{\beta} - 2 \epsilon_1 \epsilon_3 + a^2 (m_{\nu H}^2 - m_{\nu l}^2 + m_{\phi}^2)
\label{eq:gh}
\end{aligned}
\end{equation}
appear in the kernels' Dirac delta distributions.

In order to perform the angular integration in equations~(\ref{eq:C0_nuH_dcosa})-(\ref{eq:C0_phi_dcosa}), we first rewrite the Dirac delta distributions in the scattering kernels~\eqref{eq:kernels1} in terms of the angular variable of interest.  For $K_\alpha(|\mathbf{q}_1|,|\mathbf{q}_2|,\cos \alpha)$, for example, this can be accomplished using the  relation
\begin{equation}  
\delta_D(g(\cos \alpha)) =  \frac{\delta_D(\cos \alpha- \cos \alpha^*)}{\big| \frac{\mathrm{d}g}{\mathrm{d}\cos \alpha} \big|_{\cos \alpha^*}} = \, \frac{\delta_D(\cos{\alpha} - \cos{\alpha^*})}{2|\mathbf{q}_1||\mathbf{q}_2| }, 
\label{eq:deltaDcosa}
\end{equation}
where
\begin{align}
\cos{\alpha^*} =& \, \frac{2\epsilon_1 \epsilon_2 - a^2 (m_{\nu H}^2 + m_{\nu l}^2 - m_{\phi}^2)}{2|\mathbf{q}_1||\mathbf{q}_2| } \label{eq:cosa*} 
\end{align}
is the root of $g(\cos \alpha)$. Then, the scattering kernel $K_\alpha(|\mathbf{q}_1|,|\mathbf{q}_2|,\cos \alpha)$ can be immediately rewritten as
\begin{equation}
K_\alpha(|\mathbf{q}_1|,|\mathbf{q}_2|,\cos \alpha) = \frac{a^2(m_{\nu H}+m_{\nu l})^2 - a^2 m_{\phi}^2}{4 |\mathbf{q}_1| |\mathbf{q}_2|} \, \delta_D (\cos \alpha - \cos \alpha^*).\label{eq:ka}
\end{equation}
Similarly, applying the same procedure to $K_\beta(|\mathbf{q}_1|,|\mathbf{q}_3|,\cos \beta)$ yields
\begin{equation}
K_\beta(|\mathbf{q}_1|,|\mathbf{q}_3|,\cos \beta) = - \frac{a^2(m_{\nu H}+m_{\nu l})^2 - a^2 m_{\phi}^2}{4 |\mathbf{q}_1| |\mathbf{q}_3|} \, \delta_D (\cos \beta - \cos \beta^*),\label{eq:kb}
\end{equation}
where
\begin{align}
\cos{\beta^*} =& \, \frac{2\epsilon_1 \epsilon_3 - a^2 (m_{\nu H}^2 - m_{\nu l}^2 + m_{\phi}^2)}{2|\mathbf{q}_1||\mathbf{q}_3| } \label{eq:cosb*} 
\end{align}
is the root of $h(\cos \beta)$.  Note the overall minus sign in equation~(\ref{eq:kb}).

The angular integration is then trivial, and equations~(\ref{eq:C0_nuH_dcosa})-(\ref{eq:C0_phi_dcosa}) evaluate to
\begin{eqnarray}
\left(\frac{{\rm d} f_{\nu H} }{{\rm d} \tau}\right)^{(0)}_C (\mathbf{q}_1)
&=&  \frac{\mathfrak{g}^2 a^2 \Sigma}{4\pi \epsilon_1 |\mathbf{q}_1|} \int \mathrm{d}|\mathbf{q}_2| \Theta(\epsilon_1 - \epsilon_2) \,\Theta(1-\cos^2{\alpha^*}) \frac{|\mathbf{q}_2|}{\epsilon_2} \,  \Lambda,
\label{Eq:AngularMidStep}\\
\left(\frac{{\rm d} f_{\nu l} }{{\rm d} \tau}\right)^{(0)}_C (\mathbf{q}_2)
&=& -\frac{\mathfrak{g}^2 a^2 \Sigma}{4\pi \epsilon_2 |\mathbf{q}_2|} \int \mathrm{d}|\mathbf{q}_1| \Theta(\epsilon_1 - \epsilon_2) \,\Theta(1-\cos^2{\alpha^*}) \frac{|\mathbf{q}_1|}{\epsilon_1}\, \Lambda, \label{Eq:AngularMidStep2}\\
\left(\frac{{\rm d} f_{\phi} }{{\rm d} \tau}\right)^{(0)}_C (\mathbf{q}_3)
&=& -\frac{\mathfrak{g}^2 a^2\Sigma}{2\pi \epsilon_3 |\mathbf{q}_3|} \int \mathrm{d}|\mathbf{q}_1| \Theta(\epsilon_1 - \epsilon_3) \,\Theta(1-\cos^2{\beta^*}) \frac{|\mathbf{q}_1|}{\epsilon_1} \, \Lambda, \label{Eq:AngularMidStep3}
\end{eqnarray}
with
\begin{equation}
\Sigma \equiv (m_{\nu H}+m_{\nu l})^2 - m_{\phi}^2. \label{eq:sigma}
\end{equation}
Note that in order to enforce the conditions $\cos \alpha^* \in [-1,1]$ and $\cos \beta^* \in [-1,1]$, 
we have  inserted  into the corresponding integrand a Heaviside step function $\Theta(1-\cos^2\alpha^*)$ or $\Theta(1-\cos^2\beta^*)$.
  As we shall see below, these step functions will provide the upper and lower integration limits of the remaining momentum-integral.

%%%%%%%%%%%

\subsubsection{Integration limits}

As they stand now, the lower and upper momentum integration limits of equations~(\ref{Eq:AngularMidStep})-(\ref{Eq:AngularMidStep3}) are 0 and $\infty$ respectively.  The presence of the Heaviside step functions $\Theta(1-\cos^2{\alpha^*})$ or $\Theta(1-\cos^2{\beta^*})$, however, modifies these limits away from their canonical values.  

To derive the new integration limits, we use equation~(\ref{eq:cosa*}) to write out the Heaviside step function $\Theta(1-\cos^2{\alpha^*})$ explicitly, i.e.,
\begin{equation}
\begin{aligned}
	\Theta(1-\cos^2{\alpha^*}) &= \Theta\left(4|\mathbf{q}_1|^2|\mathbf{q}_2|^2 - \left(2\epsilon_1 \epsilon_2 - a^2 (m_{\nu H}^2 + m_{\nu l}^2 - m_{\phi}^2) \right)^2 \right)\\
	& = \Theta\left(|\mathbf{q}_1|-q_{1-}^{(\nu l)}\right) \, \Theta\left(q_{1+}^{(\nu l)}-|\mathbf{q}_1|\right)  \\
	& = \Theta\left(|\mathbf{q}_2|-q_{2-}^{(\nu H)}\right) \, \Theta\left(q_{2+}^{(\nu H)}-|\mathbf{q}_2|\right).
	\label{eq:limitsq1q2}
\end{aligned}
\end{equation}
Here, the second and third lines indicate that the step function can be reinterpreted as a product of two step functions in either $|\mathbf{q}_1|$ or $|\mathbf{q}_2|$ that serve to limit the integrand to nonzero values only in the region $|\mathbf{q}_1| \in [q_{1-}^{(\nu l)}, q_{1+}^{(\nu l)}]$ or $|\mathbf{q}_2| \in [q_{2-}^{(\nu H)}, q_{2+}^{(\nu H)}]$, where $q_{1\pm}^{(\nu l)}$ and $q_{2\pm}^{(\nu H)}$ are, respectively, the two positive-valued $|\mathbf{q}_1|$-roots and $|\mathbf{q}_2|$-roots of $1-\cos^2 \alpha^*$.  These roots are given by the compact expression
\begin{equation}
\begin{aligned}
	q_{i\pm}^{(j)} =&  \left| \frac{\epsilon_j \sqrt{(m_{k}^2 - m_{j}^2 - m_{i}^2)^2- 4 m_{i}^2 m_{j}^2} \pm |\mathbf{q}_j| |m_{k}^2 - m_{i}^2 - m_{j}^2|}{2 m_{j}^2} \right| \\
	=& \left| \frac{\epsilon_j \Delta(m^2_k,m^2_i,m^2_j) \pm |\mathbf{q}_j| \sqrt{\Delta^2(m^2_k,m^2_i,m^2_j)+4 m^2_i m^2_j}}{2 m^2_j} \right|,
	\label{eq:intlimits_compact}
\end{aligned}
\end{equation}
where $i \neq j \neq k$, the usual mapping $(\nu_H,\nu_l,\phi) \leftrightarrow (1,2,3)$ is understood, and 
we have introduced the K\"{a}ll\'{e}n function,
\begin{equation}
\Delta^2(m^2_{k},m^2_{i},m^2_j) \equiv  
(m_{k}-m_{i}-m_j)(m_{k}+m_{i}-m_j) (m_{k}-m_{i}+m_j)(m_{k}+m_{i}+m_j),
\end{equation}
in the second line.
  Similarly, using equation~(\ref{eq:cosb*}) to write out $\Theta(1-\cos^2{\beta^*})$ yields
\begin{equation}
\begin{aligned}
\Theta(1-\cos^2{\beta^*}) &= \Theta\left(4|\mathbf{q}_1|^2|\mathbf{q}_3|^2 - \left(2\epsilon_1 \epsilon_3 - a^2 (m_{\nu H}^2 - m_{\nu l}^2 + m_{\phi}^2) \right)^2 \right)\\
& = \Theta\left(|\mathbf{q}_1|-q_{1-}^{(\phi)} \right) \, \Theta\left(q_{1+}^{(\phi)}-|\mathbf{q}_1|\right),
\end{aligned}
\end{equation}
where the roots $q_{1\pm}^{(\phi)}$ are again given by equation~\eqref{eq:intlimits_compact}, and reduce to 
\begin{equation}
q_{1\pm}^{(\phi)} 
 \underset{m_\phi=0}{\to}  
\begin{dcases}
\infty \\
\left| \frac{4 m_{\nu H}^2 |\mathbf{q}_3|^2 - a^2 (m_{\nu H}^2 - m_{\nu l}^2)^2}{4 |\mathbf{q}_3| (m_{\nu H}^2 - m_{\nu l}^2)} \right| \\
\end{dcases} \, ,
\label{eq:barq1_limits}	
\end{equation}
in the $m_\phi=0$ limit.

Then, the background collision integrals~(\ref{Eq:AngularMidStep})-(\ref{Eq:AngularMidStep3}) can now be written as
\begin{eqnarray}
\left(\frac{{\rm d} f_{\nu H} }{{\rm d} \tau}\right)^{(0)}_C (\mathbf{q}_1)& =& \frac{\mathfrak{g}^2 a^2 \left((m_{\nu H}+m_{\nu l})^2-m_\phi^2 \right) }{4\pi \epsilon_1 |\mathbf{q}_1|} \int_{q_{2-}^{(\nu H)}}^{q_{2+}^{(\nu H)}} \mathrm{d}|\mathbf{q}_2| \, \frac{|\mathbf{q}_2|}{\epsilon_2} \, \Lambda, \label{eq:colnuH}\\
\left(\frac{{\rm d} f_{\nu l} }{{\rm d} \tau}\right)^{(0)}_C (\mathbf{q}_2)& =& - \frac{\mathfrak{g}^2 a^2 \left((m_{\nu H}+m_{\nu l})^2-m_\phi^2 \right) }{4\pi \epsilon_2 |\mathbf{q}_2|} \int_{q_{1-}^{(\nu l)}}^{q_{1+}^{(\nu l)}} \mathrm{d}|\mathbf{q}_1| \, \frac{|\mathbf{q}_1|}{\epsilon_1} \, \Lambda, \label{eq:colnul}\\
\left(\frac{{\rm d} f_{\phi} }{{\rm d} \tau}\right)^{(0)}_C (\mathbf{q}_3)& =& - \frac{\mathfrak{g}^2 a^2 \left((m_{\nu H}+m_{\nu l})^2-m_\phi^2 \right) }{2\pi \epsilon_3 |\mathbf{q}_3|} \int_{q_{1-}^{(\phi)}}^{\bar{q}_{1+}^{(\phi)}} \mathrm{d}|\mathbf{q}_1| \, \frac{|\mathbf{q}_1|}{\epsilon_1} \, \Lambda, \label{eq:colphi}
\end{eqnarray}
where we have omitted writing out the step functions $\Theta( \epsilon_1- \epsilon_2)$ etc., since they are automatically satisfied.

%%%%%%%%%%
%%%%%%%%%%%

\subsubsection{Alternative momentum-integrals}

Consider the phase space factor $\Lambda(|\mathbf{q}_1|, |\mathbf{q}_2|, |\mathbf{q}_3|)$, given in equation~(\ref{eq:lambdafactor}), in light of the results~(\ref{eq:colnuH})-(\ref{eq:colphi}).  Clearly, some of constituent terms of $\Lambda$ do not depend directly on the chosen momentum-integration variable at all, and it may in fact be more convenient, especially numerically, to integrate these terms over the ``other'' momentum instead.  For example, the $\bar{f}_{\nu H} (|\mathbf{q}_1|) \bar{f}_\phi(|\mathbf{q}_3|)$ term in $\Lambda$ can be more easily integrated numerically over $|\mathbf{q}_3|$  than  $|\mathbf{q}_2|$ when embedded in the $\nu_H$ collision integral~(\ref{eq:colnuH}).  For this reason, we also give in the following ``alternative'' versions of the integrals~(\ref{eq:colnuH})-(\ref{eq:colphi}).

Because the phase space factor $\Lambda$ in the said integrals~(\ref{eq:colnuH})-(\ref{eq:colphi}) has no angular dependence, the simplest way to derive their alternative versions in terms of the ``other'' momentum is to perform a change of integration variables while observing conservation of energy.  Nonetheless, we opt here to take a longer route, as the calculations shown below will also turn out to be useful when it comes to computing the first-order collision integrals.

Beginning again with equations~(\ref{Eq:Master})-(\ref{Eq:Master3}), but integrating out first the ``other'' momentum, it is straightforward to arrive at a set of integrals analogous to equations~(\ref{eq:C0_nuH_dcosa})-(\ref{eq:C0_phi_dcosa}):
\begin{eqnarray}
\left(\frac{{\rm d} f_{\nu H} }{{\rm d} \tau}\right)^{(0)}_C (\mathbf{q}_1)&= & -\frac{\mathfrak{g}^2}{\pi \epsilon_1} \int \mathrm{d} |\mathbf{q}_3| \, \mathrm{d}\cos{\beta} \, \Theta(\epsilon_1 - \epsilon_3)  \frac{|\mathbf{q}_3|^2}{\epsilon_3} \, K_\beta(|\mathbf{q}_1|,|\mathbf{q}_3|,\cos \beta)\, \Lambda, 
\label{eq:C0_nuH_dcosb}		\\
\left(\frac{{\rm d} f_{\nu l} }{{\rm d} \tau}\right)^{(0)}_C (\mathbf{q}_2)&= & - \frac{\mathfrak{g}^2}{\pi \epsilon_2} \int \mathrm{d} |\mathbf{q}_3| \, \mathrm{d}\cos{\gamma} \,  \Theta(\epsilon_2 + \epsilon_3)  \frac{|\mathbf{q}_3|^2}{\epsilon_3} \, K_\gamma(|\mathbf{q}_2|,|\mathbf{q}_3|,\cos \gamma)\, \Lambda, 
\label{eq:C0_nul_dcosb}	\\
\left(\frac{{\rm d} f_{\phi} }{{\rm d} \tau}\right)^{(0)}_C (\mathbf{q}_3)&= &- \frac{2 \mathfrak{g}^2}{\pi \epsilon_3} \int \mathrm{d} |\mathbf{q}_2| \, \mathrm{d}\cos{\gamma} \,  \Theta(\epsilon_2 + \epsilon_3)  \frac{|\mathbf{q}_2|^2}{\epsilon_2} \,
K_\gamma(|\mathbf{q}_2|,|\mathbf{q}_3|,\cos \gamma)\,
 \Lambda, 
\label{eq:C0_phi_dcosb}	
\end{eqnarray}
where the scattering kernel $K_\beta(|\mathbf{q}_1|,|\mathbf{q}_3|,\cos \beta)$ is given in equation~\eqref{eq:kb}, and we have introduced a third scattering kernel
\begin{equation}
\begin{aligned}
K_\gamma(|\mathbf{q}_2|,|\mathbf{q}_3|,\cos \gamma) & \equiv \delta_D\left(j(\cos \gamma)\right)\left(\epsilon_2 \epsilon_3 - |\mathbf{q}_2||\mathbf{q}_3|\cos{\gamma} + a^2 m_{\nu l}(m_{\nu H} +m_{\nu l})\right) \\
& = \frac{a^2(m_{\nu H}+m_{\nu l})^2 - a^2 m_{\phi}^2}{4 |\mathbf{q}_2| |\mathbf{q}_3|} \, \delta_D (\cos \gamma - \cos \gamma^*),\label{eq:kg}
\end{aligned}
\end{equation}
with
\begin{equation}
j(\cos \gamma)   \equiv \, -2|\mathbf{q}_2||\mathbf{q}_3|\cos{\gamma} +2 \epsilon_2 \epsilon_3 + a^2 (-m_{\nu H}^2 + m_{\nu l}^2 + m_{\phi}^2),
\end{equation}
and
\begin{equation}
\cos \gamma^* = \frac{2\epsilon_2 \epsilon_3 -a^2(m_{\nu H}^2 - m_{\nu l}^2 - m_{\phi}^2)}{2|\mathbf{q}_2||\mathbf{q}_3|}
\end{equation}
is the root of $j(\cos \gamma)$.

Following the same procedure as in section~\ref{sec:angular} leads us to
\begin{eqnarray}
\left(\frac{{\rm d} f_{\nu H} }{{\rm d} \tau}\right)^{(0)}_C (\mathbf{q}_1)
	&=& \, \frac{\mathfrak{g}^2 a^2 \Sigma}{4\pi \epsilon_1 |\mathbf{q}_1|} \int \mathrm{d}|\mathbf{q}_3| \,\Theta(\epsilon_1 - \epsilon_3) \Theta(1-\cos^2{\beta^*}) \frac{|\mathbf{q}_3|}{\epsilon_3} \,   \Lambda,
\label{Eq:AngularMidStepb}\\
\left(\frac{{\rm d} f_{\nu l} }{{\rm d} \tau}\right)^{(0)}_C (\mathbf{q}_2)
&=& \,-\frac{\mathfrak{g}^2 a^2 \Sigma}{4\pi \epsilon_2 |\mathbf{q}_2|} \int \mathrm{d}|\mathbf{q}_3| \,\Theta(\epsilon_2 + \epsilon_3) \Theta(1-\cos^2{\gamma^*}) \frac{|\mathbf{q}_3|}{\epsilon_3}\, \Lambda, \label{Eq:AngularMidStep2b}\\
\left(\frac{{\rm d} f_{\phi} }{{\rm d} \tau}\right)^{(0)}_C (\mathbf{q}_3)
&=& \,-\frac{\mathfrak{g}^2 a^2\Sigma}{2\pi \epsilon_3 |\mathbf{q}_3|} \int \mathrm{d}|\mathbf{q}_2| \,\Theta(\epsilon_2 + \epsilon_3) \Theta(1-\cos^2{\gamma^*}) \frac{|\mathbf{q}_2|}{\epsilon_2} \, \Lambda, \label{Eq:AngularMidStep3b}
\end{eqnarray}
where $\Sigma$ is given in equation~(\ref{eq:sigma}).
Again, the Heaviside step functions can be turned into momentum-integration limits, i.e.,
\begin{equation}
\begin{aligned}
\Theta(1-\cos^2{\beta^*}) &= \Theta\left(4|\mathbf{q}_1|^2|\mathbf{q}_3|^2 - \left(2\epsilon_1 \epsilon_3 - a^2 (m_{\nu H}^2 - m_{\nu l}^2 + m_{\phi}^2) \right)^2 \right)\\
& = \Theta\left(|\mathbf{q}_3|-q_{3-}^{(\nu H)}\right) \, \Theta\left(q_{3+}^{(\nu H)}-|\mathbf{q}_3|\right),
\end{aligned}
\end{equation}
and
\begin{equation}
\begin{aligned}
	\Theta(1-\cos^2{\gamma^*}) &= \Theta\left(4|\mathbf{q}_2|^2|\mathbf{q}_3|^2 - \left(2\epsilon_2 \epsilon_3 - a^2 (m_{\nu H}^2 - m_{\nu l}^2 - m_{\phi}^2) \right)^2 \right)\\
	& = \Theta\left(|\mathbf{q}_2|-q_{2-}^{(\phi)}\right) \, \Theta\left(q_{2+}^{(\phi)}-|\mathbf{q}_2|\right)  \\
	& = \Theta\left(|\mathbf{q}_3|-q_{3-}^{(\nu l)}\right) \, \Theta\left(q_{3+}^{(\nu l)}-|\mathbf{q}_3|\right).
	\label{eq:thetagamma}
\end{aligned}
\end{equation}
Again, the three sets of roots $q_{3\pm}^{(\nu H)}$, $q_{2\pm}^{(\phi)}$, and $q_{3\pm}^{(\nu l)}$ can be easily read off equation~\eqref{eq:intlimits_compact}, with $q_{2\pm}^{(\phi)}$ tending to 
\begin{equation}
q_{2\pm}^{(\phi)} 
 \underset{m_\phi=0}{\to}  
\begin{dcases}
\infty \\
\left| \frac{4 m_{\nu l}^2 |\mathbf{q}_3|^2 - a^2 (m_{\nu H}^2 - m_{\nu l}^2)^2}{4 |\mathbf{q}_3| (m_{\nu H}^2 - m_{\nu l}^2)} \right| \\
\end{dcases} \, 	\label{eq:barq2_limits}
\end{equation}
in the $m_\phi=0$ limit.

Then, it follows that equations~(\ref{Eq:AngularMidStepb})-(\ref{Eq:AngularMidStep3b}) can be written as
\begin{eqnarray}
\left(\frac{{\rm d} f_{\nu H} }{{\rm d} \tau}\right)^{(0)}_C (\mathbf{q}_1)& =& \frac{\mathfrak{g}^2 a^2 \left((m_{\nu H}+m_{\nu l})^2-m_\phi^2 \right) }{4\pi \epsilon_1 |\mathbf{q}_1|} \int_{q_{3-}^{(\nu H)}}^{q_{3+}^{(\nu H)}} \mathrm{d}|\mathbf{q}_3| \, \frac{|\mathbf{q}_3|}{\epsilon_3} \, \Lambda, \label{eq:colnuH2}\\
\left(\frac{{\rm d} f_{\nu l} }{{\rm d} \tau}\right)^{(0)}_C (\mathbf{q}_2)& =& - \frac{\mathfrak{g}^2 a^2 \left((m_{\nu H}+m_{\nu l})^2-m_\phi^2 \right) }{4\pi \epsilon_2 |\mathbf{q}_2|} \int_{q_{3-}^{(\nu l)}}^{q_{3+}^{(\nu l)}} \mathrm{d}|\mathbf{q}_3| \, \frac{|\mathbf{q}_3|}{\epsilon_3} \, \Lambda,
\label{eq:colnul2} \\
\left(\frac{{\rm d} f_{\phi} }{{\rm d} \tau}\right)^{(0)}_C (\mathbf{q}_3)& =& - \frac{\mathfrak{g}^2 a^2 \left((m_{\nu H}+m_{\nu l})^2-m_\phi^2 \right) }{2\pi \epsilon_3 |\mathbf{q}_3|} \int_{q_{2-}^{(\phi)}}^{q_{2+}^{(\phi)}} \mathrm{d}|\mathbf{q}_2| \, \frac{|\mathbf{q}_2|}{\epsilon_2} \, \Lambda, \label{eq:colphi2}
\end{eqnarray}
which are physically exactly equivalent to the collision integrals~(\ref{eq:colnuH})-(\ref{eq:colphi}).

%%%%%%%%%%%%%%%%%%%%%%
%%%%%%%%%%%%%%%%%%%%%%

\subsubsection{The $\bar{f}_{\nu H}(|\mathbf{q}_1|)$ term}

Lastly, we note the phase space factor $\Lambda$ contains a term, $- \bar{f}_{\nu H}(|\mathbf{q}_1|)$,  whose contribution to the $\nu_H$ collision integral~\eqref{eq:colnuH2} (or equation~(\ref{eq:colnuH}))  is independent of the integration variable.  Thus, the relevant final momentum-integration can be immediately performed, leading to the familiar result
\begin{equation}
\left(\frac{{\rm d} f_{\nu H} }{{\rm d} \tau}\right)^{(0)}_C (\mathbf{q}_1) =  \frac{\mathfrak{g}^2 a^2 \left((m_{\nu H}+m_{\nu l})^2-m_\phi^2 \right) }{4\pi \epsilon_1 |\mathbf{q}_1|} \left[ \cdots - \frac{|\mathbf{q}_1|}{m_{\nu H}^2} \Delta(m^2_{\nu H},m^2_{\nu l},m^2_\phi)
\bar{f}_{\nu H}(|\mathbf{q}_1|)   \right].
\label{eq:fbar1}
\end{equation}
The r.h.s.\ of equation~\eqref{eq:fbar1} is now of the form $\cdots - \Gamma_{\rm dec}^0 \left(a^2  m_{\nu H}/\epsilon_1 \right)  \bar{f}_{\nu H}$, where $\Gamma_{\rm dec}^0$ is but the  rest-frame decay rate.

%%%%%%%%%
%%%%%%%%

\subsection{First-order collision integral}

At linear order in the perturbed phase space densities, the collision integrals are given by
\begin{eqnarray}
\left(\frac{{\rm d} f_{\nu H} }{{\rm d} \tau}\right)^{(1)}_C (\mathbf{q}_1)
 &= & \, \frac{4 \mathfrak{g}^2}{\epsilon_1} \int \frac{\mathrm{d}^3\mathbf{q}_2}{(2\pi)^3 2 \epsilon_2} \int \frac{\mathrm{d}^3 \mathbf{q}_3}{(2\pi)^3 2 \epsilon_3} (2\pi)^4 \delta^{(4)}(q_1 - q_2 - q_3) \nonumber \\
 && \qquad \times (\eta_{\mu \nu} q_1^{\mu} q_2^{\nu} + a^2 m_{\nu H} m_{\nu l}) \left[ - F_{\nu H}(\mathbf{q}_1)\Omega_1(|\mathbf{q}_2|, |\mathbf{q}_3|)  \right.\nonumber\\
 && \qquad \qquad \left.+  F_{\nu l}(\mathbf{q}_2) \Omega_2(|\mathbf{q}_1|,|\mathbf{q}_3|)  +  F_{\phi}(\mathbf{q}_3) \Omega_3(|\mathbf{q}_1|,|\mathbf{q}_2|) \right],	\label{Eq:MasterPert} \\
\left(\frac{{\rm d} f_{\nu l} }{{\rm d} \tau}\right)^{(1)}_C (\mathbf{q}_2)
&= & \, - \frac{4 \mathfrak{g}^2}{\epsilon_2} \int \frac{\mathrm{d}^3\mathbf{q}_1}{(2\pi)^3 2 \epsilon_1}  \int \frac{\mathrm{d}^3 \mathbf{q}_3}{(2\pi)^3 2 \epsilon_3} (2\pi)^4 \delta^{(4)}(q_1 - q_2 - q_3)  \nonumber \\
&& \qquad \times (\eta_{\mu \nu} q_1^{\mu} q_2^{\nu} + a^2 m_{\nu H} m_{\nu l}) \left[ - F_{\nu H}(\mathbf{q}_1)\Omega_1(|\mathbf{q}_2|, |\mathbf{q}_3|)  \right.\nonumber\\
&& \qquad \qquad  \left.+  F_{\nu l}(\mathbf{q}_2) \Omega_2(|\mathbf{q}_1|,|\mathbf{q}_3|)  +  F_{\phi}(\mathbf{q}_3) \Omega_3(|\mathbf{q}_1|,|\mathbf{q}_2|) \right], \label{Eq:MasterPert2} 	 \\	
\left(\frac{{\rm d} f_{\phi} }{{\rm d} \tau}\right)^{(1)}_C (\mathbf{q}_3)
&= & \, - \frac{8 \mathfrak{g}^2}{\epsilon_3}\int \frac{\mathrm{d}^3\mathbf{q}_1}{(2\pi)^3 2 \epsilon_1}  \int \frac{\mathrm{d}^2 \mathbf{q}_2}{(2\pi)^3 2 \epsilon_2} (2\pi)^4 \delta^{(4)}(q_1 - q_2 - q_3) \nonumber \\
&& \qquad \times (\eta_{\mu \nu} q_1^{\mu} q_2^{\nu} + a^2 m_{\nu H} m_{\nu l}) \left[ - F_{\nu H}(\mathbf{q}_1)\Omega_1(|\mathbf{q}_2|, |\mathbf{q}_3|)  \right.\nonumber\\
&& \qquad \qquad  \left.+  F_{\nu l}(\mathbf{q}_2) \Omega_2(|\mathbf{q}_1|,|\mathbf{q}_3|)  +  F_{\phi}(\mathbf{q}_3) \Omega_3(|\mathbf{q}_1|,|\mathbf{q}_2|) \right],	\label{Eq:MasterPert3} 	
\end{eqnarray}
where
\begin{equation}
\begin{aligned}
\Omega_1 (|\mathbf{q}_2|, |\mathbf{q}_3|)  & = 1  -  \bar{f}_{\nu l}(|\mathbf{q}_2|)  +  \bar{f}_{\phi}(|\mathbf{q}_3|),\\
\Omega_2(|\mathbf{q}_1|, |\mathbf{q}_3|)  & =  \bar{f}_{\nu H}(|\mathbf{q}_1|)  +  \bar{f}_{\phi}(|\mathbf{q}_3|),  \\
\Omega_3(|\mathbf{q}_1|, |\mathbf{q}_2|)  & =  \bar{f}_{\nu l}(|\mathbf{q}_2|)  -  \bar{f}_{\nu H}(|\mathbf{q}_1|) 
\end{aligned}
\end{equation}
are phase space factors.  Observe that, as in the zeroth-order background case,  the combination of phase space factors is common to all three linearly perturbed collision integrals~\eqref{Eq:MasterPert}-\eqref{Eq:MasterPert3}.  Indeed, aside from these phase space factors, the background and linear collision integrals are in fact the same, and hence amenable to the same reduction procedure --- up to an angular integration --- discussed in section~\ref{sec:C0_nuH}.

Then, pasting together the results~(\ref{eq:colnuH}), \eqref{eq:C0_nuH_dcosa}, and~\eqref{eq:C0_nuH_dcosb}, we quickly obtain a reduced form of the $\nu_H$ collision integral~\eqref{Eq:MasterPert}:
\begin{equation}
\begin{aligned}
\left(\frac{{\rm d} f_{\nu H} }{{\rm d} \tau}\right)^{(1)}_C (\mathbf{q}_1) =& - \frac{\mathfrak{g}^2 a^2 \left((m_{\nu H}+m_{\nu l})^2-m_\phi^2 \right) }{4\pi \epsilon_1 |\mathbf{q}_1|} F_{\nu H}(\mathbf{q}_1) \int_{q_{2-}^{(\nu H)}}^{q_{2+}^{(\nu H)}} \mathrm{d}|\mathbf{q}_2| \, \frac{|\mathbf{q}_2|}{\epsilon_2} \, \Omega_1  \\
& + \frac{\mathfrak{g}^2}{\pi \epsilon_1} \int \mathrm{d} |\mathbf{q}_2| \, \mathrm{d}\cos{\alpha} \,  \Theta(\epsilon_1 - \epsilon_2)  \frac{|\mathbf{q}_2|^2}{\epsilon_2} \, K_\alpha(|\mathbf{q}_1|,|\mathbf{q}_2|,\cos \alpha) \, F_{\nu l}(\mathbf{q}_2) \, \Omega_2 \\
& -\frac{\mathfrak{g}^2}{\pi \epsilon_1} \int \mathrm{d} |\mathbf{q}_3| \, \mathrm{d}\cos{\beta} \,  \Theta(\epsilon_1 - \epsilon_3)  \frac{|\mathbf{q}_3|^2}{\epsilon_3} \, K_\beta(|\mathbf{q}_1|,|\mathbf{q}_3|,\cos \beta) \,  F_{\phi}(\mathbf{q}_3) \, \Omega_3, 
\label{eq:C1_nuH_1}
\end{aligned}
\end{equation}
where the first integral can be further evaluated to
\begin{equation}
\begin{aligned}
& \int_{q_{2-}^{(\nu H)}}^{q_{2+}^{(\nu H)}} \mathrm{d}|\mathbf{q}_2| \, \frac{|\mathbf{q}_2|}{\epsilon_2} \, \Omega_1 (|\mathbf{q}_2|, |\mathbf{q}_3|) = \\
& \qquad- \frac{|\mathbf{q}_1| }{m_{\nu H}^2}  \Delta(m_{\nu H}^2,m_{\nu l}^2, m_\phi^2)+ \int_{q_{2-}^{(\nu H)}}^{q_{2+}^{(\nu H)}} \mathrm{d} |\mathbf{q}_2| \frac{|\mathbf{q}_2|}{\epsilon_2} \bar{f}_{\nu l}(|\mathbf{q}_2|) - \int_{q_{3-}^{(\nu H)}}^{q_{3+}^{(\nu H)}} \mathrm{d} |\mathbf{q}_3| \frac{|\mathbf{q}_3|}{\epsilon_3} \bar{f}_{\phi}(|\mathbf{q}_3|) 
\end{aligned}   
\end{equation}
following equations~\eqref{eq:colnuH2} and~\eqref{eq:fbar1}.

Similarly, the reduced form of the $\nu_l$ collision integral~\eqref{Eq:MasterPert2}  can be constructed from equations~\eqref{eq:C0_nul_dcosa}, \eqref{eq:colnul}, and~\eqref{eq:C0_nul_dcosb}:
\begin{equation}
\begin{aligned}
\left(\frac{{\rm d} f_{\nu l} }{{\rm d} \tau}\right)^{(1)}_C (\mathbf{q}_2)=&  \, \frac{\mathfrak{g}^2}{\pi \epsilon_2} \int \mathrm{d} |\mathbf{q}_1| \, \mathrm{d}\cos{\alpha} \,  \Theta(\epsilon_1 - \epsilon_2)  \frac{|\mathbf{q}_1|^2}{\epsilon_1} \, K_\alpha (|\mathbf{q}_1|,|\mathbf{q}_2|,\cos\alpha) \, F_{\nu H}(\mathbf{q}_1) \, \Omega_1 \\
&- \frac{\mathfrak{g}^2 a^2 \left((m_{\nu H}+m_{\nu l})^2-m_\phi^2 \right) }{4\pi \epsilon_2 |\mathbf{q}_2|} F_{\nu l}(\mathbf{q}_2)
\int_{q_{1-}^{(\nu l)}}^{q_{1+}^{(\nu l)}} \mathrm{d}|\mathbf{q}_1| \, \frac{|\mathbf{q}_1|}{\epsilon_1} \, \Omega_2 \\
& - \frac{\mathfrak{g}^2}{\pi \epsilon_2} \int \mathrm{d} |\mathbf{q}_3| \, \mathrm{d}\cos{\gamma} \,  \Theta(\epsilon_2 + \epsilon_3)  \frac{|\mathbf{q}_3|^2}{\epsilon_3} \, K_\gamma(|\mathbf{q}_2|,|\mathbf{q}_3|,\cos \gamma)\, F_{\phi}(\mathbf{q}_3)\, \Omega_3,
\label{eq:C1_nul_1}
\end{aligned}
\end{equation}
where the second integral can also be written as
\begin{equation}
\int_{q_{1-}^{(\nu l)}}^{q_{1+}^{(\nu l)}} \mathrm{d}|\mathbf{q}_1| \, \frac{|\mathbf{q}_1|}{\epsilon_1} \, \Omega_2(|\mathbf{q}_1|,|\mathbf{q}_3|)=
\int_{q_{1-}^{(\nu l)}}^{q_{1+}^{(\nu l)}} \mathrm{d}|\mathbf{q}_1| \, \frac{|\mathbf{q}_1|}{\epsilon_1} \, \bar{f}_{\nu H} (|\mathbf{q}_1|) +  \int_{q_{3-}^{(\nu l)}}^{q_{3+}^{(\nu l)}} \mathrm{d}|\mathbf{q}_3| \, \frac{|\mathbf{q}_3|}{\epsilon_3} \, \bar{f}_\phi(|\mathbf{q}_3|)
\end{equation}
using equation~(\ref{eq:colnul2}).

Finally, the $\phi$ collision integral~\eqref{Eq:MasterPert3} follows from equations~equations~\eqref{eq:C0_phi_dcosa}, \eqref{eq:C0_phi_dcosb}, and \eqref{eq:colphi}:
\begin{equation}
\begin{aligned}
\left(\frac{{\rm d} f_{\phi} }{{\rm d} \tau}\right)^{(1)}_C (\mathbf{q}_3)= & \,- \frac{2 \mathfrak{g}^2}{\pi \epsilon_3} \int \mathrm{d} |\mathbf{q}_1| \, \mathrm{d}\cos{\beta} \,  \Theta(\epsilon_1 - \epsilon_3)  \frac{|\mathbf{q}_1|^2}{\epsilon_1}  \,K_\beta(|\mathbf{q}_1|,|\mathbf{q}_3|,\cos\beta) \, F_{\nu H}(\mathbf{q}_1) \, \Omega_1\\
&- \frac{2 \mathfrak{g}^2}{\pi \epsilon_3} \int \mathrm{d} |\mathbf{q}_2| \, \mathrm{d}\cos{\gamma} \,  \Theta(\epsilon_2 + \epsilon_3)  \frac{|\mathbf{q}_2|^2}{\epsilon_2} \, K_\gamma (|\mathbf{q}_2|,|\mathbf{q}_3|,\cos \gamma)\,
F_{\nu l}(\mathbf{q}_2) \, \Omega_2\\
& - \frac{\mathfrak{g}^2 a^2 \left((m_{\nu H}+m_{\nu l})^2-m_\phi^2 \right) }{2\pi \epsilon_3 |\mathbf{q}_3|}F_\phi (\mathbf{q}_3) \int_{q_{1-}^{(\phi)}}^{q_{1+}^{(\phi)}} \mathrm{d}|\mathbf{q}_1| \, \frac{|\mathbf{q}_1|}{\epsilon_1} \, \Omega_3,
\label{eq:C1_phi_1}
\end{aligned}
\end{equation}
where the last integral is equivalently
\begin{equation}
\int_{q_{1-}^{(\phi)}}^{q_{1+}^{(\phi)}} \mathrm{d}|\mathbf{q}_1| \, \frac{|\mathbf{q}_1|}{\epsilon_1} \, \Omega_3(|\mathbf{q}_1|,|\mathbf{q}_2|)=\int_{q_{2-}^{(\phi)}}^{q_{2+}^{(\phi)}} \mathrm{d}|\mathbf{q}_2| \, \frac{|\mathbf{q}_2|}{\epsilon_2} \, \bar{f}_{\nu l}(|\mathbf{q}_2|) - \int_{q_{1-}^{(\phi)}}^{q_{1+}^{(\phi)}} \mathrm{d}|\mathbf{q}_1| \, \frac{|\mathbf{q}_1|}{\epsilon_1} \, \bar{f}_{\nu H}(|\mathbf{q}_1|)
\end{equation}
following equation~(\ref{eq:colphi2}).

%%%%%%%%
%%%%%%%%

\subsection{Legendre decomposition}

In Fourier space, the perturbed phase space distribution $F_i$ is a function of the magnitudes of Fourier wave vector~$\mathbf{k}$ and comoving momentum~$\mathbf{q}_i$, and the angle between them $\hat{k} \cdot \hat{q}_i$.
We therefore express $F_i$ in term of an infinite series of Legendre polynomials $P_\ell$, i.e., 
\begin{equation}
\begin{aligned}
F_i(|\mathbf{k}|,|\mathbf{q}_i|,\hat{k}\cdot\hat{q}_i) &= \sum_{\ell = 0}^\infty (-{\rm i})^{\ell} (2\ell + 1) F_{i,\ell}(|\mathbf{k}|,|\mathbf{q}_i|) \, P_{\ell}(\hat{k}\cdot\hat{q}_i) , \\
F_{i,\ell}(|\mathbf{k}|,|\mathbf{q}_i|) &= \frac{{\rm i}^{\ell}}{2} \int_{-1}^{1} \mathrm{d}(\hat{k} \cdot \hat{q}_i) \, F_i(|\mathbf{k}|,|\mathbf{q}_i|,\hat{k} \cdot \hat{q}_i) \, P_{\ell}(\hat{k} \cdot \hat{q}_i). 
\label{eq:Legendre_dec}
\end{aligned}	 
\end{equation}
The decomposition of the l.h.s.\ Boltzmann equation into a Boltzmann hierarchy is well known~\cite{Ma:1995ey} and the result is displayed in, e.g., equation~\eqref{hierarchyPsi_nu1}.  The collision integral on the r.h.s., on the other hand, can be obtained according to
\begin{equation}
\left( \frac{{\rm d} f_{i}}{{\rm d} \tau}\right)_{C,\ell}^{(1)} 
(|\mathbf{q}_i|) = \frac{{\rm i}^\ell}{4\pi} \int \mathrm{d}\Omega_{\mathbf{k}} \,P_{\ell}(\hat{k}\cdot \hat{q}_i) \,  \left(\frac{{\rm d} f_i }{{\rm d} \tau}\right)^{(1)}_C (\mathbf{q}_i).
\label{eq:coldecompose}
\end{equation}
where ${\rm d} \Omega_\mathbf{k}$ is an incremental solid angle subtended by the wave vector $\mathbf{k}$.  

A close inspection of equations~(\ref{eq:C1_nuH_1})-\eqref{eq:C1_phi_1} shows that there are only two types of decompositions to deal with:
\begin{enumerate}
\item  Those whose angular dependence is contained in a phase space perturbation~$F_i$ that appears outside the integral.  The first, second and third terms of equations~(\ref{eq:C1_nuH_1}), (\ref{eq:C1_nul_1}) and (\ref{eq:C1_phi_1}), respectively, fall into this category.  The decomposition of these terms is trivial.  Using the first term of equation~(\ref{eq:C1_nuH_1}), the decomposition consists in a simple replacement
\begin{equation}
F_{\nu H}(|\mathbf{k}|,|\mathbf{q}_1|,\hat{k} \cdot \hat{q}_1) \int_{q_{2-}^{(\nu H)}}^{q_{2+}^{(\nu H)}} \mathrm{d}|\mathbf{q}_2| \, \frac{|\mathbf{q}_2|}{\epsilon_2} \, \Omega_1 \to F_{\nu H,\ell}(|\mathbf{k}|,|\mathbf{q}_1|) \int_{q_{2-}^{(\nu H)}}^{q_{2+}^{(\nu H)}} \mathrm{d}|\mathbf{q}_2| \, \frac{|\mathbf{q}_2|}{\epsilon_2} \, \Omega_1,
\end{equation}
when embedding the term in the relevant Boltzmann hierarchy.

\item The angular part of the remaining six terms are of the form	
\begin{equation}
\begin{aligned}
\frac{i^{\ell}}{4\pi}  \int \mathrm{d}\Omega_{\mathbf{k}} \, \int \mathrm{d}(\hat{q}_i \cdot \hat{q}_j) \, K(|\mathbf{q}_i|,|\mathbf{q}_j|,\hat{q}_i\cdot \hat{q}_j) \, F_u(|\mathbf{k}|,|\mathbf{q}_u|,\hat{k}\cdot \hat{q}_u) \, P_{\ell}(\hat{k}\cdot\hat{q}_v) \, 
\end{aligned}
\end{equation} 
where $u, v= i,j$, $i \neq j$, and $u \neq v$.
\end{enumerate}

We use the second term of~\eqref{eq:C1_nuH_1} to illustrate the decomposition of this last type of terms; generalisation to other similar terms is simple.
We begin by applying the decomposition~\eqref{eq:coldecompose} to the angular part:
\begin{equation}
\begin{aligned}
C\equiv \frac{i^{\ell}}{4\pi}  \int \mathrm{d}\Omega_{\mathbf{k}} \, \int \mathrm{d}\cos{\alpha} \, K_\alpha(|\mathbf{q}_1|,|\mathbf{q}_2|,\cos{\alpha}) \, F_u(|\mathbf{k}|,|\mathbf{q}_u|,\hat{k}\cdot \hat{q}_u) \, P_{\ell}(\hat{k}\cdot\hat{q}_v) \, ,
\end{aligned}
\label{eq:Legendre2}
\end{equation} 
where $u=2$ and $v=1$ in this case; setting $u=1$ and $v=2$ corresponds to the first term of equation~(\ref{eq:C1_nul_1}).  We keep the $u,v$ labels general, in order to demonstrate the applicability of final result under exchange of these indices.
Expanding \(F_{\nu l}(\hat{k}\cdot \hat{q}_2) \) in a Legendre series as per equation~\eqref{eq:Legendre_dec} and rewriting the Legendre polynomials in terms of spherical harmonics,
\begin{equation}
	P_{\ell}(\hat{k}\cdot\hat{q}_i) = \frac{4\pi}{2\ell + 1} \sum_{m=-\ell}^\ell Y_{\ell,m}(\hat{q}_i)Y_{\ell,m}^*(\hat{k}) \, , 
	\label{eq:PlY}
\end{equation}
we find
\begin{equation}
	\begin{aligned}
	C =&\frac{1}{4\pi}  \sum_{\ell'=0}^\infty i^\ell (-i)^{\ell'} (2\ell'+1) F_{u,\ell} (|\mathbf{k}|,|\mathbf{q}_u|)\int \mathrm{d}\Omega_{\mathbf{k}} \int \mathrm{d}\cos{\alpha} \, K_\alpha(|\mathbf{q}_1|,|\mathbf{q}_2|,\cos{\alpha})  \\
	& \qquad\times \frac{4\pi}{(2\ell+1)}\frac{4\pi}{(2\ell'+1)} \sum_{m=-\ell}^\ell \sum_{m'=-\ell'}^{\ell'}Y_{\ell,m}(\hat{q}_v)Y_{\ell,m}^*(\hat{k}) Y_{\ell',m'}(\hat{k})Y_{\ell',m'}^*(\hat{q}_u) \, . 
	\end{aligned}
\end{equation}
Two of the spherical harmonics can be eliminated by using the orthogonality condition
\begin{equation}
	\int \mathrm{d}\Omega_{\mathbf{k}} \, Y_{\ell,m}(\hat{k}) Y^*_{\ell',m'}(\hat{k}) = \delta_{\ell \ell'} \delta_{m m'} \, , 
\end{equation}
where \(\delta\) is the Kronecker delta.  This leaves us with
\begin{equation}
C =  F_{u,\ell}(|\mathbf{k}|,|\mathbf{q}_u|) \int \mathrm{d}\cos{\alpha} \, K(|\mathbf{q}_1|,|\mathbf{q}_2|,\cos{\alpha}) \, P_{\ell}(\hat{q}_1\cdot\hat{q}_2) \, ,
\end{equation}
where the last two spherical harmonics that have been regrouped in a Legendre polynomial via relation~(\ref{eq:PlY}), and we have explicitly used $u=2$ and $v=1$ in the argument of the Legendre polynomial because it is symmetric in these indices.
 Lastly, using the Dirac delta distribution contained in the scattering kernel $K_\alpha(|\mathbf{q}_1|,|\mathbf{q}_2|,\cos\alpha)$  (see equation~\eqref{eq:ka}) to evaluate the angular integration, we find
\begin{equation}
C =  \frac{a^2(m_{\nu H}+m_{\nu l})^2 - a^2m_{\phi}^2}{4 |\mathbf{q}_1| |\mathbf{q}_2|} \, F_{u,\ell}(|\mathbf{k}|,|\mathbf{q}_u|) \, P_{\ell}(\cos \alpha^*) \, ,
\label{eq:cc}
\end{equation}
where $\cos \alpha^*$ is given in equation~(\ref{eq:cosa*}).  

It is straightforward to generalise the result~(\ref{eq:cc}) to the other two pairs of $\mathbf{q}$ vectors:
\begin{equation}
\begin{aligned}
&\frac{i^{\ell}}{4\pi}  \int \mathrm{d}\Omega_{\mathbf{k}} \, \int \mathrm{d}\cos{\beta} \, K_\beta(|\mathbf{q}_1|,|\mathbf{q}_3|,\cos{\beta}) \, F_u(|\mathbf{k}|,|\mathbf{q}_u|,\hat{k}\cdot \hat{q}_u) \, P_{\ell}(\hat{k}\cdot\hat{q}_v)  \\
& \hspace{50mm}=  - \frac{a^2(m_{\nu H}+m_{\nu l})^2 - a^2m_{\phi}^2}{4 |\mathbf{q}_1| |\mathbf{q}_3|} \, F_{u,\ell}(|\mathbf{k}|,|\mathbf{q}_u|) \, P_{\ell}(\cos \beta^*) ,
\end{aligned}
\end{equation}
where $u=1,3$, and 
\begin{equation}
\begin{aligned}
&\frac{i^{\ell}}{4\pi}  \int \mathrm{d}\Omega_{\mathbf{k}} \, \int \mathrm{d}\cos{\beta} \, K_\gamma(|\mathbf{q}_2|,|\mathbf{q}_3|,\cos{\gamma}) \, F_u(|\mathbf{k}|,|\mathbf{q}_u|,\hat{k}\cdot \hat{q}_u) \, P_{\ell}(\hat{k}\cdot\hat{q}_v)  \\
& \hspace{50mm}=  \frac{a^2(m_{\nu H}+m_{\nu l})^2 - a^2m_{\phi}^2}{4 |\mathbf{q}_2| |\mathbf{q}_3|} \, F_{u,\ell}(|\mathbf{k}|,|\mathbf{q}_u|) \, P_{\ell}(\cos \gamma^*),
\end{aligned}
\end{equation}
for $u=2,3$.  Collecting all terms and bearing in mind that the conditions $\cos \alpha^* \in [-1,1]$, etc.\ generate integration limits for the remaining momentum-integral, we finally arrive at
\begin{equation}
\begin{aligned}
\left( \frac{{\rm d} f_{\nu H}}{{\rm d} \tau}\right)_{C,\ell}^{(1)} 
(|\mathbf{q}_1|)
=& \, \frac{\mathfrak{g}^2 a^2 \left((m_{\nu H}+m_{\nu l})^2-m_\phi^2 \right) }{4\pi \epsilon_1 |\mathbf{q}_1|} \left \{-F_{\nu H,\ell}(|\mathbf{q}_1|) \int_{q_{2-}^{(\nu H)}}^{q_{2+}^{(\nu H)}} \mathrm{d}|\mathbf{q}_2| \, \frac{|\mathbf{q}_2|}{\epsilon_2} \, \Omega_1 \right. \\
& \hspace{-30mm}\left.+  \int^{q_{2+}^{(\nu H)}}_{q_{2-}^{(\nu H)}} \mathrm{d} |\mathbf{q}_2|  \frac{|\mathbf{q}_2|}{\epsilon_2} \, P_\ell (\cos \alpha^*) \, F_{\nu l,\ell}(|\mathbf{q}_2|) \, \Omega_2 +\int^{q_{3+}^{(\nu H)}}_{q_{3-}^{(\nu H)}} \mathrm{d} |\mathbf{q}_3|  \,  \frac{|\mathbf{q}_3|}{\epsilon_3} \, P_\ell(\cos \beta^*) \,  F_{\phi,\ell}(|\mathbf{q}_3|) \, \Omega_3 \right\},
\label{eq:C1_nuH_1a}
\end{aligned}
\end{equation}
\begin{equation}
\begin{aligned}
\left( \frac{{\rm d} f_{\nu l}}{{\rm d} \tau}\right)_{C,\ell}^{(1)} 
(|\mathbf{q}_2|)
=&\,
 \frac{\mathfrak{g}^2 a^2 \left((m_{\nu H}+m_{\nu l})^2-m_\phi^2 \right) }{4\pi \epsilon_2 |\mathbf{q}_2|} \left\{
 \int_{q_{1-}^{(\nu l)}}^{q_{1+}^{(\nu l)}}  \mathrm{d} |\mathbf{q}_1| \,   \frac{|\mathbf{q}_1|}{\epsilon_1} \, P_\ell(\cos \alpha^*) F_{\nu H,\ell}(|\mathbf{q}_1|) \, \Omega_1 \right.\\
&\hspace{-5mm}\left.-  F_{\nu l,\ell}(|\mathbf{q}_2|)
\int_{q_{1-}^{(\nu l)}}^{q_{1+}^{(\nu l)}} \mathrm{d}|\mathbf{q}_1| \, \frac{|\mathbf{q}_1|}{\epsilon_1} \, \Omega_2-  \int_{q_{3-}^{(\nu l)}}^{q_{3+}^{(\nu l)}}  \mathrm{d} |\mathbf{q}_3|  \,   \frac{|\mathbf{q}_3|}{\epsilon_3} \, P_\ell(\cos \gamma^*) F_{\phi,\ell}(|\mathbf{q}_3|)\, \Omega_3 \right\},
\end{aligned}
\end{equation}
\begin{equation}
\begin{aligned}
\left( \frac{{\rm d} f_{\phi}}{{\rm d} \tau}\right)_{C,\ell}^{(1)} 
(|\mathbf{q}_3|)
= &\,\frac{\mathfrak{g}^2 a^2 \left((m_{\nu H}+m_{\nu l})^2-m_\phi^2 \right) }{2\pi \epsilon_3 |\mathbf{q}_3|} \left\{
 \int_{q_{1-}^{(\phi)}}^{q_{1+}^{(\phi)}} \mathrm{d} |\mathbf{q}_1|  \,  \frac{|\mathbf{q}_1|}{\epsilon_1}  \, P_\ell(\cos \beta^*) F_{\nu H,\ell}(|\mathbf{q}_1|) \, \Omega_1 \right.\\
&\hspace{-5mm}\left.- \int_{q_{2-}^{(\phi)}}^{q_{2+}^{(\phi)}} \mathrm{d} |\mathbf{q}_2| \,   \frac{|\mathbf{q}_2|}{\epsilon_2} \, P_\ell(\cos \gamma^*)
F_{\nu l,\ell}(|\mathbf{q}_2|) \, \Omega_2
 - F_{\phi,\ell} (|\mathbf{q}_3|) \int_{q_{1-}^{(\phi)}}^{q_{1+}^{(\phi)}} \mathrm{d}|\mathbf{q}_1| \, \frac{|\mathbf{q}_1|}{\epsilon_1} \, \Omega_3 \right\}.
\end{aligned}
\end{equation}
The $m_\phi=0$ limits of these expressions are given in the main text.

%%%%%%%%%%%%%
%%%%%%%%%%%%%

\section{Anisotropic stress loss under the separable ansatz}
\label{app:separable}

We show in this appendix some of the details in the derivation of equation~\eqref{eq:loss1} from the exact collision integral~\eqref{eq:dpidt}, under the assumptions of (i)~equilibrium Maxwell--Boltzmann statistics, (ii)~the separable ansatz~\eqref{eq:separable}, and (iii) same perturbation contrast in all particle species as per equation~\eqref{eq:sameperturbation}.  Justifications for these assumptions can be found in section~\ref{sec:separable}.

Firstly, we note that under the above assumptions, the total anisotropic stress of the neutrino--scalar system~\eqref{eq:a4pidef} can be expressed  as 
\begin{equation}
\begin{aligned}
a^4 \Pi_{\nu\phi} & = \frac{2 T^4_0}{\pi^2} \left[  \frac{g_{\nu H}}{2}\left(\frac{a m_{\nu H}}{T_0} \right)^2 K_2 \left(\frac{a m_{\nu H}}{T_0} \right)
e^{\mu_{\nu H}/T_0}  + g_{\nu l}  e^{\mu_{\nu l}/T_0}+ g_{\phi} e^{\mu_\phi/T_0} \right] \, {\cal F}_{\ell}   \\
& = 2 T_0 \left[a^3 \bar{n}_{\nu H} + a^3 \bar{n}_{\nu l} + a^3 \bar{n}_{\phi} \right]  {\cal F}_{\ell} ,
\end{aligned}
\end{equation}
where we have used the relation
\begin{equation}
a^3 \bar{n}_i = g_i \frac{a^2 m_i^2 T_0}{2 \pi^2} \,  K_2\left(\frac{a m_i}{T_0} \right)\,  e^{\mu_i/T_0}
\end{equation}
at the approximate second equality.  The function $K_2(x)$ is a modified Bessel function of the second kind, and $x^2 K_2(x) \to 2$ as $x \to 0$.

For the collision integral~\eqref{eq:dpidt}, the individual contributions can be computed as follows.  Under the above assumptions, reduction of the first term is straightforward:
\begin{equation}
\begin{aligned}
& \frac{1}{3 \pi^2}
\left[	g_{\nu H} \int \mathrm{d}q_1 \, q_1^2 \, \frac{q_1^2}{\epsilon_1}  \left(\frac{\mathrm{d} f_{\nu H}}{\mathrm{d}\tau}\right)_{C,  2}^{(1)} \right]  \\
& =  a \tilde{\Gamma}_{\rm dec} T_0^{-4} 
\int {\rm d} q_1 \, \frac{q_1^3}{{\epsilon}_1^2} \, e^{-{\epsilon}_1/T_0} 
\int_{{q}_{2-}^{(\nu H)}}^{{q}_{2+}^{(\nu H)}} \mathrm{d} {q}_2 
\left[- \frac{{q}_1^2}{{\epsilon}_1} 
+ 2\,  {q}_2   \,P_2 (\cos \alpha^*)   \right] (a^4 \Pi_{\nu \phi})\\
& = a \tilde{\Gamma}_{\rm dec} T_0^{-4} 
\int {\rm d} {q}_1 \, {q}_1^3 e^{-{\epsilon}_1/T_0}
\left[ \frac{3{\epsilon}_1}{2{q}_1} - \frac{q_1}{2 {\epsilon}_1}- \frac{{q}_1^3}{{\epsilon}_1^3}- \frac{3 a^2{m}_{\nu H}^2}{{\epsilon}_1 {q}_1} 
+ \frac{3 a^4 {m}_{\nu H}^4}{4 {\epsilon}_1^2 {q}_1^2} \ln\left(\frac{{\epsilon}_1+{q}_1}{{\epsilon}_1-{q}_1}\right)
\right](a^4 \Pi_{\nu \phi}) ,
\label{eq:term1}
\end{aligned}
\end{equation}
with
\begin{equation}
\tilde{\Gamma}_{\rm dec} \equiv \Gamma_{{\rm dec}}^0 \frac{a m_{\nu H}}{24  T_0} \frac{\bar{n}_{\nu H}}{\sum_i \bar{n}_i} \left[\frac{1}{2}
\left(\frac{a m_{\nu H}}{T_0} \right)^2 K_2 \left(\frac{a m_{\nu H}}{T_0} \right)
\right]^{-1}.
\end{equation}
Note that we have not expanded equation~\eqref{eq:term1} in small $a m_{\nu H}/\epsilon_1$, and we leave the integration over $q_1$ for a later stage of the calculation.

Similarly, under the same assumptions, the second term of the collision integral~\eqref{eq:dpidt} reads
\begin{equation}
\begin{aligned}
& \frac{1}{3 \pi^2}
 \left[	g_{\nu l} \int \mathrm{d}q_2 \, q_2^3 \, \left(\frac{\mathrm{d} f_{\nu l}}{\mathrm{d}\tau}\right)_{C,  2}^{(1)} \right]  
 = a \tilde{\Gamma}_{\rm dec} T_0^{-4}  
\int {\rm d} {q}_2 \, {q}_2 \\
&\times \left\{
\int_{{q}_{1-}^{(\nu l)}}^{{q}_{1+}^{(\nu l)}}  \mathrm{d} {q}_1 \,  \frac{{q}_1^3}{{\epsilon}_1^2} \, e^{-{\epsilon}_1/T_0} \left[P_2(\cos \alpha^*) -  \frac{{q}_2{\epsilon}_1}{{q}_1^2} \right]
- \int_{{q}_{3-}^{(\nu l)}}^{{q}_{3+}^{(\nu l)}}  \mathrm{d} {q}_3  \,    P_2(\cos \gamma^*) {q}_3 e^{-{\epsilon}_1/T_0} \right\} \,  (a^4 \Pi_{\nu \phi}) .
\label{eq:term2}
\end{aligned}
\end{equation}
To proceed further, we note that the integration order of the two double integrals can be easily swapped via the relations between Heaviside step functions derived in appendix~\ref{app:CollisionIntegralReduction}. The first double integral can rearranged using equation~\eqref{eq:limitsq1q2}  into
\begin{equation}
\int {\rm d} q_2  
\int_{q_{1-}^{(\nu l)}}^{q_{1+}^{(\nu l)}}  \mathrm{d} q_1  = \int {\rm d} q_1  
\int_{q_{2-}^{(\nu H)}}^{q_{2+}^{(\nu H)}}  \mathrm{d} q_2.
\label{eq:rearrange1}
\end{equation}
To rearrange the second double integral, we first change the integration variable from $q_3$ to $q_1$ via the relation $q_3 = \epsilon_1-q_2$, so that the rest of the rearrangement  proceeds again via a relation between the Heaviside step functions~\eqref{eq:limitsq1q2} and~\eqref{eq:thetagamma}:
\begin{equation}
\begin{aligned}
\left. \Theta(q_3 - {q}_{3-}^{(\nu l)} ) \Theta({q}_{3+}^{(\nu l)}-q_3)\right|_{q_3 \to \epsilon_1 - q_2} &= \left.  \Theta(1-\cos\gamma^*) \right|_{q_3 \to \epsilon_1 - q_2} \\
& =  \Theta(1-\cos \alpha^*) \\
& = \Theta(q_2 - {q}_{2-}^{(\nu H)} ) \Theta({q}_{2+}^{(\nu H)}-q_2),
\label{eq:rearrange2}
\end{aligned}
\end{equation}
where the notation $\left. P_2(\cos \gamma^*)   \right|_{q_3 \to \epsilon_1-q_2}$ indicates that all occurrences of $q_3$ in the scattering kernel should be replaced with $\epsilon_1-q_2$.

Then, substituting equations~\eqref{eq:rearrange1} and~\eqref{eq:rearrange2} into equation~\eqref{eq:term2}, we find
 \begin{equation}
 \begin{aligned}
 & \frac{1}{3 \pi^2}
  \left[	g_{\nu l} \int \mathrm{d}q_2 \, q_2^3 \, \left(\frac{\mathrm{d} f_{\nu l}}{\mathrm{d}\tau}\right)_{C,  2}^{(1)} \right]  \\
&  = a \tilde{\Gamma}_{\rm dec} T_0^{-4}   \int  \mathrm{d} q_1 \, \frac{{q}_1^3}{{\epsilon}_1^2} \,  e^{-{\epsilon}_1/T_0}  \\
&\qquad  \times 
 \int^{q_{2+}^{(\nu H)}}_{q_{2-}^{(\nu H)}} {\rm d} {q}_2 \, {q}_2  \left[P_2(\cos \alpha^*)-  \frac{{q}_2{\epsilon}_1}{{q}_1^2}  -  
 \frac{{\epsilon}_1({\epsilon}_1-{q}_2)}{{q}_1^2}    \left.P_2(\cos \gamma^*)\right|_{q_3 \to \epsilon1-q_2} \right] \,  (a^4 \Pi_{\nu \phi}) \\
 & =  a \tilde{\Gamma}_{\rm dec} T_0^{-4}   
 \int  \mathrm{d} {q}_1 \, {q}_1^3 \,  e^{-{\epsilon}_1/T_0}
 \left[ \frac{{\epsilon}_1}{4 {q}_1} - \frac{{q}_1}{4 {\epsilon}_1}
 - \frac{3 a^4 {m}_{\nu H}^4}{8 {\epsilon}_1^2 {q}_1^2} \ln\left(\frac{{\epsilon}_1+{q}_1}{{\epsilon}_1-{q}_1} \right)  \right]
 (a^4 \Pi_{\nu \phi}).
 \label{eq:term2a}
 \end{aligned}
 \end{equation}
Lastly, we note that the third term of the  collision integral~\eqref{eq:dpidt} must equal the second term, 
\begin{equation}
\frac{1}{3 \pi^2}
 \left[	g_{\nu l} \int \mathrm{d}q_2 \, q_2^3 \, \left(\frac{\mathrm{d} f_{\nu l}}{\mathrm{d}\tau}\right)_{C,  2}^{(1)} \right] 
 = \frac{1}{3 \pi^2}
\left[	g_{\phi} \int \mathrm{d}q_3 \, q_3^3 \, \left(\frac{\mathrm{d} f_{\phi}}{\mathrm{d}\tau}\right)_{C,  2}^{(1)} \right] ,
\label{eq:equal}
\end{equation}
because of the assumption that $m_{\nu l} = m_\phi=0$.

Combining together equations~\eqref{eq:term1}, \eqref{eq:term2a}, and \eqref{eq:equal}, we finally arrive at
\begin{equation}
\begin{aligned}
\left(\frac{\mathrm{d}(a^4\Pi_{\nu \phi})}{\mathrm{d}\tau}\right)_C = 
a \tilde{\Gamma}_{\rm dec} T_0^{-4}   
\int  \mathrm{d} {q}_1 \, {q}_1^3 \,  e^{-{\epsilon}_1/T_0}
\left[ \frac{2{\epsilon}_1}{{q}_1} - \frac{{q}_1}{ {\epsilon}_1}- \frac{{q}_1^3}{{\epsilon}_1^3} -  \frac{3 a^2{m}_{\nu H}^2}{{\epsilon}_1 {q}_1} 
\right](a^4 \Pi_{\nu \phi}),
\end{aligned}
\end{equation}
which is identically equation~\eqref{eq:loss1}.

%%%%%%%%%%%%
%%%%%%%%%%%%%

\section{Number, energy and momentum conservation}
\label{appendix:conservation_laws}

%%%%%%%%%%%
\begin{figure}
	\begin{minipage}[b]{0.517\linewidth}
		\centering
		\includegraphics[width=\textwidth]{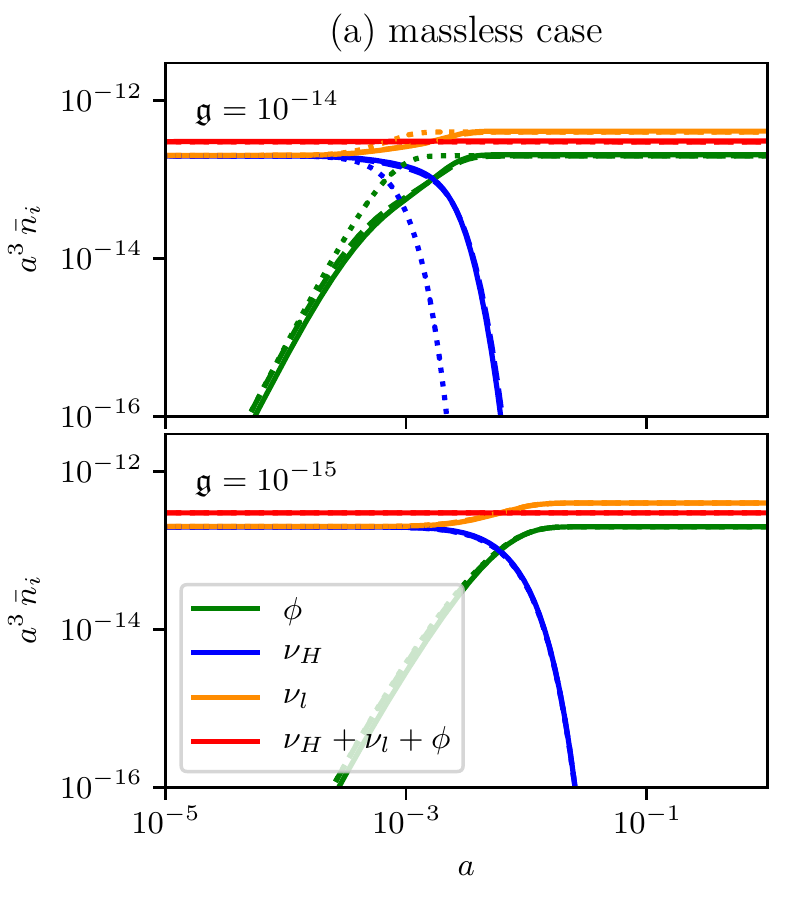}
	\end{minipage}
	\begin{minipage}[b]{0.463\linewidth}
		\centering
		\includegraphics[width=\textwidth]{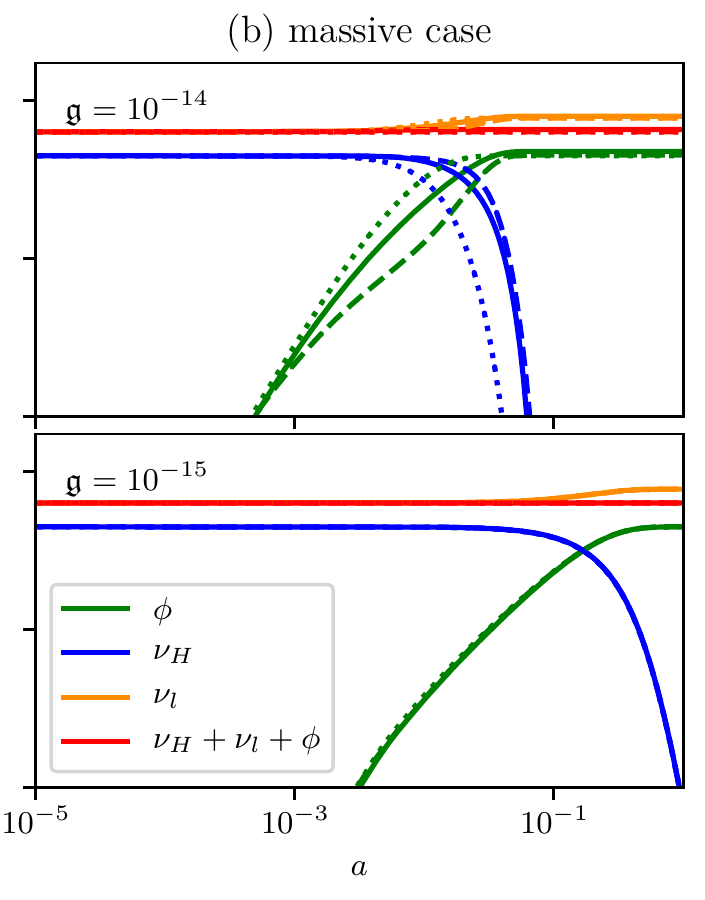}
	\end{minipage}
	\caption{Numerical demonstration of $n_{\nu H}+ \frac{1}{2} (n_{\phi}+ n_{\nu l})$ conservation (red line) for (a)~the massless $\nu_l$ case considered in section~\ref{sec:Massless} with $m_{\nu H}=0.3$ eV (left), and (b)~the massive $\nu_l$ case, studied in section~\ref{sec:Massive} assuming $m_{\nu l}=0.1$~eV (right).  Solid lines include all contributions~(dec+inv+qs), dashed lines only decay and inverse decay~(dec+inv), and dotted lines only decay (dec). }
	\label{fig:number}
\end{figure}
%%%%%%%%%

In this appendix we demonstrate analytically in the decay-only limit that our Boltzmann equations~\eqref{eq:background_Boltzmann1}-\eqref{eq:background_Boltzmann3} conserve energy, momentum and number density. 

Conservation of these quantities imply the requirements
\begin{enumerate}
\item Number density conservation:
\begin{equation}
g_{\nu H} \! \! \int \!\! \mathrm{d}q_1 \,  q_1^2  \left( \frac{{\rm d} f_{\nu H}}{{\rm d} \tau}\right)_{C,\ell=0}^{(0,1)} 
 + \frac{1}{2} \left[ g_{\nu l} \!\! \int \!\!  \mathrm{d}q_2 \,  q_2^2  \left( \frac{{\rm d} f_{\nu l}}{{\rm d} \tau}\right)_{C,\ell=0}^{(0,1)} 
 + g_{\phi} \!\! \int \!\! \mathrm{d}q_3 \, q_3^2 \left( \frac{{\rm d} f_{\nu l}}{{\rm d} \tau}\right)_{C,\ell=0}^{(0,1)} \right]=0,
\label{eq:number_conservation}
\end{equation} 
which applies to the background and perturbed $\ell=0$ collision integrals.

\item Energy conservation:
\begin{equation}
g_{\nu H} \! \! \int \!\! \mathrm{d}q_1\,  q_1^2 \epsilon_1 \left( \frac{{\rm d} f_{\nu H}}{{\rm d} \tau}\right)_{C,\ell=0}^{(0,1)} 
+  g_{\nu l} \!\! \int \!\!  \mathrm{d}q_2 \,  q_2^2 \epsilon_2 \left( \frac{{\rm d} f_{\nu l}}{{\rm d} \tau}\right)_{C,\ell=0}^{(0,1)} 
+ g_{\phi} \!\! \int \!\! \mathrm{d}q_3 \, q_3^2  \epsilon_3 \left( \frac{{\rm d} f_{\nu l}}{{\rm d} \tau}\right)_{C,\ell=0}^{(0,1)} =0,
\label{eq:energy_conservation}
\end{equation} 
which, again, must hold at both the background and perturbed $\ell=0$ level.

\item Momentum conservation:
\begin{equation}
g_{\nu H} \!  \int \!\mathrm{d}q_1\,  q_1^3 \left( \frac{{\rm d} f_{\nu H}}{{\rm d} \tau}\right)_{C,\ell=1}^{(1)} 
+  g_{\nu l} \! \int \!  \mathrm{d}q_2 \,  q_2^3 \left( \frac{{\rm d} f_{\nu l}}{{\rm d} \tau}\right)_{C,\ell=1}^{(1)} 
+ g_{\phi} \! \int \! \mathrm{d}q_3 \, q_3^3\left( \frac{{\rm d} f_{\nu l}}{{\rm d} \tau}\right)_{C,\ell=1}^{(1)} =0,
\label{eq:momentum_conservation}
\end{equation} 
which, in this form, applies to the perturbed $\ell=1$ collision integrals.  Momentum conservation of the background collision integrals is trivially fulfilled because $\int \mathrm{d} {q}_i \, q_i^2 \int {\rm d} \Omega_{\mathbf{q}_i} \mathbf{q}_i\,  f(q_i ) =0$ if $f(q_i )$ has no directional dependence. 

\end{enumerate}
For simplicity, we restrict our analytical calculations  to the case of  $m_{\nu l}=0$.  Numerical solutions of   equations~\eqref{eq:background_Boltzmann1}-\eqref{eq:background_Boltzmann3} --- including all effects --- demonstrate however that conditions~\eqref{eq:number_conservation} and~\eqref{eq:energy_conservation} are indeed well satisfied on the background level. See, respectively, figures~\ref{fig:number} and~\ref{fig:energy_massless}.

The first terms in \eqref{eq:number_conservation},  \eqref{eq:energy_conservation} and \eqref{eq:momentum_conservation} are trivially given respectively by 
\begin{align}
2\int \mathrm{d}q_1\,  q_1^2 \,  \left( \frac{{\rm d} f_{\nu H}}{{\rm d} \tau}\right)_{C}^{(0)} &=  - 2\frac{a^2 m_{\nu H}}{\tau_0} \int \mathrm{d}q_1 \frac{q_1^2}{\epsilon_1} \, \bar{f}_{\nu H}(q_1),  \hspace{-0.5cm} &(\text{number})
\label{eq:number_RHS} \\
2\int \mathrm{d}q_1\,  q_1^2 \epsilon_1 \,  \left( \frac{{\rm d} f_{\nu H}}{{\rm d} \tau}\right)_{C}^{(0)} &= - 2\frac{a^2 m_{\nu H}}{\tau_0} \int \mathrm{d}q_1 q_1^2 \, \bar{f}_{\nu H}(q_1) = - \frac{2 \pi^2 a^5 m_{\nu H}}{\tau_0} \bar{n}_{\nu H}, \hspace{-0.5cm} &(\text{energy})
\label{eq:energy_RHS} \\
2\int \mathrm{d} q_1\,  q_1^3 \,  \left( \frac{{\rm d} f_{\nu H}}{{\rm d} \tau}\right)_{C,\ell=1}^{(1)} 
 &=  -2 \frac{a^2 m_{\nu H}}{\tau_0} \int \mathrm{d}q_1 \frac{q_1^3}{\epsilon_1} \, F_{\nu H,1}(q_1),  \hspace{-0.5cm} &(\text{momentum})
\label{eq:momentum_RHS}
\end{align}
where $\tau_0 \equiv 4 \pi/(\mathfrak{g} a^2 m^2_{\nu H})$ is the rest-frame lifetime of $\nu_H$ in the limit under consideration, and 
 equations~\eqref{eq:number_RHS} and~\eqref{eq:energy_RHS} can be easily generalised to the linear-order perturbed  level with the  replacement $\bar{f}_{\nu H} \rightarrow F_{\nu H,0}$.

The second and third terms in \eqref{eq:number_conservation} --- and similarly the corresponding pairs in~\eqref{eq:energy_conservation} and \eqref{eq:momentum_conservation} --- are identical in the $m_{\nu l} = m_\phi = 0$ limit: they differ superficially only in the labelling of an integration variable.  We therefore express their sum as a multiple of  the second term.
Considering first the background part of equation~\eqref{eq:number_conservation}, we note that the double integral of this sum can be considerably simplified by a change of integration limits as follows:
\begin{equation}
\begin{aligned}
2\int \mathrm{d }q_2\,  q_2^2 \, \left( \frac{{\rm d} f_{\nu l}}{{\rm d} \tau}\right)_{C}^{(0)} 
&= 2\frac{a^2 m_{\nu H}}{\tau_0}   \int  \mathrm{d} q_2 \int \mathrm{d} q_1 \, \frac{q_1}{\epsilon_1} \, \bar{f}_{\nu H}(q_1) \, \Theta(q_1 - q_{1-}^{(\nu l)} ) \Theta( q_{1+}^{(\nu l)}-1 )\\\
&= 2\frac{a^2 m_{\nu H}}{\tau_0}   \int \mathrm{d} q_1 \, \frac{q_1}{\epsilon_1} \, \bar{f}_{\nu H}(q_1) \int \mathrm{d} q_2 \,
\Theta(q_2 - q_{2-}^{(\nu H)}) \Theta(q_{2+}^{(\nu H)} -q_2)
\\
&= 2\frac{a^2 m_{\nu H}}{\tau_0} \int^\infty_0 \mathrm{d}q_1 \, \frac{q_1^2}{\epsilon_1} \, \bar{f}_{\nu H}(q_1),   
\label{eq:number_RHSnul} 
\end{aligned}
\end{equation}
where, at the second equality, we have made use of the properties of the Heaviside step function to convert the integration limits on $q_1$ to limits on $q_2$ via equation~\eqref{eq:limitsq1q2}, 
and proceeded to evaluate the $q_2$-integral at the third equality.  Adding together equations~\eqref{eq:number_RHS}  and~\eqref{eq:number_RHSnul}
it is immediately clear that number density conservation~\eqref{eq:number_conservation} is well satisfied at the background level.  The proof of number density conservation at the linear-order perturbed level proceeds in the exactly same manner, but for the replacement $\bar{f}_{\nu H} \rightarrow F_{\nu H,0}$.

Similarly, the sum of the second and third terms of equation~\eqref{eq:energy_conservation} can be expressed as
\begin{equation}
\begin{aligned}
4 \int \mathrm{d} q_2\,  q_2^3 \, \left( \frac{{\rm d} f_{\nu l}}{{\rm d} \tau}\right)_{C}^{(0)} 
& = 4\frac{a^2 m_{\nu H}}{\tau_0}   \int \mathrm{d} q_1 \frac{q_1}{\epsilon_1} \bar{f}_{\nu H}(q_1) \int_{q_{2-}^{(\nu H)}}^{q_{2-}^{(\nu H)}} \mathrm{d} q_2 \, q_2 \\
&= 2 \frac{a^2 m_{\nu H}}{\tau_0} \int \mathrm{d}q_1 \, q_1^2 \, \bar{f}_{\nu H}(q_1) =   \frac{2 \pi^2 a^5 m_{\nu H}}{\tau_0} \bar{n}_{\nu H},
\end{aligned}
\label{eq:energy_prove}
\end{equation}
which, when added together with equation~\eqref{eq:energy_RHS}, demonstrates energy conservation~\eqref{eq:energy_conservation} at the background level, and we note again that the linear-order proof is identical, save for the replacement $\bar{f}_{\nu H} \rightarrow F_{\nu H,0}$.
Finally, rewriting the sum of the second and third terms of equations~\eqref{eq:momentum_conservation}  as
\begin{equation}
\begin{aligned}
4 \int \mathrm{d}q_2\,  q_2^3 \, \left( \frac{{\rm d} f_{\nu l}}{{\rm d} \tau}\right)_{C,\ell=1}^{(1)} 
 &= 4 \frac{a^2 m_{\nu H}}{\tau_0} \int \mathrm{d} q_1 \, \frac{q_1}{\epsilon_1} F_{\nu H,1}(q_1) \int_{q_{2+}^{(\nu H)}}^{q_{2-}^{(\nu H)}} \mathrm{d} q_2 \, q_2 P_1 \left( \frac{2 \epsilon_1 q_2-a^2m_{\nu H}^2}{2 q_1 q_2} \right) \\
&=2 \frac{a^2 m_{\nu H}}{\tau_0} \int \mathrm{d}q_1 \frac{q_1^3}{\epsilon_1} F_{\nu H,1}(q_1)
\end{aligned}
\label{eq:momentum_prove}
\end{equation} 
demonstrates momentum conservation in conjunction with equation~\eqref{eq:momentum_RHS}.

 %%%%%%%%%%%%%
 %%%%%%%%%%%

 \section{Non-relativistic decay into massless daughters}
 \label{app:nonrel}
 
 We derive in this appendix the collisional terms of equations~\eqref{eq:delta_1} and~\eqref{eq:theta_1}; the gravity and free-streaming terms of these equations are well-known and have been derived elsewhere (e.g.,~\cite{Ma:1995ey}).
 
 Beginning with equation~\eqref{eq:delta_2} we note that the collisional parts of the equations are equivalently
  \begin{eqnarray}
 \int \mathrm{d}q \, q^3 \left(2\dot{F}_{\nu l, 0}+ \dot{F}_{\phi, 0} \right) & = &  \cdots + 4 \int \mathrm{d} q\,  q^3 \, \left( \frac{{\rm d} f_{\nu l}}{{\rm d} \tau}\right)_{C,\ell=0}^{(1)} \simeq \cdots +   \frac{2 \pi^2 a^5}{\tau_0} \delta \rho_{\nu H},\\
\int \mathrm{d}q \, q^3 \left(2\dot{\bar{f}}_{\nu l}+\dot{\bar{f}}_{\phi} \right) & = &  4 \int \mathrm{d} q\,  q^3 \, \left( \frac{{\rm d} f_{\nu l}}{{\rm d} \tau}\right)_{C}^{(0)} \simeq +   \frac{2 \pi^2 a^5}{\tau_0} \bar{\rho}_{\nu H},
 \end{eqnarray}
following from the r.h.s.\ of equation~(\ref{eq:energy_prove}) along with the approximation $\rho_{\nu H} \simeq m_{\nu H} n_{\nu H}$ in the non-relativistic limit.  Similarly, the collisional part of equation~\eqref{eq:theta_2}
 can be deduced from the r.h.s.\ of equation~\eqref{eq:momentum_prove}, i.e.,
  \begin{equation}
  \int \mathrm{d}q \, q^3 \left(2 \dot{F}_{\nu l, 1}+ \dot{F}_{\phi, 1}\right)  =    \cdots +  4 \int \mathrm{d} q\,  q^3 \, \left( \frac{{\rm d} f_{\nu l}}{{\rm d} \tau}\right)_{C,\ell=1}^{(1)} \simeq \cdots + \frac{2 \pi^2 a^5}{k  \tau_0} \bar{\rho}_{\nu H} \theta_{\nu H},\\
 \end{equation}
 where we have used the definition~\eqref{eq:deltatheta} and  the approximations $ \epsilon_{\nu H} \simeq a m_{\nu H}$ and $\bar{P}_{\nu H} \simeq 0$.
 Then, noting that 
\begin{equation}
\int {\rm d} q \, q^3 \,\left( 2 \bar{f}_{\nu_l}+ \bar{f}_{\phi}  \right)= 2 \pi^2 a^4 \bar{\rho}_{\nu l+\phi},
\end{equation}
we immediately find for equation~\eqref{eq:delta_2}
\begin{equation}
\frac{\int \mathrm{d}q \, q^3 \left(2\dot{F}_{\nu l, 0}+ \dot{F}_{\phi, 0} \right)}{\int \mathrm{d}q \, q^3 \left(2\bar{f}_{\nu_l} +  \bar{f}_\phi\right)} - \frac{\int \mathrm{d}q \, q^3 \left(2\dot{\bar{f}}_{\nu l}+\dot{\bar{f}}_{\phi} \right)}{\int \mathrm{d}q \, q^3 \left(2\bar{f}_{\nu_l} +  \bar{f}_{\phi} \right)}  \delta_{\nu l + \phi} \simeq \cdots - \frac{a}{ \tau_0} \frac{\bar{\rho}_{\nu H}}{\bar{\rho}_{\nu l+\phi}} \left( \delta_{\nu l+\phi}-\delta_{\nu H} \right),
\end{equation}
and for equation~\eqref{eq:theta_2}
 \begin{equation}
  \frac{3k}{4}  \frac{\int \mathrm{d}q \, q^3 \left(2 \dot{F}_{\nu l, 1}+ \dot{F}_{\phi, 1}\right)}{\int \mathrm{d}q \, q^3 \left(2\bar{f}_{\nu_l}+  \bar{f}_{\phi} \right)} - \frac{\int \mathrm{d}q \, q^3 \left(2\dot{\bar{f}}_{\nu l}+\dot{\bar{f}}_{\phi} \right)}{\int \mathrm{d}q \, q^3 \left(2 \bar{f}_{\nu_l}+ \bar{f}_{\phi}\right)}  \theta_{\nu l + \phi} \simeq \cdots	- \frac{a}{ \tau_0} \frac{3 \bar{\rho}_{\nu H}}{4 \bar{\rho}_{\nu l+\phi}} \left( \frac{4}{3} \theta_{\nu l+\phi} - \theta_{\nu H}  \right).
 	\end{equation}
 These are the collisional terms of equations~\eqref{eq:delta_1} and~\eqref{eq:theta_1}.
 
Finally, if we wish to apply the non-relativistic limit to equation~\eqref{eq:momentum_prove},  this must be done \textit{after} integrating over the integral kernel, i.e., after the second line of equation~\eqref{eq:momentum_prove}:
 \begin{equation}
 \int \mathrm{d} q_2\,  q_2^3 \, \left( \frac{{\rm d} f_{\nu l}}{{\rm d} \tau}\right)_{C,\ell=1}^{(1)} \simeq  \frac{a}{2\tau_0} \int \mathrm{d}q_1 q_1^3 F_{\nu H,1} (q_1) = \frac{a^5}{2 \tau_0} \frac{1}{4 \pi k}\bar{\rho}_{\nu H} \theta_{\nu H} .
 \label{eq:momentum_nonrel}
 \end{equation}
  If instead the non-relativistic limit had been applied  directly to the interaction kernel in the first line of equation~\eqref{eq:momentum_prove}, we would have obtained  $P_1 \left(\frac{2\epsilon_1 q_2-a^2m_{\nu H}^2}{2 q_1 q_2} \right) \simeq 0$ (assuming $q_2 \simeq a m_{\nu H}/2$ as in equation~\eqref{eq:qprime}).  This would have led us to the erroneous conclusion that equation~\eqref{eq:momentum_nonrel} is vanishing in the non-relativistic limit, which in fact violates momentum conservation as well as gauge invariance.  This we believe is the source of error in reference~\cite{Kaplinghat:1999xy}.

%%%%%%%%%%%%%%%%%%%%%%%%%
%%%%%%%%%%%%%%%%%%%%%%%%%

\bibliographystyle{utcaps}
\bibliography{Literature}

%%%%%%%%%%%%%%%%%%%%%%%%%%%%%%%%%%%%%%%%%%%

%%%%%%%%%%%%%%%%%%%%%%%%%%%%%%%%%%%%%%%%%%%%%%%%%%%%%%%%%%%%%%%%%%%%%%%%%%%%%%%%%%%%%

\end{document}